\documentclass[english]{article}
\usepackage[LGR,T1]{fontenc}
\usepackage[latin9]{inputenc}
\usepackage{textcomp}
\usepackage{amsmath}
\usepackage{amssymb}
\usepackage{accents}
\usepackage{graphicx}

\makeatletter

\DeclareRobustCommand{\greektext}{%
  \fontencoding{LGR}\selectfont\def\encodingdefault{LGR}}
\DeclareRobustCommand{\textgreek}[1]{\leavevmode{\greektext #1}}
\ProvideTextCommand{\~}{LGR}[1]{\char126#1}

\newcommand{\docedilla}[2]{\underaccent{#1\mathchar'30}{#2}}
\newcommand{\cedilla}[1]{\mathpalette\docedilla{#1}}

\newcommand{\lyxaddress}[1]{
	\par {\raggedright #1
	\vspace{1.4em}
	\noindent\par}
}

\makeatother

\usepackage{babel}
\begin{document}
\title{Stochastic Lagrangians for Statistical Dynamics}
\author{Massimo Materassi}
\maketitle

\lyxaddress{Institute for Complex Systems of the National Research Council (CNR-ISC),
Florence, Italy.\\
E-mail: massimo.materassi@isc.cnr.it, massimomaterassi27@gmail.com.\\
Web: www.materassiphysics.com.}
\begin{abstract}
The concept of stochastic Lagrangian and its use in statistical dynamics
is illustrated theoretically, and with some examples.

Dynamical variables undergoing stochastic differential equations are
stochastic processes themselves, and their realization probability
functional within a given time interval arises from the interplay
between the deterministic parts of dynamics and noise statistics.
The stochastic Lagrangian is a tool to formulate realization probabilities
via functional integrals, once the statistics of noises involved in
the stochastic dynamical equations is known. In principle, it allows
to highlight the invariance properties of the statistical dynamics
of the system.

In this work, after a review of the stochastic Lagrangian formalism,
some applications of it to physically relevant cases are illustrated.\medskip{}


\tableofcontents{}

\newpage{}
\end{abstract}

\section{Introduction\label{sec:Introduction}}

After Isaac Newton's big work about ``the motion of bodies'' \cite{Newton.Principia},
it was understood that physical systems are generally governed by
\emph{equations of motion}, expressing the variability with time of
\emph{the state of the system} in terms of \emph{the state of what
acts on the system}. When one writes $\vec{F}=m\vec{a}$, the time
variability $\frac{d\vec{p}}{dt}$ of ``the state of motion'' of
the pointlike particle $\vec{p}$ (namely, its momentum) is put in
relationship with ``the force'' $\vec{F}$ as
\begin{equation}
\frac{d\vec{p}}{dt}=\vec{F},\label{eq:000}
\end{equation}
and this force \emph{is} a function of the state of what acts on the
system. As an example, think about Kepler's problem, in which the
point of mass $m$ at position $\vec{x}$ undergoes the gravitational
force exerted by the presence of another body of mass $M$ at position
$\vec{x}_{\odot}$, according to the law
\begin{equation}
\frac{d\vec{p}}{dt}=\frac{GMm}{\left|\vec{x}_{\odot}-\vec{x}\right|^{3}}\left(\vec{x}_{\odot}-\vec{x}\right):\label{eq:Kepler.F.ma}
\end{equation}
the state of what acts on the pointlike particle is described indeed
by the force $\vec{F}=\frac{GMm}{\left|\vec{x}_{\odot}-\vec{x}\right|^{3}}\left(\vec{x}_{\odot}-\vec{x}\right)$,
containing the value of the gravitational mass $M$ of the second
body and its relative position with respect to the point particle
of momentum $\vec{p}$, i.e. $\vec{x}_{\odot}-\vec{x}$. By the way,
notice that in the force the state of the system itself appears as
$\vec{x}$, rendering (\ref{eq:000}) and (\ref{eq:Kepler.F.ma})
proper differential equations for the state of the system $\psi=\left(\vec{x},\vec{p}\right)$.
As Newton's equation $\vec{F}=m\vec{a}$ is re-written considering
also the relationship between momentum and velocity, one writes
\begin{equation}
\begin{cases}
 & \dfrac{d\vec{x}}{dt}=\dfrac{\vec{p}}{m},\\
\\
 & \dfrac{d\vec{p}}{dt}=\vec{F}\left(\vec{x},\vec{p},t,X_{\mathrm{env}}\right):
\end{cases}\label{eq:F.ma.full}
\end{equation}
here it is stressed that the ``force'' is a function of the system
state, of time $t$ and of the ``state of the environment'' acting
on the system, indicated as $X_{\mathrm{env}}$. A system as (\ref{eq:F.ma.full})
is properly a system of \emph{ordinary differential equations} (ODE),
in which the variability of the state $\psi$, of the portion of universe
we are intersted in, is put in relationship with how this portion
of universe interacts with the environment. Similar systems of ODEs
may be written as:
\begin{equation}
\dfrac{d\psi}{dt}=\varphi\left(\psi,t,X_{\mathrm{env}}\right).\label{eq:deterministic.ODE.for.psi}
\end{equation}
In such a formulation, the system is described by a state $\psi\in\mathbb{V}$,
where $\mathbb{V}$ is a certain mathematical ambient through which
the state $\psi$ moves as time flows; $\mathbb{V}$ is referred to
as the \emph{phase space} of the system.

Calculus teaches us that when ODEs as (\ref{eq:deterministic.ODE.for.psi})
are equipped with some \emph{initial condition }$\psi\left(t_{0}\right)=\psi_{0}$,
the problem
\begin{equation}
\begin{cases}
 & \dfrac{d\psi}{dt}=\varphi\left(\psi,t,X_{\mathrm{env}}\right),\\
\\
 & \psi\left(t_{0}\right)=\psi_{0}
\end{cases}\label{eq:Cauchy.Problem}
\end{equation}
admits a unique solution $\psi\left(t\right)=\Psi_{\varphi}\left(t,\psi_{0}\right)$
for all $t\ge t_{0}$. To be honest, this happens only if the Cauchy
Problem (\ref{eq:Cauchy.Problem}) has the expression $\varphi\left(\psi,t,X_{\mathrm{env}}\right)$
that has particularly favorable conditions with respect to $\psi$,
which we assume to happen ``always'' (however, see \cite{Cystoseira.2019}
for a not-that-problematic counterexample). From here on, the statement
\begin{equation}
(\ref{eq:Cauchy.Problem})\thinspace\implies\thinspace\psi\left(t\right)=\Psi_{\varphi}\left(t,\psi\left(t_{0}\right)\right)\ \forall\ t\ge t_{0}\label{eq:evolution.map}
\end{equation}
means that there exists a (suitably regular) map $\Psi_{\varphi}:\mathbb{R}\times\mathbb{V}\mapsto\mathbb{V}$
depending on the dynamics $\varphi$, associating the state $\psi\left(t\right)$
to the initial condition $\psi\left(t_{0}\right)$ in an injective
way. This map $\Psi_{\varphi}$ is what one calls \emph{evolution}.
As a note, let us introduce here the \emph{system velocity space}
$\mathbb{W}$, or dynamical flow space, so that $\varphi\left(\psi,t,X_{\mathrm{env}}\right)\in\mathbb{W}$
and $\varphi:\mathbb{V}\times\mathbb{R}\times\mathbb{V}_{\mathrm{env}}\mapsto\mathbb{W}$
(here $\mathbb{V}_{\mathrm{env}}$ is the phase space of the environment
acting on the system).

\emph{Uniqueness} of the solution of initial value problems (\ref{eq:Cauchy.Problem})
nourishes the \emph{Deterministic Paradigm} (DP), according to which
once a system's initial conditions are given, its future history will
be completely determined, as long as what acts on it is known for
all the future times. This is strongly accepted all through the Classical
Physics, and after all it is true also in Quantum Mechanics \cite{Diracs.book},
just considering the ODE (\ref{eq:deterministic.ODE.for.psi}) to
be Schrödinger equation $i\hbar\frac{d}{dt}\left|\psi\right\rangle =H\left|\psi\right\rangle $,
according to which the motion of the quantum state $\left|\psi\right\rangle $
is a perfectly deterministic trajectory through the quantum state
Hilbert space $\mathbb{H}_{S}$.

The intelligent criticism to the DP (that actually turns out to be
a generalization of the DP itself) must be based on the observation
that, in order for (\ref{eq:Cauchy.Problem}) to have a unique solution,
the quantities $X_{\mathrm{env}}$ and $\psi_{0}$ appearing there,
and the full mathematical construction of $\varphi\left(\psi,t,X_{\mathrm{env}}\right)$,
\emph{must be known perfectly}, i.e. with \emph{no uncertainty}.

For instance, as the initial condition $\psi_{0}$ is not known perfectly,
instead one knows just $\psi_{0}$ to belong to some subset $A_{0}\subseteq\mathbb{V}$
of the phase space $\mathbb{V}$, as
\begin{equation}
\begin{cases}
 & \dfrac{d\psi}{dt}=\varphi\left(\psi,t,X_{\mathrm{env}}\right),\\
\\
 & \psi\left(t_{0}\right)\in A_{0},
\end{cases}\label{eq:coarse.grained.initial.conditions}
\end{equation}
one must admit the state of the system at time $t\ge t_{0}$ to be
\emph{any state} $\Psi\left(t,\psi\left(t_{0}\right)\right)$ for
\emph{any} $\psi\left(t_{0}\right)\in A_{0}$, i.e.:
\[
(\ref{eq:coarse.grained.initial.conditions})\thinspace\implies\thinspace\psi\left(t\right)\in\Psi_{\varphi}\left(t,A_{0}\right)\ \forall\ t\ge t_{0},
\]
being $A_{\ensuremath{0}}$ a finite size set\footnote{Stating anything about the ``size'' of a set in $\mathbb{V}$ has
not sense until a proper definition of ``size'', or better ``measure'',
is defined on $\mathbb{V}$, which hasn't been done, and won't be
done, here. Of course, if $\mathbb{V}$ is a metric space, the built-in
distance $d:\mathbb{V}\times\mathbb{V}\mapsto\mathbb{R}^{+}$ may
be used to define the size of $A\subseteq\mathbb{V}$, as $\ell\left(A\right)=\max\left(d\left(x,y\right)\right)\thinspace/\thinspace x,y\in A$,
and this is also true if a measure is defined on $\mathbb{V}$, as
when it is treated as the sample space of some probability. In the
latter case, if one is able to define a probability density on $\mathbb{V}$,
a physically useful measure of $A$ could be the Shannon entropy associated
to this probability relative to all the points in $A$: again, this
is beyond the scope of the present exposition.}, and so the set $\Psi_{\varphi}\left(t,A_{0}\right)$. Put in a simpler
way, with a coarse grained knowledge of the initial condition as $\psi\left(t_{0}\right)\in A_{0}$
instead of the sharp $\psi\left(t_{0}\right)=\psi_{0}$, one has to
get content with a coarse grained knowledge of the state $\psi\left(t\right)$
at later times. The uncertainty in the initial conditions will render
uncertain the evolution of the system, setting a natural limit to
the DP due to our finite precision and to how fast a finite size initial
condition set $A_{0}$ may be deformed by the dynamics in (\ref{eq:Cauchy.Problem}):
the whole \emph{querelle} between the DP and \emph{chaos theory} comes
precisely from the capacity of non-linear dynamics to deform the initial
condition set $A_{0}$ and diffuse it all over extended and complicated
regions of the phase space, rendering the evolution unpredictable
to an arbitrarily high precision.

In this work we will examine a different ``limit of the DP'': namely,
we will deal with what happens when some elements of the mathematical
expression of $\varphi$ are known \emph{only to some statistical
extent}, i.e. when this mathematical function of $\psi$, $t$ and
$X_{\mathrm{env}}$ has terms whose exact values is unknown, and of
which one can only state they appear according to some given \emph{probability
distribution}. As a simple example, consider $\psi$ to be a real
variable, and the dynamics $\varphi$ to read
\begin{equation}
\varphi=K\psi^{2}+\gamma\left(t\right)\ /\ \gamma\left(t\right)\sim P_{t}:\Xi\mapsto\mathbb{R}^{+},\ \int_{\Xi}\mathcal{P}_{t}\left(\gamma\right)d\gamma=1,\ \forall\thinspace t>t_{0},\label{eq:Phi.simple.example}
\end{equation}
where $K$ is a coefficient, and the time-dependent term $\gamma\left(t\right)$
is an addendum that is ``extracted from some real set $\Xi$'' at
each $t>t_{0}$, according to the probability distribution function
$P_{t}$. A situation as that written in (\ref{eq:Phi.simple.example})
represents the case in which the dynamics $\varphi$ is obtained,
at each time, by a first ``completely known'' term $K\psi^{2}$
plus some term $\gamma$ about which one only knows that it can be
a value within the set $\Xi$: as nothing sharper can be stated on
$\gamma$, \emph{the value $\gamma\left(t\right)$ comes randomly
within} $\Xi$ at each different time, with a probability $P_{t}\left(\gamma^{*}\right)\delta$
to fall in any interval $\left[\gamma^{*}-\frac{\delta}{2},\gamma^{*}+\frac{\delta}{2}\right]$.

Terms as the $\gamma\left(t\right)$ in (\ref{eq:Phi.simple.example})
are referred to as \emph{stochastic terms} or, more simply, \emph{noises}.
In general, ``noise terms'' may appear in a variety of ways in the
dynamics $\varphi$, due to different ``physical'' reasons: typically,
as one distinguishes the dynamical variables $\psi$ assigning the
state of the system from ``everything else'' encoded in $X_{\mathrm{env}}$,
noise terms will sensibly describe the degree of uncertainty about
$X_{\mathrm{env}}$, which is the other possible source of uncertainty
in (\ref{eq:Cauchy.Problem}), next to the initial conditions. For
instance, in (\ref{eq:Phi.simple.example}) one might imagine that
$\psi$ self-evolves with $K\psi^{2}$, and undergoes the action of
the ``random kicks'' $\gamma$.

Considering that noise terms typically come from what one refers to
as ``environment'', the general form of (\ref{eq:Phi.simple.example})
may read:
\begin{equation}
\varphi=\varphi\left(\psi,t,X_{\mathrm{env}}\left(\gamma\left(t\right)\right)\right)\ /\ \gamma\left(t\right)\sim P_{t}:\Xi\mapsto\mathbb{R}^{+},\ \int_{\Xi}P_{t}\left(\gamma\right)d\gamma=1,\ \forall\thinspace t>t_{0}.\label{eq:001.Phi.of.gamma}
\end{equation}
In the presence of noise terms, the equation (\ref{eq:deterministic.ODE.for.psi})
is named \emph{stochastic dynamical equation} (SDE). The stochastic
version of (\ref{eq:Cauchy.Problem}) will read:
\begin{equation}
\begin{cases}
 & \dfrac{d\psi}{dt}=\varphi\left(\psi,t,X_{\mathrm{env}}\left(\gamma\left(t\right)\right)\right),\\
\\
 & \psi\left(t_{0}\right)=\psi_{0},\\
\\
 & \gamma\left(t\right)\sim P_{t}:\Xi\mapsto\mathbb{R}^{+}.
\end{cases}\label{eq:Cauchy.problem.with.SDE}
\end{equation}

\begin{figure}
\centering{}\includegraphics[scale=0.06]{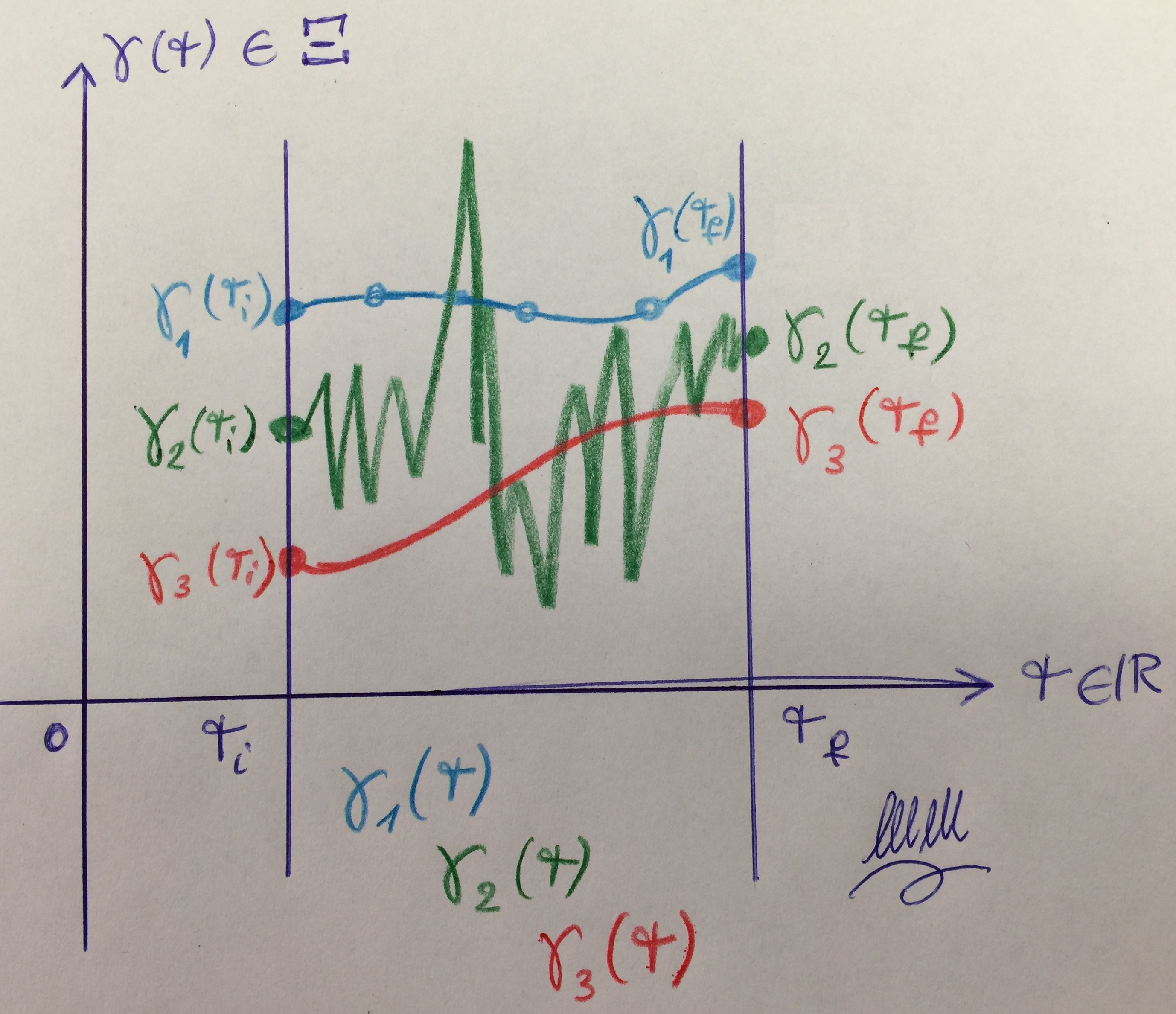}\caption{\label{fig:gamma.histories}Three possible realizations of the noise
$\gamma\left(t\right)$ between $t_{\mathrm{i}}$ and $t_{\mathrm{f}}$:
as $\gamma\left(t\right)$ can be any possible continuous curve in
$\Xi$, here we have drawn two rather regular curves $\gamma_{1}\left(t\right)$
and $\gamma_{3}\left(t\right)$, and a less regular, sharp-cornered
one $\gamma_{2}\left(t\right)$.}
\end{figure}

\begin{figure}
\centering{}\includegraphics[scale=0.06]{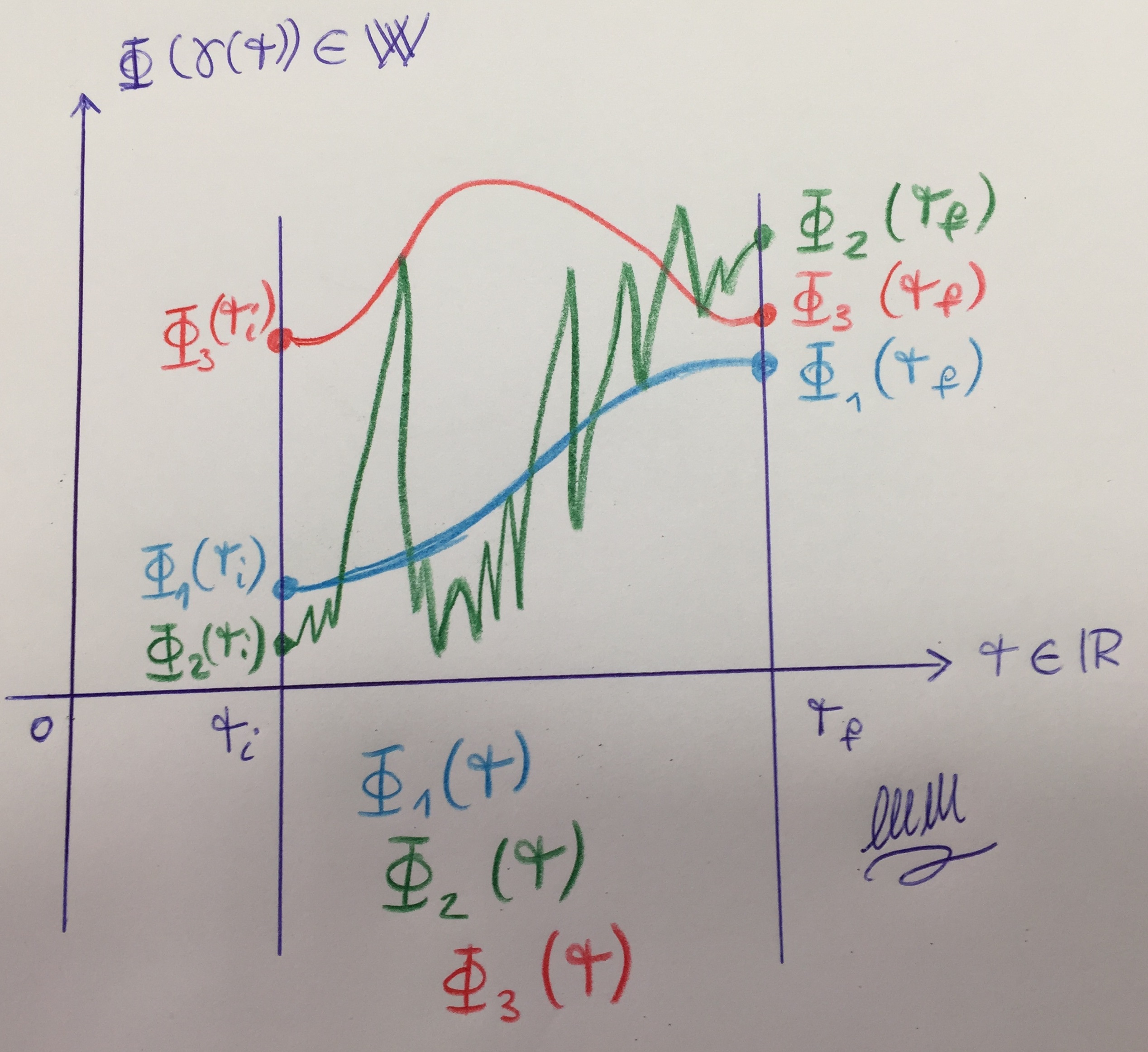}\caption{\label{fig:Phi.histories}The cartoon of the three realizations of
the process $\varphi\left(\psi,t,X_{\mathrm{env}}\left(\gamma\left(t\right)\right)\right)$
corresponding to the realizations $\gamma_{1}\left(t\right)$, $\gamma_{2}\left(t\right)$
and $\gamma_{3}\left(t\right)$ cited in Figure \ref{fig:gamma.histories}.}
\end{figure}

\begin{figure}
\centering{}\includegraphics[scale=0.06]{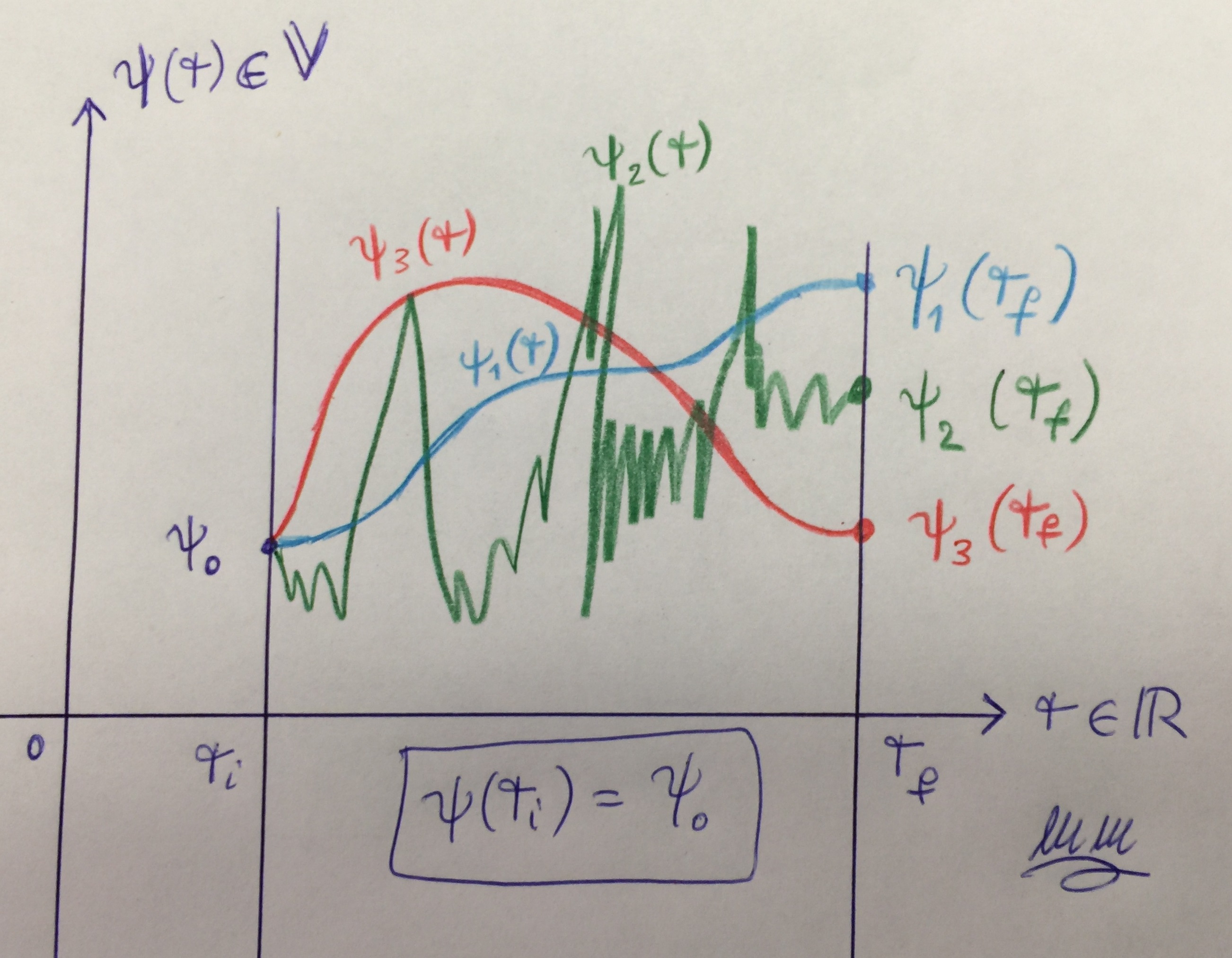}\caption{\label{fig:Psi.histories}The cartoon of the three solutions $\psi_{i=1,2,3}\left(t\right)$
of a stochastic Cauchy problem as (\ref{eq:Cauchy.problem.with.SDE})
corresponding to the realizations $\gamma_{1}\left(t\right)$, $\gamma_{2}\left(t\right)$
and $\gamma_{3}\left(t\right)$ in Figure \ref{fig:gamma.histories}.}
\end{figure}

The most relevant fact passing from (\ref{eq:Cauchy.Problem}) to
(\ref{eq:Cauchy.problem.with.SDE}) is that \emph{the uniqueness of
the solution is lost}, even in the presence of perfect knowledge of
the initial condition $\psi\left(t_{0}\right)=\psi_{0}$. Indeed,
depending on what point in $\Xi$ is picked at each time $t$ to play
the role of $\gamma\left(t\right)$, one has \emph{different possible
curves}
\[
\gamma=\gamma\left(t\right),\ \gamma\in\mathcal{C}\left(\mathbb{R},\Xi\right),
\]
hence \emph{a different ``histories'' }$\varphi\left(\psi,t,X_{\mathrm{env}}\left(\gamma\left(t\right)\right)\right)$
of the dynamics of the system. Since the noises $\gamma$ may describe
\emph{any} continuous trajectory for $t\in\left[t_{\mathrm{i}},t_{\mathrm{f}}\right]$,
as illustrated in the cartoon of Figure \ref{fig:gamma.histories},
also the corresponding dynamics $\varphi\left(\psi,t,X_{\mathrm{env}}\left(\gamma\left(t\right)\right)\right)$
will have \emph{any} continuous shape as a curve in $\mathbb{W}$,
see Figure \ref{fig:Phi.histories}. This explains in a pictorial
way the fail of uniqueness of the solution to the stochastic Cauchy
problem (\ref{eq:Cauchy.problem.with.SDE}): the randomness transits
from noises $\gamma$ to dynamics $\varphi$, and from dynamics to
the solution of the Cauchy problem $\psi\left(t\right)$, see Figure
\ref{fig:Psi.histories} for the final step of the pictorial explanation.
Note that, even if all of the three solutions to (\ref{eq:Cauchy.problem.with.SDE})
start at the same initial value, they develop very differently and
the uniqueness invoked by DP is lost.

The random nature of each term $\gamma\left(t\right)$ in (\ref{eq:001.Phi.of.gamma})
and (\ref{eq:Cauchy.problem.with.SDE}) renders it useful to consider
the sample space $\Sigma\left(t_{\mathrm{i}},t_{\mathrm{f}}\right)$
of all the possible continuous trajectories $\gamma\left(t\right)$
in $\Xi$ with $t\in\left[t_{\mathrm{i}},t_{\mathrm{f}}\right]$,
i.e. the sample space of all the possible realizations of the stochastic
process $\gamma:\mathbb{R}\mapsto\Xi$. The set $\Sigma\left(t_{\mathrm{i}},t_{\mathrm{f}}\right)$
is infinite dimensional, namely $\Sigma\left(t_{\mathrm{i}},t_{\mathrm{f}}\right)\subseteq\mathcal{C}\left(\mathbb{R},\Xi\right)$;
a probability measure defined on it, should be possibly based upon
the functional counterpart of what an ordinary probability distribution
function is for finite dimensional sample spaces, $\mathcal{P}\left[\gamma;t_{\mathrm{i}},t_{\mathrm{f}}\right)$.
In this script
\[
\mathcal{P}\left[\gamma;t_{\mathrm{i}},t_{\mathrm{f}}\right),
\]
the square bracket left to $\gamma$ means that $\mathcal{P}$ depends
on the infinite number of values $\gamma\left(t\right)$ (one per
each $t\in\left[t_{\mathrm{i}},t_{\mathrm{f}}\right]$), while the
round bracket right after $t_{\mathrm{f}}$ indicates that $\mathcal{P}$
depends also on the two real variables $t_{\mathrm{i}}$ and $t_{\mathrm{f}}$.
The $\mathcal{P}\left[\gamma;t_{\mathrm{i}},t_{\mathrm{f}}\right)$
is referred to as \emph{realization probability functional} (RPF)
of $\gamma$, and its normalization condition reads:
\begin{equation}
\int_{\Sigma\left(t_{\mathrm{i}},t_{\mathrm{f}}\right)}\mathcal{P}\left[\gamma;t_{\mathrm{i}},t_{\mathrm{f}}\right)\left[d\gamma\right]=1.\label{eq:RPF.normalization.condition}
\end{equation}
In this (\ref{eq:RPF.normalization.condition}) the symbol $\left[d\gamma\right]$
indicates the \emph{functional integral on }$\Sigma\left(t_{\mathrm{i}},t_{\mathrm{f}}\right)$
of the stochastic terms, namely the continuous product:
\[
\int_{\Sigma\left(t_{\mathrm{i}},t_{\mathrm{f}}\right)}\left[d\gamma\right]...=\prod_{t\in\left[t_{\mathrm{i}},t_{\mathrm{f}}\right]}\int_{\Xi}d\gamma\left(t\right)...
\]
(see \cite{Mate.SESF.2019} for a thorough explanation of this). About
the RPF $\mathcal{P}\left[\gamma;t_{\mathrm{i}},t_{\mathrm{f}}\right)$,
it must be stressed that this will be expressable as the continuous
product of all the $P_{t}\left(\gamma\right)^{dt}$ (see (\ref{eq:001.Phi.of.gamma})
and (\ref{eq:Cauchy.problem.with.SDE})), for $t\in\left[t_{\mathrm{i}},t_{\mathrm{f}}\right]$,
\emph{only for time-$\delta$-correlated noises}, see § \ref{subsec:delta.correlated}:
assigning the functional $\mathcal{P}\left[\gamma;t_{\mathrm{i}},t_{\mathrm{f}}\right)$
is then more powerful than giving all the time-local PDFs $P_{t}\left(\gamma\right)$,
because $\mathcal{P}\left[\gamma;t_{\mathrm{i}},t_{\mathrm{f}}\right)$
contains \emph{any} type of correlation among noises, that the collection
of the PDFs does not, and indeed the latter is equivalent to the RPF
only for the $\delta$-correlated stochastic variables. In fully mathematical
terms, one could write:
\[
P_{t}\left(\gamma\right)=\int_{t'\neq t}\left[d\gamma\left(t'\right)\right]\mathcal{P}\left[\gamma;t_{\mathrm{i}},t_{\mathrm{f}}\right)\ \forall\ t\in\left[t_{\mathrm{i}},t_{\mathrm{f}}\right].
\]

In what follows, the program is to start with some particular form
of the SDE in (\ref{eq:Cauchy.problem.with.SDE}), in which one will
understand how to pass from the ensemble statistics of noises $\gamma\left(t\right)$
to that of the system realizations $\psi\left(t\right)$, considering
the interplay between the deterministic and the stochastic features
of $\varphi$:
\begin{equation}
\mathcal{P}\left[\gamma;t_{\mathrm{i}},t_{\mathrm{f}}\right)\overset{\varphi}{\mapsto}\mathcal{A}\left[\psi;t_{\mathrm{i}},t_{\mathrm{f}}\right),\label{eq:P2A.thanks.Phi}
\end{equation}
where $\mathcal{A}\left[\psi;t_{\mathrm{i}},t_{\mathrm{f}}\right)$
is the RPF of the stochastic process $\psi:\left[t_{\mathrm{i}},t_{\mathrm{f}}\right]\mapsto\mathbb{V}$,
while the script $\overset{\varphi}{\mapsto}$ means that this pass
is done thanks to the form of $\varphi$.

For the types of SDE we are going to work with, the scheme to perform
the transition (\ref{eq:P2A.thanks.Phi}) includes the introduction
of the concept of \emph{stochastic Lagrangian} (SL), that will be
introduced in § \ref{sec:FFSD}. Then, in § \ref{sec:Examples} the
machinery will be applied to some particular case of physical relevance.

\section{Functional Formalism for Statistical Dynamics\label{sec:FFSD}}

The mathematical quantity $\psi$, that describes the state of the
system, is in general given by a certain number of components, indicated
as $\psi^{I}$, so locally $\mathbb{V}$ will be described as some
$\mathbb{R}^{n}$ (or $\mathbb{C}^{n}$) space, as $I=1,...,n$. Remarkably,
an infinite dimensional state may well be necessary, e.g. in a classical
continuum theory, in which the system is described by the local properties
of the continuum, so that the index $I$ of $\psi$ will rather be
some continuous position $\vec{x}$ in the three dimensional space.
The functional formalism is described here for a general \emph{finite
dimensional} system, while in \cite{MateConso.SMHD,Mate.Marsiglia.SMHD.2D,Mate.Conso.Tetrad.01,Mate.Conso.Tetrad.02}
an infinite dimensional system, namely magneto-hydrodynamics, is considered.

Let's go back to the SDE in (\ref{eq:Cauchy.problem.with.SDE}), and
consider a particular class of functions $\varphi\left(\psi,t,X_{\mathrm{env}}\left(\gamma\left(t\right)\right)\right)$
of the noises $\gamma$. Let us assume that the noises enter the dynamical
equation (DE) of $\psi$ both in an additive form, as (\ref{eq:Phi.simple.example}),
and in a multiplicative way. In particular, as we assume the phase
space $\mathbb{V}$ to be $n$-dimensional, let us consider \emph{two
stochastic $n$-vectors}, one of components $f^{I}$ and the other
of components $g_{J}$, whose statistics is assigned through their
RPF $\mathcal{P}\left[f,g;t_{\mathrm{i}},t_{\mathrm{f}}\right)$,
and appearing in the SDE of $\psi$ as:
\begin{equation}
\dfrac{d\psi^{I}}{dt}=\Lambda\left(\psi\right)+g_{J}\left(t\right)\Gamma^{JI}\left(\psi\right)+f^{I}\left(t\right).\label{eq:Mate.000.DE.psi}
\end{equation}
In (\ref{eq:Mate.000.DE.psi}) one has the co-existence of deterministic
expressions, depending on $\psi$
\[
\Lambda\in\mathcal{C}^{\infty}\left(\mathbb{V},\mathrm{T}\left(\mathbb{V}\right)\right),\ \Gamma\in\mathcal{C}^{\infty}\left(\mathbb{V},\mathrm{T}\left(\mathbb{V}\right)\otimes\mathrm{T}\left(\mathbb{V}\right)\right),
\]
with that of the \emph{noise terms }$f$ and $g$, of which one supposes
to know the statistics via $\mathcal{P}\left[f,g;t_{\mathrm{i}},t_{\mathrm{f}}\right)$.

Equation (\ref{eq:Mate.000.DE.psi}) might appear rather particular
and simple, as a form of SDE: it is clearly more complicated than
the simplest case (\ref{eq:Phi.simple.example}), but one may well
think more involuted expressions, as for instance
\[
\dfrac{d\psi}{dt}=\Lambda\left(\psi\right)+G\exp\left(\frac{A\psi^{2}+\gamma\left(t\right)}{B\cos\left(\gamma\left(t\right)\right)}\right)...,
\]
being $\mathbb{V}=\mathbb{R}$ for simplicity. The point is that:
\begin{itemize}
\item the ``simple'' form (\ref{eq:Mate.000.DE.psi}) is the one allowing
for the formalism of stochastic Lagrangian, that is of interest here;
\item however, as well explained in \cite{MateConso.SMHD,Mate.Marsiglia.SMHD.2D,Mate.Conso.Tetrad.01,Mate.Conso.Tetrad.02},
and in \cite{Chang.FSOC}, important realistic systems as space plasmas
may be described by (\ref{eq:Mate.000.DE.psi}).
\end{itemize}
Our work will then concentrate on the stochastic Cauchy problem:
\begin{equation}
\begin{cases}
 & \dfrac{d\psi^{I}}{dt}=\Lambda\left(\psi\right)+g_{J}\left(t\right)\Gamma^{JI}\left(\psi\right)+f^{I}\left(t\right),\\
\\
 & \psi\left(t_{\mathrm{i}}\right)=\psi_{\mathrm{i}},\\
\\
 & \left(f\left(t\right),g\left(t\right)\right)\sim\mathcal{P}\left[f,g;t_{\mathrm{i}},t_{\mathrm{f}}\right):\thinspace\Sigma\left(t_{\mathrm{i}},t_{\mathrm{f}}\right)\mapsto\mathbb{R}^{+}.
\end{cases}\label{eq:Mate.001.Stochastic.Cauchy.Problem}
\end{equation}
As the paper \cite{Phythian.77} by Phythian is the first paper introducing
this formalism, we will refer to (\ref{eq:Mate.000.DE.psi}) as Langevin-Phythian
Equation (LPE) (as far as the Author is aware of, the formalism of
the stochastic Lagrangian\footnote{Please note that the binomial ``stochastic Lagrangian'' is used,
in the fluid dynamics literature, as a couple of adjectives characterizing
the Lagrangian, i.e. material, description of a fluid in the presence
of stochastic terms, while here ``stochastic Lagrangian'' is not
a couple of adjectives, because they mean ``the Lagrangian function
of a stochastic theory'', hence ``stochastic'' is the attribution
of the noun ``Lagrangian''. Indeed, the term ``Lagrangian'' means
the same thing as in Quantum Field Theory, not in Fluid Dynamics!} has been introduced in \cite{Phythian.77} and in \cite{Jouvet.Phythian}).

Before going ahead, notice that in the SDE (\ref{eq:Mate.000.DE.psi}),
and in the system (\ref{eq:Mate.001.Stochastic.Cauchy.Problem}),
noises seem to appear ``directly'' without the mediation of the
environmental variables $X_{\mathrm{env}}$ in (\ref{eq:Cauchy.problem.with.SDE}):
this is only an appearance, as the reader may get convinced of going
through \cite{MateConso.SMHD}. In that case, e.g., noise terms are
identified with the terms $-\vec{\partial}\times\left(\zeta\cdot\vec{J}\right)$,
$\frac{\vec{J}}{\rho}$ and $-\frac{\vec{\partial}p}{\rho}$, being
$\zeta$ the plasma conductivity tensor, $\vec{J}$ the electric current,
$\rho$ the plasma mass density and $p$ the plasma pressure: as the
state of the system $\psi=\left(\vec{V},\vec{B}\right)$ included,
there, the plasma bulk velocity and the magnetic induction vector,
the quantities $\zeta$, $\vec{J}$, $\rho$ and $p$ could be understood
as ``environmental variables'' forcing $\psi$. Following this suggestion,
one forms noises with quantities determined by the microscopic nature
of the continuum, the \emph{microscopic stochastically treated degrees
of freedom} (\textgreek{m}STDoF, see also \cite{Mate.Metriplectic.Entropy}
for this concept) of which are the ``environment'' for the otherwise
isolated fluid variable system.

The dynamics governed by conditions (\ref{eq:Mate.001.Stochastic.Cauchy.Problem})
will produce \emph{a multi-history evolution} for $\psi$: as pictorially
indicated in Figure \ref{fig:Psi.histories}, the system history between
$t_{\mathrm{i}}$ and $t_{\mathrm{f}}$ is ``statistically distributed''
along many elements of $\Sigma\left(t_{\mathrm{i}},t_{\mathrm{f}}\right)$,
according to the RPF $\mathcal{A}\left[\psi;t_{\mathrm{i}},t_{\mathrm{f}}\right)$
that will depend on the noise RPF $\mathcal{P}\left[f,g;t_{\mathrm{i}},t_{\mathrm{f}}\right)$
via
\begin{equation}
\mathcal{P}\left[f,g;t_{\mathrm{i}},t_{\mathrm{f}}\right)\overset{\Lambda,\Gamma}{\mapsto}\mathcal{A}\left[\psi;t_{\mathrm{i}},t_{\mathrm{f}}\right),\label{eq:Mate.002.P2A}
\end{equation}
that is the version of (\ref{eq:P2A.thanks.Phi}) adapted to the dynamics
(\ref{eq:Mate.001.Stochastic.Cauchy.Problem}). The theoretical program
we want to pursue here is to calculate the map just mimicked in (\ref{eq:Mate.002.P2A}),
i.e. to obtain a (closed as possible) expression of the RPF of $\psi$
from that of the RPF of the noises $f$ and $g$, and the deterministic
parts $\Lambda$ and $\Gamma$ of (\ref{eq:Mate.001.Stochastic.Cauchy.Problem})
(precisely, the \emph{competition between chance and necessity} about
which Haken speaks in his book \cite{Haken.Synergetics.I} about ``Synergetics'').

Once the RPF $\mathcal{A}\left[\psi;t_{\mathrm{i}},t_{\mathrm{f}}\right)$
is given, the following program may be realized:
\begin{enumerate}
\item evaluate any statistical quantity $\left\langle F\right\rangle $,
for any functional $F\left[\psi\right]$, on the ensemble $\Sigma\left(t_{\mathrm{i}},t_{\mathrm{f}}\right)$
of trajectories through $\mathbb{V}$ admitted for the dynamics (\ref{eq:Mate.001.Stochastic.Cauchy.Problem});\label{enu:esamble-statistics-any-F}
\item calculate the transition probability $\mathcal{P}_{\psi_{\mathrm{i}}\mapsto\psi_{\mathrm{f}}}\left(t_{\mathrm{i}},t_{\mathrm{f}}\right)$
for the system from an initial condition $\psi_{\mathrm{i}}=\psi\left(t_{\mathrm{i}}\right)$
and a final one $\psi_{\mathrm{f}}=\psi\left(t_{\mathrm{f}}\right)$.\label{enu:transition-probability}
\end{enumerate}
Both any $\left\langle F\right\rangle $ and the transition probability
$\mathcal{P}_{\psi_{\mathrm{i}}\mapsto\psi_{\mathrm{f}}}\left(t_{\mathrm{i}},t_{\mathrm{f}}\right)$
may be expressed in terms of $\mathcal{A}\left[\psi;t_{\mathrm{i}},t_{\mathrm{f}}\right)$.
The quantity $\left\langle F\right\rangle $ in Point \ref{enu:esamble-statistics-any-F}
is calculated as
\begin{equation}
\left\langle F\right\rangle =\int_{C\left(\left[t_{\mathrm{i}},t_{\mathrm{f}}\right],\mathbb{V}\right)}\left[d\psi\right]F\left[\psi\right]\mathcal{A}\left[\psi;t_{\mathrm{i}},t_{\mathrm{f}}\right),\label{eq:Mate.003.average.F}
\end{equation}
where it is intended that $F\left[\psi\right]$ depends on the whole
trajectory $\psi\left(t\right)$ for $t\in\left[t_{\mathrm{i}},t_{\mathrm{f}}\right]$:
actually, the prescription (\ref{eq:Mate.003.average.F}) may work
also for a time-local \emph{function} $F\left(\psi\right)$, that
can always be expressed as an integration with the presence of some
$\delta\left(t-\hat{t}\right)$, provided $\hat{t}$ is the instant
of interest to calculate the function.

The transition probability $\mathcal{P}_{\psi_{\mathrm{i}}\mapsto\psi_{\mathrm{f}}}\left(t_{\mathrm{i}},t_{\mathrm{f}}\right)$
may be calculated as the integration of $\mathcal{A}\left[\psi;t_{\mathrm{i}},t_{\mathrm{f}}\right)$
on all the possible configurations $\psi\left(t\right)\thinspace/\thinspace t\in\left(t_{\mathrm{i}},t_{\mathrm{f}}\right)$,
while the initial and final configurations $\psi\left(t_{\mathrm{i}}\right)$
and $\psi\left(t_{\mathrm{f}}\right)$ are not integrated on, and
kept fixed to the values of interest $\psi_{\mathrm{i}}$ and $\psi_{\mathrm{f}}$
respectively. One may write
\begin{equation}
\mathcal{P}_{\psi_{\mathrm{i}}\mapsto\psi_{\mathrm{f}}}\left(t_{\mathrm{i}},t_{\mathrm{f}}\right)=\lim_{\epsilon\rightarrow0}\int_{C\left(\left[t_{\mathrm{i}}+\epsilon,t_{\mathrm{f}}-\epsilon\right],\mathbb{V}\right)}\left[d\psi\right]\left.\mathcal{A}\left[\psi;t_{\mathrm{i}},t_{\mathrm{f}}\right)\right|_{\psi\left(t_{\mathrm{i}}\right)=\psi_{\mathrm{i}},\thinspace\psi\left(t_{\mathrm{f}}\right)=\psi_{\mathrm{f}}}.\label{eq:Mate.004.transition.probability.1}
\end{equation}
This may be also written as
\begin{equation}
\mathcal{P}_{\psi_{\mathrm{i}}\mapsto\psi_{\mathrm{f}}}\left(t_{\mathrm{i}},t_{\mathrm{f}}\right)=\left.\left\langle 1\right\rangle \right|_{\psi\left(t_{\mathrm{i}}\right)=\psi_{\mathrm{i}},\thinspace\psi\left(t_{\mathrm{f}}\right)=\psi_{\mathrm{f}}}.\label{eq:Mate.005.transition.probability.2}
\end{equation}

\subsection{Stochastic Lagrangian\label{subsec:Stochastic-Lagrangian}}

The program stated in Points \ref{enu:esamble-statistics-any-F} and
\ref{enu:transition-probability} before needs the knowledge of $\mathcal{A}\left[\psi;t_{\mathrm{i}},t_{\mathrm{f}}\right)$:
in this work the RPF of $\psi$ is represented through a time-local
function $\mathcal{\mkern2mu\mathchar'40\mkern-7mu L}\left(\frac{d\psi}{dt},\psi\right)$,
so that
\begin{equation}
\begin{cases}
 & \mathcal{A}\left[\psi;t_{\mathrm{i}},t_{\mathrm{f}}\right)=N_{0}\left(t_{\mathrm{i}},t_{\mathrm{f}}\right)\exp\left[-i\int_{t_{\mathrm{i}}}^{t_{\mathrm{f}}}\mathcal{\mkern2mu\mathchar'40\mkern-7mu L}\left(\frac{d\psi}{dt}\left(t\right),\psi\left(t\right)\right)dt\right],\\
\\
 & \dfrac{1}{N_{0}\left(t_{\mathrm{i}},t_{\mathrm{f}}\right)}\overset{\mathrm{def}}{=}\int_{C\left(\left[t_{\mathrm{i}},t_{\mathrm{f}}\right],\mathbb{V}\right)}\left[d\psi\right]e^{-i\int_{t_{\mathrm{i}}}^{t_{\mathrm{f}}}\mathcal{\mkern2mu\mathchar'40\mkern-7mu L}\left(\frac{d\psi}{dt}\left(t\right),\psi\left(t\right)\right)dt}.
\end{cases}\label{eq:Mate.006.A.of.L}
\end{equation}
Clearly, the term $N_{0}\left(t_{\mathrm{i}},t_{\mathrm{f}}\right)$
is a normalization factor, so that the condition
\begin{equation}
\int_{C\left(\left[t_{\mathrm{i}},t_{\mathrm{f}}\right],\mathbb{V}\right)}\left[d\psi\right]\mathcal{A}\left[\psi;t_{\mathrm{i}},t_{\mathrm{f}}\right)=1\label{eq:Mate.007.normalization.A.psi}
\end{equation}
holds. The time-local function $\mathcal{\mkern2mu\mathchar'40\mkern-7mu L}\left(\frac{d\psi}{dt},\psi\right)$
is referred to as \emph{stochastic Lagrangian} (SL).

Since the relationship between this $\mkern2mu \mathcal{\mkern2mu\mathchar'40\mkern-7mu L}\left(\frac{d\psi}{dt},\psi\right)$
and the RPF $\mathcal{A}$ of the stochastic process is the same that
one has between the classical Lagrangian function and the quantum
amplitude in Quantum Mechanics \cite{Feynman.Hibbs.book}, also here
one may expect the ``symmetries'' of $\mathcal{\mkern2mu\mathchar'40\mkern-7mu L}$
to turn into invariances of the system statistics. One has to add
that, for real $\psi$, as transition probabilities $\mathcal{P}_{\psi_{\mathrm{i}}\mapsto\psi_{\mathrm{f}}}\left(t_{\mathrm{i}},t_{\mathrm{f}}\right)$
must be real positive numbers, the quantity $\mathcal{\mkern2mu\mathchar'40\mkern-7mu L}\left(\frac{d\psi}{dt},\psi\right)$
must be \emph{an imaginary number}, as it happens indeed in real cases
\cite{MateConso.SMHD,Mate.Conso.Tetrad.01}.

About the stochastic Lagrangian $\mathcal{\mkern2mu\mathchar'40\mkern-7mu L}\left(\frac{d\psi}{dt},\psi\right)$
one has to stress that it is a function \emph{intrinsically different}
from what one calls ``Lagrangian'' in Analytical Mechanics: indeed,
\emph{this} Lagrangian $\mathcal{\mkern2mu\mathchar'40\mkern-7mu L}\left(\frac{d\psi}{dt},\psi\right)$,
contains the maximum time-derivative appearing in the ODE of the system
(\ref{eq:Mate.000.DE.psi}), that is an \emph{intrinsically first
order ODE}, while \emph{the} Lagrangian of Analytical Mechanics $L\left(\frac{d\underline{q}}{dt},\underline{q}\right)$
does not contain the maximum order time-derivative of the evolution
equation of the system, as Euler-Lagrange equations $\frac{d}{dt}\left(\frac{\partial L}{\partial\underline{\dot{q}}}\right)-\frac{\partial L}{\partial\underline{q}}=0$
are\emph{ second order ODEs} (here the dot means time-derivative,
as $\dot{\underline{q}}\overset{\mathrm{def}}{=}\frac{d\underline{q}}{dt}$
and $\dot{\psi}\overset{\mathrm{def}}{=}\frac{d\psi}{dt}$). In §
\ref{subsec:Point-Particle-with-noise}, where the case of a point
particle undergoing classical and stochastic forces is treated, the
difference between $\mathcal{\mkern2mu\mathchar'40\mkern-7mu L}$
and the \emph{would-be}-Lagrangian $L$ of the point prticle without
noise is apparent: indeed, where the traditional Lagrangian would
be $L\left(\dot{\vec{x}},\vec{x}\right)$, a function of position
and velocity, the stochastic Lagrangian is some more complicated function
$\mathcal{\mkern2mu\mathchar'40\mkern-7mu L}\left(\dot{\vec{x}},\dot{\vec{p}},\vec{x},\vec{p}\right)$,
depending on position, velocity, momentum and momentum derivative.
Last difference we need to stress between the Analytical Mechanics
Lagrangian $L$ and the stochastic Lagrangian $\mathcal{\mkern2mu\mathchar'40\mkern-7mu L}$
is that, while $L$ contains ``all the Physics'' of the system \emph{only
for conservative systems}, $\mathcal{\mkern2mu\mathchar'40\mkern-7mu L}$
encodes the whole Physics of \emph{any stochastic system in the form
(\ref{eq:Mate.000.DE.psi})}, regardless it is conservative (Hamiltonian,
see § \ref{subsec:Hamiltonian-Systems}), or dissipative (metriplectic,
see § \ref{subsec:Metriplectic-Systems}).

It is also of use to give the definition of \emph{stochastic action}
$\mathcal{\cedilla{S}}\left[\psi,t_{\mathrm{i}},t_{\mathrm{f}}\right)$
as the time-integral of the Lagrangian $\mathcal{\mkern2mu\mathchar'40\mkern-7mu L}$:
\begin{equation}
\mathcal{\cedilla{S}}\left[\psi,t_{\mathrm{i}},t_{\mathrm{f}}\right)\overset{\mathrm{def}}{=}\int_{t_{\mathrm{i}}}^{t_{\mathrm{f}}}dt\mathcal{\mkern2mu\mathchar'40\mkern-7mu L}\left(\frac{d\psi}{dt},\psi\right),\label{eq:action}
\end{equation}
so that $\mathcal{A}=N_{0}\exp\left(-i\mathcal{\cedilla{S}}\right)$.
The \emph{caveat} to make no confusion between this $\mathcal{\cedilla{S}}$
and the action of Analytical Mechanics is the same one as that of
not confusing the stochastic $\mathcal{\mkern2mu\mathchar'40\mkern-7mu L}$
and the mechanical Lagrangian $L$ discussed just now.

The construction of the SL is described in \cite{Phythian.77}, and
inspired by the previous literature cited therein. In particular,
thanks to the mathematical nature of the LPE (\ref{eq:Mate.000.DE.psi}),
it is possible to introduce the kernel $\mathcal{A}\left[\psi;t_{\mathrm{i}},t_{\mathrm{f}}\right)$
starting from the definition of an ensemble statistical average $\left\langle F\right\rangle $,
based on \emph{noise statistics}

\[
\left\langle F\right\rangle \overset{\mathrm{def}}{=}\left\langle F\right\rangle _{f,g},
\]
as noises represent the only element giving to (\ref{eq:Mate.001.Stochastic.Cauchy.Problem})
a stochastic character. The definition of the average $\left\langle F\right\rangle _{f,g}$
is obviously
\[
\left\langle F\right\rangle _{f,g}\overset{\mathrm{def}}{=}\int_{\Sigma\left(t_{\mathrm{i}},t_{\mathrm{f}}\right)}F\left[\psi\right]\mathcal{P}\left[f,g;t_{\mathrm{i}},t_{\mathrm{f}}\right)\left[df\right]\left[dg\right]:
\]
our program is to obtain an expression of $\mathcal{A}$ so that
\begin{equation}
\int_{\Sigma\left(t_{\mathrm{i}},t_{\mathrm{f}}\right)}\left[df\right]\int_{\Sigma\left(t_{\mathrm{i}},t_{\mathrm{f}}\right)}\left[dg\right]F\left[\psi\right]\mathcal{P}\left[f,g;t_{\mathrm{i}},t_{\mathrm{f}}\right)=\int_{C\left(\left[t_{\mathrm{i}},t_{\mathrm{f}}\right],\mathbb{V}\right)}\left[d\psi\right]F\left[\psi\right]\mathcal{A}\left[\psi;t_{\mathrm{i}},t_{\mathrm{f}}\right)\ \forall\thinspace F.\label{eq:Mate.008}
\end{equation}

The first step taken in \cite{Phythian.77} to obtain $\mathcal{A}$
satisfying (\ref{eq:Mate.008}), is to consider $n$ auxiliary variables
$\chi_{I=1,...,n}$ that in a sense represent the Fourier momenta
conjugated with the additive noises $f^{I=1,...,n}$ in the functional
space $\Sigma\left(t_{\mathrm{i}},t_{\mathrm{f}}\right)$. Then, an
auxiliary kernel, depending on $\psi$ and $\chi$ is defined as:
\begin{equation}
\begin{cases}
 & A\left[\psi,\chi;t_{\mathrm{i}},t_{\mathrm{f}}\right)=\\
\\
 & =A_{0}\left(t_{\mathrm{i}},t_{\mathrm{f}}\right)C\left[\chi,\Gamma;t_{\mathrm{i}},t_{\mathrm{f}}\right)e^{-i\int_{t_{\mathrm{i}}}^{t_{\mathrm{f}}}dt\left[\dot{\psi}^{I}\chi_{I}-\Lambda^{I}\left(\psi\right)\chi_{I}-\frac{i}{2}\frac{\partial}{\partial\psi^{I}}\Lambda^{I}\left(\psi\right)\right]},\\
\\
\\
 & C\left[\chi,\Gamma;t_{\mathrm{i}},t_{\mathrm{f}}\right)=\\
\\
 & =\left\langle e^{i\int_{t_{\mathrm{i}}}^{t_{\mathrm{f}}}dt\left[f^{I}\chi_{I}+g_{I}\Gamma^{IJ}\left(\psi\right)\chi_{J}+g_{I}\frac{\partial}{\partial\psi^{J}}\Gamma^{IJ}\left(\psi\right)\right]}\right\rangle _{f,g}.
\end{cases}\label{eq:kernel.definition}
\end{equation}
In (\ref{eq:kernel.definition}) the quantity $A_{0}\left(t_{\mathrm{i}},t_{\mathrm{f}}\right)$
is a normalization factor, so that
\[
\int_{C\left(\left[t_{\mathrm{i}},t_{\mathrm{f}}\right],\mathrm{T}\left(\mathbb{V}^{*}\right)\right)}\left[d\chi\right]\int_{C\left(\left[t_{\mathrm{i}},t_{\mathrm{f}}\right],\mathbb{V}\right)}\left[d\psi\right]A\left[\psi,\chi;t_{\mathrm{i}},t_{\mathrm{f}}\right)=1;
\]
the factor $C\left[\chi,\Gamma;t_{\mathrm{i}},t_{\mathrm{f}}\right)$
is the term in which the noise statistics ends up being encoded, as
\[
C\left[\chi,\Gamma;t_{\mathrm{i}},t_{\mathrm{f}}\right)=\int_{\Sigma\left(t_{\mathrm{i}},t_{\mathrm{f}}\right)}\left[df\right]\int_{\Sigma\left(t_{\mathrm{i}},t_{\mathrm{f}}\right)}\left[dg\right]e^{i\int_{t_{\mathrm{i}}}^{t_{\mathrm{f}}}dt\left[f^{I}\chi_{I}+g_{I}\Gamma^{IJ}\left(\psi\right)\chi_{J}+g_{I}\frac{\partial}{\partial\psi^{J}}\Gamma^{IJ}\left(\psi\right)\right]}\mathcal{P}\left[f,g;t_{\mathrm{i}},t_{\mathrm{f}}\right).
\]

Once the auxiliary kernel $A\left[\psi,\chi;t_{\mathrm{i}},t_{\mathrm{f}}\right)$
is constructed, the definition of the physical kernel $\mathcal{A}\left[\psi;t_{\mathrm{i}},t_{\mathrm{f}}\right)$
reads simply:
\begin{equation}
\mathcal{A}\left[\psi;t_{\mathrm{i}},t_{\mathrm{f}}\right)\overset{\mathrm{def}}{=}\int_{C\left(\left[t_{\mathrm{i}},t_{\mathrm{f}}\right],\mathrm{T}\left(\mathbb{V}^{*}\right)\right)}\left[d\chi\right]A\left[\psi,\chi;t_{\mathrm{i}},t_{\mathrm{f}}\right).\label{eq:A.chi.psi.2.A.psi}
\end{equation}

The calculation indicated in (\ref{eq:A.chi.psi.2.A.psi}) is nothing
but obvious, its feasibility is definitely not for grant. Indeed,
summing over all the possible histories $\chi\left(t\right)\thinspace/\thinspace t\in\left[t_{\mathrm{i}},t_{\mathrm{f}}\right]$
is still matter of being able to do a functional integration, that
we know is not an easy task in general. In § \ref{subsec:delta.correlated}
we will see this integration is rather tractable if noises $f$ and
$g$ are Gaussian fluctuations $\delta$-correlated in time, as in
(\ref{eq:Mate.009}).

The great effort is then re-casting the RPF defined as (\ref{eq:A.chi.psi.2.A.psi})
and (\ref{eq:kernel.definition}) in the Lagrangian form (\ref{eq:Mate.006.A.of.L}).

\subsection{$\delta$-correlated Gaussian noise\label{subsec:delta.correlated}}

The example in which the calculations indicated in (\ref{eq:kernel.definition})
and (\ref{eq:A.chi.psi.2.A.psi}) are completely, and rather easily,
feasible, is that in which $f\left(t\right)$ and $g\left(t\right)$
are time-$\delta$-correlated noises with $t$-local Gaussian PDF.
Let us assume that the two noises $f$ and $g$ have $t$-local probability
density functions 
\begin{equation}
\begin{cases}
 & P_{t}\left(f\right)=\sqrt{\frac{2^{n-1}\left\Vert \lambda\left(t\right)\right\Vert }{\pi}}e^{-\lambda_{IJ}\left(t\right)\left(f^{I}-f_{0}^{I}\right)\left(f^{J}-f_{0}^{J}\right)},\\
\\
 & Q_{t}\left(g\right)=\sqrt{\frac{2^{n-1}\left\Vert \mu\left(t\right)\right\Vert }{\pi}}e^{-\mu_{IJ}\left(t\right)\left(g^{I}-g_{0}^{I}\right)\left(g^{J}-g_{0}^{J}\right)}.
\end{cases}\label{eq:Mate.009}
\end{equation}
As noises taken at different times are independent of each other,
the whole RPF is a continuous product of the time-local PDFs, i.e.:
\begin{align}
\begin{array}{c}
\mathcal{P}\left[f,g;t_{\mathrm{i}},t_{\mathrm{f}}\right)={\displaystyle \prod_{t\in\left[t_{\mathrm{i}},t_{\mathrm{f}}\right]}}\left(\frac{2^{2\left(n-1\right)}\left\Vert \lambda\left(t\right)\right\Vert \left\Vert \mu\left(t\right)\right\Vert }{\pi^{2}}\right)^{\frac{dt}{2}}*\\
\\
*e^{-\int_{t_{\mathrm{i}}}^{t_{\mathrm{f}}}dt\left[\lambda_{IJ}\left(t\right)\left(f^{I}\left(t\right)-f_{0}^{I}\left(t\right)\right)\left(f^{J}\left(t\right)-f_{0}^{J}\left(t\right)\right)+\mu_{IJ}\left(t\right)\left(g^{I}\left(t\right)-g_{0}^{I}\left(t\right)\right)\left(g^{J}\left(t\right)-g_{0}^{J}\left(t\right)\right)\right]}
\end{array}\label{eq:Mate.010.RPF.delta-correlated.G}
\end{align}
(in the expression (\ref{eq:Mate.010.RPF.delta-correlated.G}) the
emergence of the integration $\int_{t_{\mathrm{i}}}^{t_{\mathrm{f}}}dt...$
in the argument of the exponential, as well as the power $\frac{dt}{2}$
in the normalization factor, come from the continuous product, see
\cite{Mate.SESF.2019}). About the coefficients $\lambda$ and $\mu$
in (\ref{eq:Mate.009}) and (\ref{eq:Mate.010.RPF.delta-correlated.G}),
one has to note that their relationship with the standard deviation
$\sigma_{f}$ and $\sigma_{g}$ of noises reads:
\begin{equation}
\lambda=\frac{1}{2\sigma_{f}^{2}},\ \mu=\frac{1}{2\sigma_{g}^{2}}\label{eq:J.0007.mu.lambda.vs.sigma}
\end{equation}
(all the quantities in (\ref{eq:J.0007.mu.lambda.vs.sigma}) are $\mathbb{R}^{n,n}$
matrices). A further assumption necessary for (\ref{eq:Mate.010.RPF.delta-correlated.G})
to represent the correct RPF of those noises, is that $f$ and $g$
are \emph{uncorrelated with each other}, i.e.:
\[
\left\langle f^{I}\left(t\right)g^{J}\left(t'\right)\right\rangle =0\ \forall\thinspace t,t',I,J.
\]

The RPF (\ref{eq:Mate.010.RPF.delta-correlated.G}) gives rise to
the average over noises:
\begin{align*}
\left\langle F\right\rangle _{f,g} & =\left(\prod_{t\in\left[t_{\mathrm{i}},t_{\mathrm{f}}\right]}\left(\frac{2^{2\left(n-1\right)}\left\Vert \lambda\left(t\right)\right\Vert \left\Vert \mu\left(t\right)\right\Vert }{\pi^{2}}\right)^{\frac{dt}{2}}\int_{\mathbb{R}^{n}}df\left(t\right)\int_{\mathbb{R}^{n}}dg\left(t\right)\right)*\\
\\
 & *F\left[\psi\right]e^{-\int_{t_{\mathrm{i}}}^{t_{\mathrm{f}}}dt\left[\lambda_{IJ}\left(t\right)\left(f^{I}\left(t\right)-f_{0}^{I}\left(t\right)\right)\left(f^{J}\left(t\right)-f_{0}^{J}\left(t\right)\right)+\mu_{IJ}\left(t\right)\left(g^{I}\left(t\right)-g_{0}^{I}\left(t\right)\right)\left(g^{J}\left(t\right)-g_{0}^{J}\left(t\right)\right)\right]}.
\end{align*}

Without loss of generality one may put
\begin{equation}
\lambda_{IJ}=\lambda^{\left(I\right)}\delta_{IJ},\ \mu_{IJ}=\mu^{\left(I\right)}\delta_{IJ},\label{eq:Mate.011}
\end{equation}
so that the factor $C\left[\chi,\Gamma;t_{\mathrm{i}},t_{\mathrm{f}}\right)$
defined in (\ref{eq:kernel.definition}) is calculated as follows:
\begin{equation}
\begin{array}{c}
C\left[\chi,\Gamma;t_{\mathrm{i}},t_{\mathrm{f}}\right)=2^{\left(n-1\right)}e^{-\int_{t_{\mathrm{i}}}^{t_{\mathrm{f}}}dt\sum_{I=1}^{n}\left[\frac{1}{4\mu^{\left(I\right)}}\frac{\partial}{\partial\psi^{J}}\Gamma^{IJ}\left(\psi\right)\frac{\partial}{\partial\psi^{H}}\Gamma^{IH}\left(\psi\right)+ig_{0}^{I}\frac{\partial}{\partial\psi^{J}}\Gamma^{IJ}\left(\psi\right)\right]}*\\
\\
*e^{\int_{t_{\mathrm{i}}}^{t_{\mathrm{f}}}dt\sum_{I=1}^{n}\left[-\frac{1}{4\lambda^{\left(I\right)}}\chi_{I}^{2}-\frac{\Gamma^{IJ}\left(\psi\right)\Gamma^{IH}\left(\psi\right)}{4\mu^{\left(I\right)}}\chi_{J}\chi_{H}-\left(\frac{\Gamma^{IJ}\left(\psi\right)}{2\mu^{\left(I\right)}}\frac{\partial}{\partial\psi^{H}}\Gamma^{IH}\left(\psi\right)+ig_{0}^{I}\Gamma^{IJ}\left(\psi\right)\right)\chi_{J}\right]}.
\end{array}\label{eq:Mate.012}
\end{equation}
(in these scripts, repeated nearby indices mean to be summed over).
This (\ref{eq:Mate.012}) is usefully re-written separating the functional
depending only on $\psi$ from the one depending \emph{also} on $\chi$:
\begin{equation}
\begin{cases}
 & C\left[\chi,\Gamma;t_{\mathrm{i}},t_{\mathrm{f}}\right)=J\left[\psi;t_{\mathrm{i}},t_{\mathrm{f}}\right)M\left[\psi,\chi;t_{\mathrm{i}},t_{\mathrm{f}}\right),\\
\\
 & J\left[\psi;t_{\mathrm{i}},t_{\mathrm{f}}\right)=2^{\left(n-1\right)}e^{-\int_{t_{\mathrm{i}}}^{t_{\mathrm{f}}}dt\sum_{I=1}^{n}\left[\frac{1}{4\mu^{\left(I\right)}}\frac{\partial}{\partial\psi^{J}}\Gamma^{IJ}\left(\psi\right)\frac{\partial}{\partial\psi^{H}}\Gamma^{IH}\left(\psi\right)+ig_{0}^{I}\frac{\partial}{\partial\psi^{J}}\Gamma^{IJ}\left(\psi\right)\right]},\\
\\
 & M\left[\psi,\chi;t_{\mathrm{i}},t_{\mathrm{f}}\right)=e^{\int_{t_{\mathrm{i}}}^{t_{\mathrm{f}}}dt\sum_{I=1}^{n}\left[-\frac{1}{4\lambda^{\left(I\right)}}\chi_{I}^{2}-\frac{\Gamma^{IJ}\left(\psi\right)\Gamma^{IH}\left(\psi\right)}{4\mu^{\left(I\right)}}\chi_{J}\chi_{H}-\left(\frac{\Gamma^{IJ}\left(\psi\right)}{2\mu^{\left(I\right)}}\frac{\partial}{\partial\psi^{H}}\Gamma^{IH}\left(\psi\right)+ig_{0}^{I}\Gamma^{IJ}\left(\psi\right)\right)\chi_{J}\right]}.
\end{cases}\label{eq:Mate013.C.deltacorr.J.times.M}
\end{equation}

We are now in the position of calculating $A\left[\psi,\chi;t_{\mathrm{i}},t_{\mathrm{f}}\right)$
by using this expression in (\ref{eq:kernel.definition}), as done
in § \ref{subsec:A.2.A.chi.psi.A.psi}, with the important (and not
generality-losing) assumption (\ref{eq:Mate.011}). Moreover, a suitable
symmetric $\mathrm{T}\left(\mathbb{V}\right)\otimes\mathrm{T}\left(\mathbb{V}\right)$
tensor
\begin{equation}
\tau^{JH}\left(\lambda,\mu,\psi\right)\overset{\mathrm{def}}{=}\frac{\delta^{JH}}{4\lambda^{\left(J\right)}}+\sum_{I=1}^{n}\frac{\Gamma^{IJ}\left(\psi\right)\Gamma^{IH}\left(\psi\right)}{4\mu^{\left(I\right)}}\label{eq:Mate.014.def.tau}
\end{equation}
may be defined, that is easily shown to be positive definite
\begin{equation}
\left\Vert \tau\right\Vert >0,\label{eq:Mate.022}
\end{equation}
where the symbol $\left\Vert \tau\right\Vert $ is the determinant
of the matrix of $\tau$. Moreover, a vector $\beta$ in $\mathbb{V}$
is defined as: 
\begin{equation}
\beta^{J}\left(\mu,\dot{\psi},\psi\right)\overset{\mathrm{def}}{=}\sum_{I=1}^{n}\left[\frac{\Gamma^{IJ}\left(\psi\right)}{2\mu^{\left(I\right)}}\frac{\partial}{\partial\psi^{H}}\Gamma^{IH}\left(\psi\right)+ig_{0}^{I}\Gamma^{IJ}\left(\psi\right)+i\delta^{IJ}\left(\dot{\psi}^{I}-\Lambda^{I}\left(\psi\right)\right)\right],\label{eq:Mate.015.def.beta}
\end{equation}
so that all in all one remains with:
\begin{equation}
\begin{array}{c}
A\left[\psi,\chi;t_{\mathrm{i}},t_{\mathrm{f}}\right)\overset{(\ref{eq:Mate.015.def.beta})}{=}\\
\\
\overset{(\ref{eq:Mate.015.def.beta})}{=}A_{0}\left(t_{\mathrm{i}},t_{\mathrm{f}}\right)J\left[\psi;t_{\mathrm{i}},t_{\mathrm{f}}\right)e^{-\frac{1}{2}\int_{t_{\mathrm{i}}}^{t_{\mathrm{f}}}dt\frac{\partial}{\partial\psi^{I}}\Lambda^{I}\left(\psi\right)}e^{-\int_{t_{\mathrm{i}}}^{t_{\mathrm{f}}}dt\left[\tau^{JH}\left(\lambda,\mu,\psi\right)\chi_{J}\chi_{H}+\beta^{J}\left(\mu,\dot{\psi},\psi\right)\chi_{J}\right]},
\end{array}\label{eq:Mate.016}
\end{equation}
that is basically something Gaussian in the $\chi$ variables\footnote{The fact that this $A\left[\psi,\chi;t_{\mathrm{i}},t_{\mathrm{f}}\right)$
is Gaussian in $\chi$ is not at all a surprise: indeed, the parts
depending on $\chi$ in (\ref{eq:kernel.definition}) are: the factor
\[
e^{-i\int_{t_{\mathrm{i}}}^{t_{\mathrm{f}}}dt\left[\dot{\psi}^{I}\chi_{I}-\Lambda^{I}\left(\psi\right)\chi_{I}-\frac{i}{2}\frac{\partial}{\partial\psi^{I}}\Lambda^{I}\left(\psi\right)\right]},
\]
the exponential of a linear composition of the $\chi$s, and the $C\left[\chi,\Gamma;t_{\mathrm{i}},t_{\mathrm{f}}\right)$,
i.e. the functional Fourier transform of the Gaussian RPF $\mathcal{P}\left[f,g;t_{\mathrm{i}},t_{\mathrm{f}}\right)$,
i.e. a Gaussian again.}. The result (\ref{eq:Mate.019}) in § \ref{subsec:A.2.A.chi.psi.A.psi}
is interesting because, as the definition of $J$ in (\ref{eq:Mate013.C.deltacorr.J.times.M})
is considered
\begin{equation}
\begin{array}{c}
\mathcal{A}\left[\psi;t_{\mathrm{i}},t_{\mathrm{f}}\right)=N_{0}\left(t_{\mathrm{i}},t_{\mathrm{f}}\right)e^{-\int_{t_{\mathrm{i}}}^{t_{\mathrm{f}}}dt\sum_{I,J=1}^{n}\left[\frac{1}{4\mu^{\left(I\right)}}\frac{\partial\Gamma^{IJ}\left(\psi\right)}{\partial\psi^{J}}\frac{\partial\Gamma^{IJ}\left(\psi\right)}{\partial\psi^{J}}+ig_{0}^{I}\frac{\partial\Gamma^{IJ}\left(\psi\right)}{\partial\psi^{J}}+\frac{1}{2}\frac{\partial\Lambda^{I}\left(\psi\right)}{\partial\psi^{I}}\right]}*\\
\\
*e^{\frac{1}{4}\int_{t_{\mathrm{i}}}^{t_{\mathrm{f}}}dt\beta\left(\mu\left(t\right),\dot{\psi}\left(t\right),\psi\left(t\right)\right)\cdot\tau^{-1}\left(\lambda\left(t\right),\mu\left(t\right),\psi\left(t\right)\right)\cdot\beta^{\mathrm{T}}\left(\mu\left(t\right),\dot{\psi}\left(t\right),\psi\left(t\right)\right)},
\end{array}\label{eq:Mate.023}
\end{equation}
it is possible to define a total SL
\begin{equation}
\begin{array}{c}
\mathcal{\mkern2mu\mathchar'40\mkern-7mu L}\left(\lambda,\mu,\dot{\psi},\psi\right)=\frac{1}{4}i\beta\left(\mu,\dot{\psi},\psi\right)\cdot\tau^{-1}\left(\lambda,\mu,\psi\right)\cdot\beta^{\mathrm{T}}\left(\mu,\dot{\psi},\psi\right)+\\
\\
+{\displaystyle \sum_{I,J=1}^{n}}\left[g_{0}^{I}\frac{\partial\Gamma^{IJ}\left(\psi\right)}{\partial\psi^{J}}-\frac{i}{4\mu^{\left(I\right)}}\frac{\partial\Gamma^{IJ}\left(\psi\right)}{\partial\psi^{J}}\frac{\partial\Gamma^{IJ}\left(\psi\right)}{\partial\psi^{J}}-\frac{i}{2}\frac{\partial\Lambda^{I}\left(\psi\right)}{\partial\psi^{I}}\right].
\end{array}\label{eq:Mate.020.SL.delta.correlated.Gaussian}
\end{equation}
This is used as:
\begin{equation}
\begin{cases}
 & \mathcal{A}\left[\psi;t_{\mathrm{i}},t_{\mathrm{f}}\right)=N_{0}\left(t_{\mathrm{i}},t_{\mathrm{f}}\right)e^{-i\int_{t_{\mathrm{i}}}^{t_{\mathrm{f}}}dt\mathcal{\mkern2mu\mathchar'40\mkern-7mu L}\left(\lambda,\mu,\dot{\psi},\psi\right)},\\
\\
 & N_{0}\left(t_{\mathrm{i}},t_{\mathrm{f}}\right)\overset{\mathrm{def}}{=}2^{n-1}A_{0}\left(t_{\mathrm{i}},t_{\mathrm{f}}\right).
\end{cases}\label{eq:J.0009.A.psi.via.L}
\end{equation}
The SL of the problem (\ref{eq:Mate.001.Stochastic.Cauchy.Problem})
with $f$ and $g$ time-$\delta$-correlated Gaussian noises, so that
$\mathcal{A}=N_{0}e^{-i\int_{t_{\mathrm{i}}}^{t_{\mathrm{f}}}\mkern2mu\mathchar'40\mkern-7mu Ldt}$,
is the function (\ref{eq:Mate.020.SL.delta.correlated.Gaussian}).

In what follows, relevant examples of system stirred by time-$\delta$-correlated
Gaussian noises will be described, making reference to the SL (\ref{eq:Mate.020.SL.delta.correlated.Gaussian}).

\section{Examples\label{sec:Examples}}

In this Section some mechanical examples are reported, to show few,
physically relevant cases in which the construction of $\mathcal{A}$
and $\mathcal{\mkern2mu\mathchar'40\mkern-7mu L}$ is performed completely.
These are: the case of a pointlike particle of classical mechanics,
that of Hamiltonian and metriplectic systems, and that of a general
Leibniz dynamics. Only in the very simple case of the particle undergoing
a deterministic viscous force, plus noise, the functional measure
$\left[d\psi\right]$ appearing in (\ref{eq:Mate.003.average.F})
and (\ref{eq:Mate.004.transition.probability.1}) is calculated explicitly.
For sake of feasibility, all the noises in the examples are supposed
to be time-$\delta$-correlated and Gaussian.

From the examples mentioned here, the reader should learn at least
a general idea about the meaning and use of SLs; however, a more complex
and practical application of this formulation may be found in papers
\cite{MateConso.SMHD,Mate.Marsiglia.SMHD.2D,Mate.ISSS.Aquila,Mate.Conso.Tetrad.01,Mate.Conso.Tetrad.02,Chang.FSOC,Mate.SESF.2019},
where the same formulation described here is applied to space physics
and geophysical examples.

\subsection{Point Particle with noise\label{subsec:Point-Particle-with-noise}}

Consider a point particle described by its position $\vec{x}$ and
its momentum $\vec{p}$, with the relationship $\vec{p}=m\frac{d\vec{x}}{dt}$.
Consider also Newton's DE
\[
\vec{F}\left(\vec{x},\vec{p},t\right)=\frac{d\vec{p}}{dt},
\]
where the force vector $\vec{F}\left(\vec{x},\vec{p},t\right)$ may
be given by ``any'' force law. In general, the convenient thing
is to consider an $\vec{F}$ given by the sum of a pure \emph{deterministic},
\emph{smooth} addendum $\vec{\Phi}\left(\vec{x},\vec{p},t\right)$,
plus a noise terms $\vec{\xi}\left(t\right)$, the statistics of which
is going to be specified in a moment:
\[
\vec{\Phi}\left(\vec{x},\vec{p},t\right)+\vec{\xi}\left(t\right)=\frac{d\vec{p}}{dt}.
\]
The equations of motion read:
\begin{equation}
\begin{cases}
 & \dfrac{d\vec{x}}{dt}=\dfrac{\vec{p}}{m},\\
\\
 & {\displaystyle \frac{d\vec{p}}{dt}}=\vec{\Phi}\left(\vec{x},\vec{p},t\right)+\vec{\xi}\left(t\right).
\end{cases}\label{eq:J.0014.ODE.punto.base}
\end{equation}
In equations (\ref{eq:J.0014.ODE.punto.base}), one has six components
of the state vector
\begin{equation}
\psi=\left(\begin{array}{c}
\vec{x}\\
\vec{p}
\end{array}\right),\label{eq:J.0005}
\end{equation}
while the independent additive noises are only three ones, the $\vec{\xi}$
components. In order to be cautiously compliant with the scheme of
\cite{Phythian.77}, let us introduce an auxiliary noise term $\vec{\eta}\left(t\right)$,
so that the ODEs (\ref{eq:J.0014.ODE.punto.base}) become
\begin{equation}
\begin{cases}
 & \dfrac{d\vec{x}}{dt}=\dfrac{\vec{p}}{m}+\vec{\eta}\left(t\right),\\
\\
 & {\displaystyle \frac{d\vec{p}}{dt}}=\vec{\Phi}\left(\vec{x},\vec{p},t\right)+\vec{\xi}\left(t\right):
\end{cases}\label{eq:Mate.004.ODE.con.rumore.eta}
\end{equation}
the complete equivalence between these (\ref{eq:Mate.004.ODE.con.rumore.eta})
with the ODEs (\ref{eq:J.0014.ODE.punto.base}) will be \emph{weakly
restored}, in the sense that the noises $\vec{\eta}$ will be supposed
to have their PDF $\delta$-like peaked in zero identically, i.e.
for all times $t$.

The first thing to do in order to apply what described in § \ref{subsec:delta.correlated}
to the system (\ref{eq:Mate.004.ODE.con.rumore.eta}), one has to
construct the Lagrangian (\ref{eq:Mate.020.SL.delta.correlated.Gaussian}).
As no multiplicative noise exists in the example (\ref{eq:Mate.004.ODE.con.rumore.eta}),
one can adapt what obtained before to systems with \emph{pure additive
noise} just putting everywhere:
\begin{equation}
\Gamma\left(\psi\right)=0\thinspace\forall\thinspace\psi\in\mathbb{V}.\label{eq:J.0001.Gamma.eq.0}
\end{equation}
This reduces the SL to:
\[
\mathcal{\mkern2mu\mathchar'40\mkern-7mu L}\left(\lambda,\dot{\psi},\psi\right)=\frac{i}{4}\beta^{\mathrm{T}}\left(\dot{\psi},\psi\right)\cdot\tau^{-1}\left(\lambda\right)\cdot\beta\left(\dot{\psi},\psi\right)-\frac{i}{2}\frac{\partial\Lambda^{I}\left(\psi\right)}{\partial\psi^{I}}.
\]
Also, the expressions of $\beta$ and $\tau$ must be modified according
to (\ref{eq:J.0001.Gamma.eq.0}): the definition of $\beta$ in (\ref{eq:Mate.015.def.beta})
reads:
\begin{equation}
\beta^{J}\left(\dot{\psi},\psi\right)=i\left(\dot{\psi}^{J}-\Lambda^{J}\left(\psi\right)\right),\label{eq:J.0002.beta.additive.noise}
\end{equation}
while that of $\tau$ in (\ref{eq:Mate.014.def.tau}) now becomes:
\begin{equation}
\tau^{JH}\left(\lambda\right)=\frac{\delta^{JH}}{4\lambda^{\left(J\right)}}.\label{eq:J.0003.tau.additive.noise}
\end{equation}
The tensor $\tau$ is trivially inverted
\[
\left(\tau^{-1}\right)^{JH}\left(\lambda\right)=4\lambda^{\left(J\right)}\delta^{JH},
\]
so the expression of $\mathcal{\mkern2mu\mathchar'40\mkern-7mu L}\left(\lambda,\dot{\psi},\psi\right)$
reads:
\begin{equation}
\mathcal{\mkern2mu\mathchar'40\mkern-7mu L}\left(\lambda,\dot{\psi},\psi\right)=-i\sum_{J=1}^{n}\left[\lambda^{\left(J\right)}\left(\dot{\psi}^{J}-\Lambda^{J}\left(\psi\right)\right)^{2}+\frac{1}{2}\frac{\partial\Lambda^{J}\left(\psi\right)}{\partial\psi^{J}}\right].\label{eq:J.0004.SL.additive.noise}
\end{equation}

The set of necessary functions to write the SL in (\ref{eq:J.0004.SL.additive.noise})
is completed with the $\mathrm{T}\left(\mathbb{V}\right)$-vector
$\Lambda$ collecting the deterministic part of the dynamics, that
from (\ref{eq:Mate.004.ODE.con.rumore.eta}) reads:
\begin{equation}
\Lambda=\left(\begin{array}{c}
\dfrac{\vec{p}}{m}\\
\vec{\Phi}\left(\vec{x},\vec{p},t\right)
\end{array}\right).\label{eq:J.0006}
\end{equation}
From (\ref{eq:J.0006}) one has:
\[
{\displaystyle \frac{1}{2}}{\displaystyle \sum_{J=1}^{n}}{\displaystyle \frac{\partial\Lambda^{J}\left(\psi\right)}{\partial\psi^{J}}}={\displaystyle \frac{1}{2}}{\displaystyle \sum_{h=1}^{3}}{\displaystyle \frac{\partial\Phi^{h}}{\partial p^{h}}}\left(\vec{x},\vec{p},t\right).
\]

\noindent The SL reads:
\[
\mathcal{\mkern2mu\mathchar'40\mkern-7mu L}\left(\lambda,\dot{\psi},\psi\right)=-i\sum_{J=1}^{n}\lambda^{\left(J\right)}\left(\dot{\psi}^{J}-\Lambda^{J}\left(\psi\right)\right)^{2}-{\displaystyle \frac{i}{2}}{\displaystyle \sum_{h=1}^{3}}{\displaystyle \frac{\partial\Phi^{h}}{\partial p^{h}}}.
\]
The ``kinetic'' addendum $-i\sum_{J=1}^{n}\lambda^{\left(J\right)}\left(\dot{\psi}^{J}-\Lambda^{J}\left(\psi\right)\right)^{2}$
is constructed as follows:
\[
\begin{array}{c}
-i{\displaystyle \sum_{J=1}^{n}}\lambda^{\left(J\right)}\left(\dot{\psi}^{J}-\Lambda^{J}\left(\psi\right)\right)^{2}=\\
\\
=-i{\displaystyle \sum_{k=1}^{3}}\lambda_{\eta}^{\left(k\right)}\left(\dot{x}^{k}-\dfrac{p^{k}}{m}\right)^{2}-i{\displaystyle \sum_{h=1}^{3}}\lambda_{\xi}^{\left(h\right)}\left[\dot{p}^{h}-\Phi^{h}\left(\vec{x},\vec{p},t\right)\right]^{2}.
\end{array}
\]
Considering (\ref{eq:J.0007.mu.lambda.vs.sigma}), one may conclude:
\begin{equation}
\begin{array}{c}
\mathcal{\mkern2mu\mathchar'40\mkern-7mu L}\left(\lambda,\dot{\psi},\psi\right)=-i\left\{ {\displaystyle \sum_{k=1}^{3}}\frac{1}{2\sigma^{2}\left(\eta^{k}\right)}\left(\dot{x}^{k}-\dfrac{p^{k}}{m}\right)^{2}+\right.\\
\\
\left.+{\displaystyle \sum_{h=1}^{3}}\frac{1}{2\sigma^{2}\left(\xi^{h}\right)}\left[\dot{p}^{h}-\Phi^{h}\left(\vec{x},\vec{p},t\right)\right]^{2}+\frac{1}{2}{\displaystyle \sum_{h=1}^{3}}\frac{\partial\Phi^{h}}{\partial p^{h}}\right\} ,
\end{array}\label{eq:J.0008}
\end{equation}
being $\sigma\left(\eta^{k}\right)$ and $\sigma\left(\xi^{h}\right)$
the standard deviations or the respective noise components.

When (\ref{eq:J.0008}) is used in (\ref{eq:J.0009.A.psi.via.L}),
one obtains:
\begin{equation}
\mathcal{A}\left[\psi;t_{\mathrm{i}},t_{\mathrm{f}}\right)=N_{0}\left(t_{\mathrm{i}},t_{\mathrm{f}}\right)e^{-\sum_{k=1}^{3}\int_{t_{\mathrm{i}}}^{t_{\mathrm{f}}}dt\left\{ \frac{1}{2\sigma^{2}\left(\eta^{k}\right)}\left(\dot{x}^{k}-\frac{p^{k}}{m}\right)^{2}+\frac{1}{2\sigma^{2}\left(\xi^{h}\right)}\left[\dot{p}^{h}-\Phi^{h}\left(\vec{x},\vec{p},t\right)\right]^{2}+\frac{1}{2}\frac{\partial\Phi^{h}}{\partial p^{h}}\right\} }.\label{eq:J.0010.SL.Brown.001.eta}
\end{equation}
In order to reproduce the system (\ref{eq:J.0014.ODE.punto.base})
one has to consider an identically zero noise $\vec{\eta}$, which
is realized in (\ref{eq:J.0010.SL.Brown.001.eta}) ``simply'' considering
the $\delta$-family\footnote{In the formulae (\ref{eq:delta-family}) and (\ref{eq:J.0011.SL.Brown.002}),
and in similar formulae below, the ``limit'' is to be intended in
the distributional sense, or weak limit: $\lim_{\alpha\rightarrow\alpha_{0}}\varphi\left(x,\alpha\right)=\varphi_{0}\left(x\right)$
means
\[
\lim_{\alpha\rightarrow\alpha_{0}}\int dx\varphi\left(x,\alpha\right)f\left(x\right)=\int dx\varphi_{0}\left(x\right)f\left(x\right)\ \forall\ f,
\]
with $f$ being a function in the suitable space rendering each $\int dx\varphi\left(x,\alpha\right)f\left(x\right)$
and $\int dx\varphi_{0}\left(x\right)f\left(x\right)$ finite.}
\begin{equation}
\underset{\sigma\rightarrow0}{\lim}\frac{1}{\sqrt{2\pi}\sigma}e^{-\frac{\left(y-y_{0}\right)^{2}}{2\sigma^{2}}}=\delta\left(y-y_{0}\right),\label{eq:delta-family}
\end{equation}
so that we are interested in:
\begin{equation}
\begin{array}{c}
\mathcal{A}_{\xi}\left[\psi;t_{\mathrm{i}},t_{\mathrm{f}}\right)\overset{\mathrm{def}}{=}{\displaystyle \lim_{\sigma\left(\vec{\eta}\right)\rightarrow0}}\mathcal{A}\left[\psi;t_{\mathrm{i}},t_{\mathrm{f}}\right)=\\
\\
=\mathcal{N}_{0}\left(t_{\mathrm{i}},t_{\mathrm{f}}\right)\delta\left[\frac{d\vec{x}}{dt}-\frac{\vec{p}}{m}\right]e^{-\sum_{h=1}^{3}\int_{t_{\mathrm{i}}}^{t_{\mathrm{f}}}dt\left\{ \frac{1}{2\sigma^{2}\left(\xi^{h}\right)}\left[\dot{p}^{h}-\Phi^{h}\left(\vec{x},\vec{p},t\right)\right]^{2}+\frac{1}{2}\frac{\partial\Phi^{h}}{\partial p^{h}}\right\} },
\end{array}\label{eq:J.0011.SL.Brown.002}
\end{equation}
being $\mathcal{N}_{0}\left(t_{\mathrm{i}},t_{\mathrm{f}}\right)$
the suitable, necessary normalization constant, generally different
from the $N_{0}\left(t_{\mathrm{i}},t_{\mathrm{f}}\right)$ in (\ref{eq:J.0010.SL.Brown.001.eta}).
Note that in (\ref{eq:J.0011.SL.Brown.002}) the symbol $\delta\left[\frac{d\vec{x}}{dt}-\frac{\vec{p}}{m}\right]$
is \emph{a functional} $\delta$, being effective at any time $t\in\left[t_{\mathrm{i}},t_{\mathrm{f}}\right]$,
so that the prescription $\frac{d\vec{x}}{dt}=\frac{\vec{p}}{m}$
will be enforced as the integration over $\left[d^{3}x\right]$, or
over $\left[d^{3}p\right]$, takes place.

The example of $\vec{\Phi}$ one might consider reads:
\begin{equation}
\vec{\Phi}\left(\vec{x},\vec{p}\right)=-\vec{\partial}V\left(\vec{x}\right)+\frac{q}{m}\vec{p}\times\vec{B}\left(\vec{x}\right)-\frac{\zeta}{m}\vec{p}.\label{eq:J.0012}
\end{equation}
This force is the sum of a gradient force $-\vec{\partial}V\left(\vec{x}\right)$,
as in classical gravitation, elasticity or electrostatics, a Lorenz
force $\frac{q}{m}\vec{p}\times\vec{B}\left(\vec{x}\right)$ mimicking
the effect of a magnetic field $\vec{B}$ on a particle of electric
chage $q$, and a dissipative, viscous friction force $-\frac{\zeta}{m}\vec{p}$.
Considering the calculations in § \ref{subsec:Point.Particle.SL},
one may write
\begin{equation}
{\displaystyle \frac{1}{2}}{\displaystyle \sum_{h=1}^{3}}{\displaystyle \frac{\partial\Phi^{h}}{\partial p^{h}}}\overset{(\ref{eq:J.0012})}{=}-{\displaystyle \frac{3}{2}}{\displaystyle \frac{\zeta}{m}}\label{eq:zio.001}
\end{equation}
and this (\ref{eq:zio.001}) is inserted into (\ref{eq:J.0011.SL.Brown.002}),
obtaining:
\[
\begin{array}{c}
\mathcal{A}_{\xi}\left[\psi;t_{\mathrm{i}},t_{\mathrm{f}}\right)=\\
\\
=\mathcal{N}_{0}\left(t_{\mathrm{i}},t_{\mathrm{f}}\right)e^{\frac{3\zeta}{2m}\left(t_{\mathrm{f}}-t_{\mathrm{i}}\right)}\delta\left[\frac{d\vec{x}}{dt}-\frac{\vec{p}}{m}\right]e^{\left\{ -\int_{t_{\mathrm{i}}}^{t_{\mathrm{f}}}dt\underset{h}{\sum}\frac{1}{2\sigma^{2}\left(\xi^{h}\right)}\left[\dot{p}^{h}-\Phi^{h}\left(\vec{x},\vec{p},t\right)\right]^{2}\right\} }.
\end{array}
\]
The factor $\mathcal{N}_{0}\left(t_{\mathrm{i}},t_{\mathrm{f}}\right)e^{\frac{3\zeta}{2m}\left(t_{\mathrm{f}}-t_{\mathrm{i}}\right)}$
may be put all together defining $M_{0}\left(\zeta,m;t_{\mathrm{i}},t_{\mathrm{f}}\right)\overset{\mathrm{def}}{=}\mathcal{N}_{0}\left(t_{\mathrm{i}},t_{\mathrm{f}}\right)e^{\frac{3\zeta}{2m}\left(t_{\mathrm{f}}-t_{\mathrm{i}}\right)}$,
so we may write:
\begin{equation}
\begin{cases}
 & \mathcal{A}_{\xi}\left[\psi;t_{\mathrm{i}},t_{\mathrm{f}}\right)=M_{0}\left(\zeta,m;t_{\mathrm{i}},t_{\mathrm{f}}\right)\delta\left[\frac{d\vec{x}}{dt}-\frac{\vec{p}}{m}\right]e^{-i\int_{t_{\mathrm{i}}}^{t_{\mathrm{f}}}dt\mathcal{\mkern2mu\mathchar'40\mkern-7mu L}_{\xi}\left(\dot{\vec{p}},\vec{p},\vec{x}\right)},\\
\\
 & \mathcal{\mkern2mu\mathchar'40\mkern-7mu L}_{\xi}\left(\dot{\vec{p}},\vec{p},\vec{x}\right)\overset{\mathrm{def}}{=}\underset{h}{\sum}\frac{-i}{2\sigma^{2}\left(\xi^{h}\right)}\left[\dot{p}^{h}-\Phi^{h}\left(\vec{x},\vec{p},t\right)\right]^{2},
\end{cases}\label{eq:J.0014.RPF.for.pointlike.particle}
\end{equation}
(in (\ref{eq:J.0014.RPF.for.pointlike.particle}) the components $\Phi^{h}$
are those forming the vector $\vec{\Phi}$ in (\ref{eq:J.0012}))

Some comments are needed about this result: the exponential
\[
\exp\left\{ -\int_{t_{\mathrm{i}}}^{t_{\mathrm{f}}}dt\underset{h}{\sum}\frac{1}{2\sigma^{2}\left(\xi^{h}\right)}\left[\dot{p}^{h}-\Phi^{h}\left(\vec{x},\vec{p},t\right)\right]^{2}\right\} 
\]
will, again, converge to a Dirac $\delta$ functional of the argument
$\dot{p}^{h}-\Phi^{h}\left(\vec{x},\vec{p},t\right)$ as $\sigma\left(\xi^{h}\right)$
tends to zero, i.e. as the noise vector $\vec{\xi}$ is distributed
according to a more and more peaked Gaussian, provided a suitable
term $\mathbb{O}\left(\frac{1}{\sigma\left(\xi^{h}\right)}\right)$
is admitted in the normalization factor $M_{0}\left(\zeta,m;t_{\mathrm{i}},t_{\mathrm{f}}\right)$.
In other words: in the ``classical limit'' $\sigma\left(\xi^{h}\right)\rightarrow0$,
in which one has the certainty of having zero $\vec{\xi}$, the RPF
\emph{à la Phythian} of the stochastic trajectories admitted by (\ref{eq:J.0014.ODE.punto.base})
converges to the certainty that the trajectory will obey the ``silent\footnote{``Silent'' here means ``without noise'', \emph{classical}, fully
deterministic.} ODEs'' $\frac{d\vec{x}}{dt}=\frac{\vec{p}}{m}$ and $\frac{d\vec{p}}{dt}=\vec{\Phi}$.

Before going to more\label{paragraph.measure} examples of stochastic
systems, we would like to repeat, through the formalism introduced
here, the calculations reported in § 6.1 of the book \cite{Feynman.Hibbs.book},
and make \emph{an explicit calculation of the functional measure}
$\left[d\psi\right]$ in the case of the point particle undergoing
additive noise. We will do this for the simplified case in which $\vec{\Phi}$
is a \emph{pure viscous force}, i.e. $\vec{\Phi}=-\frac{\zeta}{m}\vec{p}$,
so the ODEs of the system read:
\begin{equation}
\begin{cases}
 & \dfrac{d\vec{x}}{dt}=\dfrac{\vec{p}}{m},\\
\\
 & {\displaystyle \frac{d\vec{p}}{dt}}=-\dfrac{\zeta}{m}\vec{p}+\vec{\xi}\left(t\right).
\end{cases}\label{eq:Browniano.viscoso}
\end{equation}
The kernel in use reads:
\[
\mathcal{A}\left[\vec{x},\vec{p};t_{\mathrm{i}},t_{\mathrm{f}}\right)=\tilde{N}_{0}\left(\zeta;t_{\mathrm{i}},t_{\mathrm{f}}\right)\delta\left[\frac{d\vec{x}}{dt}-\frac{\vec{p}}{m}\right]e^{-\sum_{i=1}^{3}\lambda_{\left(i\right)}\int_{t_{\mathrm{i}}}^{t_{\mathrm{f}}}dt\left\{ \dot{p}_{i}^{2}+\frac{2\zeta}{m}\dot{p}_{i}p_{i}+\frac{\mu^{2}}{m^{2}}p_{i}^{2}\right\} }.
\]
The term $\frac{2\zeta}{m}\lambda_{\left(i\right)}\dot{p}_{i}p_{i}$
in the integral can be simplified:
\[
\int_{t_{\mathrm{i}}}^{t_{\mathrm{f}}}dt\frac{2\zeta}{m}\lambda_{\left(i\right)}\dot{p}_{i}p_{i}=\frac{\zeta}{m}\lambda_{\left(i\right)}\left[p_{i}^{2}\left(t_{\mathrm{f}}\right)-p_{i}^{2}\left(t_{\mathrm{i}}\right)\right],
\]
so one re-writes:
\begin{equation}
\begin{array}{c}
\mathcal{A}\left[\vec{x},\vec{p};t_{\mathrm{i}},t_{\mathrm{f}}\right)=\\
\\
=\tilde{N}_{0}\left(\zeta;t_{\mathrm{i}},t_{\mathrm{f}}\right)e^{-\frac{\zeta}{m}\sum_{i=1}^{3}\lambda_{\left(i\right)}\left[p_{i}^{2}\left(t_{\mathrm{f}}\right)-p_{i}^{2}\left(t_{\mathrm{i}}\right)\right]}\delta\left[\frac{d\vec{x}}{dt}-\frac{\vec{p}}{m}\right]e^{-\int_{t_{\mathrm{i}}}^{t_{\mathrm{f}}}dt\sum_{i=1}^{3}\left(\lambda_{\left(i\right)}\dot{p}_{i}^{2}+\lambda_{\left(i\right)}\frac{\zeta^{2}}{m^{2}}p_{i}^{2}\right)}.
\end{array}\label{eq:A.moto.Browniano.02}
\end{equation}

This kernel is the central quantity allowing for the discussion of
the statistical dynamics of the system (\ref{eq:Browniano.viscoso}),
as it represents the statisical weight of the realization
\[
\vec{x}\left(t\right),\ \vec{p}\left(t\right)
\]
chosen in the integrand
\begin{equation}
\mathcal{\mkern2mu\mathchar'40\mkern-7mu L}_{p}\left(\dot{\vec{p}},\vec{p}\right)\overset{\mathrm{def}}{=}-i\sum_{i=1}^{3}\left(\lambda_{\left(i\right)}\dot{p}_{i}^{2}+\lambda_{\left(i\right)}\frac{\zeta^{2}}{m^{2}}p_{i}^{2}\right).\label{eq:L.stocastica.Browniano.mu}
\end{equation}
The current use of the functional $\mathcal{A}\left[\vec{x},\vec{p};t_{\mathrm{i}},t_{\mathrm{f}}\right)$
needs the determination of the constant $\tilde{N}_{0}\left(\zeta;t_{\mathrm{i}},t_{\mathrm{f}}\right)$,
that is prescribed in order to obtain:
\[
\int\left[d\vec{x}\right]\int\left[d\vec{p}\right]\mathcal{A}\left[\vec{x},\vec{p};t_{\mathrm{i}},t_{\mathrm{f}}\right)=1,
\]
so that its value is determined as:
\begin{equation}
\tilde{N}_{0}^{-1}\left(\zeta;t_{\mathrm{i}},t_{\mathrm{f}}\right)=\int\left[d\vec{x}\right]\int\left[d\vec{p}\right]e^{-\frac{\zeta}{m}\sum_{i=1}^{3}\lambda_{\left(i\right)}\left[p_{i}^{2}\left(t_{\mathrm{f}}\right)-p_{i}^{2}\left(t_{\mathrm{i}}\right)\right]}\delta\left[\frac{d\vec{x}}{dt}-\frac{\vec{p}}{m}\right]e^{-i\int_{t_{\mathrm{i}}}^{t_{\mathrm{f}}}dt\mathcal{\mkern2mu\mathchar'40\mkern-7mu L}_{p}\left(\dot{\vec{p}},\vec{p}\right)}.\label{eq:A.moto.Browniano.03}
\end{equation}
As in the kernel in (\ref{eq:A.moto.Browniano.02}), and hence in
(\ref{eq:A.moto.Browniano.03}), no dependence on $\vec{x}$ appears
in the integrand, the integration
\[
\int\left[d\vec{x}\right]\delta\left[\frac{d\vec{x}}{dt}-\frac{\vec{p}}{m}\right]=1
\]
can be considered, simply turning (\ref{eq:A.moto.Browniano.03})
into:
\begin{equation}
\tilde{N}_{0}^{-1}\left(\zeta;t_{\mathrm{i}},t_{\mathrm{f}}\right)=\int\left[d\vec{p}\right]e^{-\frac{\zeta}{m}\lambda_{\left(i\right)}\left[p_{i}^{2}\left(t_{\mathrm{f}}\right)-p_{i}^{2}\left(t_{\mathrm{i}}\right)\right]}e^{-i\int_{t_{\mathrm{i}}}^{t_{\mathrm{f}}}dt\mathcal{\mkern2mu\mathchar'40\mkern-7mu L}_{p}\left(\dot{\vec{p}},\vec{p}\right)}.\label{eq:A.moto.Browniano.04}
\end{equation}
The way out to find a closed expression for
\[
\int\left[d\vec{p}\right]e^{-\frac{\zeta}{m}\lambda_{\left(i\right)}\left[p_{i}^{2}\left(t_{\mathrm{f}}\right)-p_{i}^{2}\left(t_{\mathrm{i}}\right)\right]}e^{-i\int_{t_{\mathrm{i}}}^{t_{\mathrm{f}}}dt\mathcal{\mkern2mu\mathchar'40\mkern-7mu L}_{p}\left(\dot{\vec{p}},\vec{p}\right)}
\]
is to make use of time-discretization as done by Feynman and Hibbs
in their book \cite{Feynman.Hibbs.book}, in which they subdivide
the interval $\left[t_{\mathrm{i}},t_{\mathrm{f}}\right]$ into $\epsilon$-long
pieces
\begin{equation}
\begin{array}{c}
t_{\mathrm{f}}-t_{\mathrm{i}}=N\epsilon,\ \epsilon=t_{k+1}-t_{k},\\
\\
t_{0}=t_{\mathrm{i}},\ t_{N}=t_{\mathrm{f}},\ \psi_{0}=\psi_{\mathrm{i}},\ \psi_{0}=\psi_{\mathrm{f}},\\
\\
\psi\left(t_{k}\right)\mapsto\dfrac{1}{2}\left[\psi\left(t_{k+1}\right)+\psi\left(t_{k}\right)\right],\\
\\
\psi\left(t_{k}\right)\mapsto\dfrac{1}{\epsilon}\left[\psi\left(t_{k+1}\right)-\psi\left(t_{k}\right)\right].
\end{array}\label{eq:FH.tempi}
\end{equation}
The full integration $\int\left[d\vec{p}\right]$ may be represented
as
\begin{equation}
\int\left[d\vec{p}\right]...\overset{\mathrm{def}}{=}\lim_{\begin{array}{c}
\epsilon\rightarrow0\\
N\rightarrow+\infty
\end{array}}{\displaystyle \prod_{k=0}^{N}}\mathcal{W}_{p}\left(k\right)\left({\displaystyle \prod_{i=1}^{3}}\int_{\mathbb{R}}dp_{i}\left(t_{k}\right)\right)...,\label{eq:FH.integrali}
\end{equation}
while the integration in the exponential in (\ref{eq:A.moto.Browniano.04})
is discretized as:
\begin{equation}
\mathcal{\cedilla{S}}_{p}\overset{\mathrm{def}}{=}-i\int_{t_{\mathrm{i}}}^{t_{\mathrm{f}}}dt\left(\lambda_{\left(i\right)}\dot{p}_{i}^{2}+\lambda_{\left(i\right)}\frac{\zeta^{2}}{m^{2}}p_{i}^{2}\right)\cong-i\lim_{\begin{array}{c}
\epsilon\rightarrow0\\
N\rightarrow+\infty
\end{array}}\sum_{k=0}^{N-1}\epsilon\lambda_{\left(i\right)}\left(\dot{p}_{i}^{2}\left(k\right)+\frac{\zeta^{2}}{m^{2}}p_{i}^{2}\left(k\right)\right).\label{eq:FH.001}
\end{equation}
Defining in a closed form the functional measure $\int\left[d\vec{p}\right]$
precisely means finding sensible expressions for the factors $\mathcal{W}_{p}\left(k\right)$
in (\ref{eq:FH.integrali}).

Feynman and Hibbs suggest the way to interpret the quantity $\dot{p}_{i}\left(k\right)$
as:
\begin{equation}
\dot{p}_{i}\left(k\right)=\frac{1}{\epsilon}\left[p_{i}\left(k+1\right)-p_{i}\left(k\right)\right],\label{eq:FH.p.dot}
\end{equation}
while the quantity $p_{i}\left(k\right)$ is better approximated as
\begin{equation}
p_{i}\left(k\right)\mapsto\frac{1}{2}\left[p_{i}\left(k+1\right)+p_{i}\left(k\right)\right].\label{eq:FH.p.mid}
\end{equation}
The replacements (\ref{eq:FH.p.dot}) and (\ref{eq:FH.p.mid}) are
peformed, and calculations are done in § \ref{subsec:Point.Particle.SL}.
The integral in (\ref{eq:FH.001}) reads:
\begin{equation}
\begin{array}{c}
\mathcal{\cedilla{S}}_{p}\cong-i{\displaystyle \sum_{k=0}^{N-1}}\dfrac{\lambda_{\left(i\right)}}{\epsilon}\left[\left({\displaystyle \frac{4m^{2}+\epsilon^{2}\zeta^{2}}{4m^{2}}}\right)p_{i}^{2}\left(k+1\right)+\right.\\
\\
\left.+\left({\displaystyle \frac{4m^{2}+\epsilon^{2}\zeta^{2}}{4m^{2}}}\right)p_{i}^{2}\left(k\right)+\left({\displaystyle \frac{\zeta^{2}\epsilon^{2}-4m^{2}}{2m^{2}}}\right)p_{i}\left(k+1\right)p_{i}\left(k\right)\right].
\end{array}\label{eq:Sstoca.001}
\end{equation}
In anticipation of calculating the limit $\epsilon\rightarrow0$ invoked
in (\ref{eq:FH.integrali}), the expressions calculated in (\ref{eq:mea1.001})
are reported neglecting $o\left(\epsilon\right)$, so that one may
re-write:
\begin{equation}
{\displaystyle \frac{4m^{2}+\epsilon^{2}\zeta^{2}}{4m^{2}}}\overset{o\left(\epsilon\right)}{=}1,\ \ \frac{\zeta^{2}\epsilon^{2}-4m^{2}}{2m^{2}}\overset{o\left(\epsilon\right)}{=}-2,\label{eq:L.k.o.epsilon}
\end{equation}
giving rise to the following approximation for the term $\mathcal{\cedilla{S}}_{p}$
in (\ref{eq:Sstoca.001}):
\begin{equation}
\mathcal{\cedilla{S}}_{p}\overset{o\left(\epsilon\right)}{\cong}-i\sum_{k=0}^{N-1}\dfrac{\lambda_{\left(i\right)}}{\epsilon}\left[p_{i}^{2}\left(k+1\right)+p_{i}^{2}\left(k\right)-2p_{i}\left(k+1\right)p_{i}\left(k\right)\right].\label{eq:Sstoca.002}
\end{equation}

Due to (\ref{eq:Sstoca.002}), one has the following normalization:
\begin{equation}
\begin{array}{c}
\tilde{N}_{0}^{-1}\left(\zeta;t_{\mathrm{i}},t_{\mathrm{f}}\right)={\displaystyle \lim_{\begin{array}{c}
\epsilon\rightarrow0\\
N\rightarrow+\infty
\end{array}}}\mathcal{W}_{p}\left(N\right)\int_{\mathbb{R}^{3}}d^{3}p\left(t_{\mathrm{f}}\right)e^{-\frac{\zeta}{m}\lambda_{\left(i\right)}p_{i}^{2}\left(t_{\mathrm{f}}\right)}*\\
\\
*{\displaystyle \prod_{k=0}^{N-1}}\mathcal{W}_{p}\left(k\right){\displaystyle \prod_{i=1}^{3}}\int_{\mathbb{R}}dp_{i}\left(t_{k}\right)e^{\frac{\zeta}{m}\lambda_{\left(i\right)}p_{i}^{2}\left(t_{\mathrm{i}}\right)}e^{-\frac{\lambda_{\left(i\right)}}{\epsilon}\left[p_{i}^{2}\left(k+1\right)+p_{i}^{2}\left(k\right)-2p_{i}\left(k+1\right)p_{i}\left(k\right)\right]}.
\end{array}\label{eq:Sstoca.003}
\end{equation}
All the integrations $\int_{\mathbb{R}}dp_{i}\left(t_{k}\right)$
are ``the same''
\begin{equation}
\mathcal{I}_{k}\overset{\mathrm{def}}{=}\mathcal{W}_{p}\left(k\right){\displaystyle \prod_{i=1}^{3}}\int_{\mathbb{R}}dp_{i}\left(t_{k}\right)e^{-\frac{\lambda_{\left(i\right)}}{\epsilon}\left[p_{i}^{2}\left(k+1\right)+p_{i}^{2}\left(k\right)-2p_{i}\left(k+1\right)p_{i}\left(k\right)\right]},\label{eq:I.k.def}
\end{equation}
apart from the integration over $p_{i}\left(t_{\mathrm{i}}\right)$,
because it contains also the factor $e^{\frac{\zeta}{m}\lambda_{\left(i\right)}p_{i}^{2}\left(t_{\mathrm{i}}\right)}$
inherited from (\ref{eq:A.moto.Browniano.02}). So, it is useful to
separate the integration over the initial momentum $p_{i}\left(t_{\mathrm{i}}\right)$
from all the other ones in (\ref{eq:Sstoca.003}):
\begin{equation}
\begin{cases}
 & \tilde{N}_{0}^{-1}\left(\zeta;t_{\mathrm{i}},t_{\mathrm{f}}\right)=\lim_{\begin{array}{c}
\epsilon\rightarrow0\\
N\rightarrow+\infty
\end{array}}\mathcal{W}_{p}\left(N\right)\mathcal{W}_{p}\left(0\right)\int_{\mathbb{R}}d^{3}p\left(t_{\mathrm{f}}\right)e^{-\frac{\zeta}{m}\lambda_{\left(i\right)}p_{i}^{2}\left(t_{\mathrm{f}}\right)}{\displaystyle \prod_{k=1}^{N-1}}\mathcal{I}_{k}\mathcal{J}_{0},\\
\\
 & \mathcal{J}_{0}\overset{\mathrm{def}}{=}\mathcal{W}_{p}\left(0\right){\displaystyle \prod_{i=1}^{3}}\int_{\mathbb{R}}d^{3}p\left(t_{\mathrm{i}}\right)e^{\frac{\zeta}{m}\lambda_{\left(i\right)}p_{i}^{2}\left(t_{\mathrm{i}}\right)}e^{-\frac{\lambda_{\left(i\right)}}{\epsilon}\left[p_{i}^{2}\left(1\right)+p_{i}^{2}\left(t_{\mathrm{i}}\right)-2p_{i}\left(1\right)p_{i}\left(t_{\mathrm{i}}\right)\right]}.
\end{cases}\label{eq:Sstoca.004}
\end{equation}
The quantity $\mathcal{J}_{0}$ reads:
\[
\mathcal{J}_{0}=\mathcal{W}_{p}\left(0\right)\sqrt{\frac{\pi^{3}\epsilon^{3}}{\left\Vert \lambda\right\Vert }}.
\]

\noindent The integral $\mathcal{I}_{k}$ instead reads:
\begin{equation}
\mathcal{I}_{k}=\mathcal{W}_{p}\left(k\right)\sqrt{\dfrac{\pi^{3}\epsilon^{3}}{\left\Vert \lambda\right\Vert }}.\label{eq:I.k.valore}
\end{equation}

It also the case to calculate $\mathcal{I}_{0}$ as:
\[
\mathcal{I}_{0}=\mathcal{W}_{p}\left(k\right){\displaystyle \prod_{i=1}^{3}}\int_{\mathbb{R}}dp_{i}\left(t_{\mathrm{i}}\right)e^{-\underset{i}{\sum}\frac{\lambda_{\left(i\right)}}{\epsilon}\left[p_{i}^{2}\left(1\right)+p_{i}^{2}\left(0\right)-2p_{i}\left(1\right)p_{i}\left(0\right)\right]}
\]

All in all, one may write:
\begin{equation}
\tilde{N}_{0}^{-1}\left(\zeta;t_{\mathrm{i}},t_{\mathrm{f}}\right)=\lim_{\begin{array}{c}
\epsilon\rightarrow0\\
N\rightarrow+\infty
\end{array}}\mathcal{W}_{p}\int_{\mathbb{R}}d^{3}p\left(t_{\mathrm{f}}\right)\int_{\mathbb{R}}d^{3}p\left(t_{\mathrm{i}}\right)e^{-\frac{\zeta}{m}\underset{i}{\sum}\lambda_{\left(i\right)}\left[p_{i}^{2}\left(t_{\mathrm{f}}\right)-p_{i}^{2}\left(t_{\mathrm{i}}\right)\right]}\left(\dfrac{\pi^{3}\epsilon^{3}}{\left\Vert \lambda\right\Vert }\right)^{\frac{N-1}{2}},\label{eq:Sstoca.005}
\end{equation}
where one has given the definition: 
\[
\mathcal{W}_{p}\overset{\mathrm{def}}{=}{\displaystyle \prod_{k=0}^{N}}\mathcal{W}_{p}\left(k\right).
\]

Once the quantity $\tilde{N}_{0}^{-1}\left(\zeta;t_{\mathrm{i}},t_{\mathrm{f}}\right)$
has been calculated as done in § \ref{subsec:J0-and-Ik}, and represented
as in (\ref{eq:Sstoca.005}), one may write:
\begin{equation}
\begin{cases}
 & \mathcal{A}\left[\vec{x},\vec{p};t_{\mathrm{i}},t_{\mathrm{f}}\right)=\tilde{N}_{0}\left(\zeta;t_{\mathrm{i}},t_{\mathrm{f}}\right)e^{-\frac{\zeta}{m}\underset{i}{\sum}\lambda_{\left(i\right)}\left[p_{i}^{2}\left(t_{\mathrm{f}}\right)-p_{i}^{2}\left(t_{\mathrm{i}}\right)\right]}\delta\left[{\displaystyle \frac{d\vec{x}}{dt}-\frac{\vec{p}}{m}}\right]e^{-i\mathcal{\cedilla{S}}_{p}},\\
\\
 & \mathcal{\cedilla{S}}_{p}\overset{o\left(\epsilon\right)}{\cong}-i\sum_{k=0}^{N-1}\dfrac{\lambda_{\left(i\right)}}{\epsilon}\left[p_{i}^{2}\left(k+1\right)+p_{i}^{2}\left(k\right)-2p_{i}\left(k+1\right)p_{i}\left(k\right)\right],\\
\\
 & \tilde{N}_{0}\left(\zeta;t_{\mathrm{i}},t_{\mathrm{f}}\right)={\displaystyle \lim_{\begin{array}{c}
\epsilon\rightarrow0\\
N\rightarrow+\infty
\end{array}}}\dfrac{e^{\frac{\zeta}{m}\underset{i}{\sum}\lambda_{\left(i\right)}\left[p_{i}^{2}\left(t_{\mathrm{f}}\right)-p_{i}^{2}\left(t_{\mathrm{i}}\right)\right]}}{{\displaystyle \prod_{k=0}^{N}}\mathcal{W}_{p}\left(k\right)\mathcal{V}_{p}\left(t_{\mathrm{f}}\right)\mathcal{V}_{p}\left(t_{\mathrm{i}}\right)}\left(\dfrac{\left\Vert \lambda\right\Vert }{\pi^{3}\epsilon^{3}}\right)^{\frac{N-1}{2}},
\end{cases}\label{eq:zio.003}
\end{equation}
where the definition $\mathcal{V}_{p}\left(t\right)\overset{\mathrm{def}}{=}\int d^{3}p\left(t\right)$
is intended. If $\tilde{N}_{0}\left(\zeta;t_{\mathrm{i}},t_{\mathrm{f}}\right)$
in inserted in the first expression in (\ref{eq:zio.003}), one has
\begin{equation}
\begin{cases}
 & \mathcal{A}\left[\vec{x},\vec{p};t_{\mathrm{i}},t_{\mathrm{f}}\right)=\delta\left[{\displaystyle \frac{d\vec{x}}{dt}-\frac{\vec{p}}{m}}\right]{\displaystyle \lim_{\begin{array}{c}
\epsilon\rightarrow0\\
N\rightarrow+\infty
\end{array}}}\dfrac{\left\Vert \lambda\right\Vert ^{\frac{N-1}{2}}e^{-i\mathcal{\cedilla{S}}_{p}\left(N,\epsilon\right)}}{{\displaystyle \left(\pi^{3}\epsilon^{3}\right)^{\frac{N-1}{2}}\prod_{k=0}^{N}}\mathcal{W}_{p}\left(k\right)\mathcal{V}_{p}\left(t_{\mathrm{f}}\right)\mathcal{V}_{p}\left(t_{\mathrm{i}}\right)},\\
\\
 & \mathcal{\cedilla{S}}_{p}\overset{o\left(\epsilon\right)}{\cong}-i\sum_{k=0}^{N-1}\dfrac{\lambda_{\left(i\right)}}{\epsilon}\left[p_{i}^{2}\left(k+1\right)+p_{i}^{2}\left(k\right)-2p_{i}\left(k+1\right)p_{i}\left(k\right)\right]:
\end{cases}\label{eq:zio.004}
\end{equation}
all in all, (\ref{eq:zio.004}) implies also:
\begin{equation}
\begin{array}{c}
\int\left[d\vec{x}\right]\int\left[d\vec{p}\right]\delta\left[{\displaystyle \frac{d\vec{x}}{dt}-\frac{\vec{p}}{m}}\right]e^{-i\mathcal{\cedilla{S}}_{p}}=\\
\\
={\displaystyle \lim_{\begin{array}{c}
\epsilon\rightarrow0\\
N\rightarrow+\infty
\end{array}}}\left(\frac{\pi^{3}\epsilon^{3}}{\left\Vert \lambda\right\Vert }\right)^{\frac{N-1}{2}}{\displaystyle \prod_{k=0}^{N}}\mathcal{W}_{p}\left(k\right)\mathcal{V}_{p}\left(t_{\mathrm{f}}\right)\mathcal{V}_{p}\left(t_{\mathrm{i}}\right).
\end{array}\label{eq:zio.005.integrale.exp(-S.stocha)}
\end{equation}
The relationship (\ref{eq:zio.005.integrale.exp(-S.stocha)}) may
be of use when ensemble calculations are to be performed via $\mathcal{A}\left[\vec{x},\vec{p};t_{\mathrm{i}},t_{\mathrm{f}}\right)$.

Going back to (\ref{eq:zio.003}), if one wants to keep finite $\tilde{N}_{0}\left(\zeta;t_{\mathrm{i}},t_{\mathrm{f}}\right)$,
the factor $\prod_{k=0}^{N}\mathcal{W}_{p}\left(k\right)$ must be
chosen accordingly. Provided one defines
\begin{equation}
\mathcal{W}_{p}\left(k\right)\overset{\mathrm{def}}{=}\dfrac{e^{\frac{\zeta}{mN}\underset{i}{\sum}\lambda_{\left(i\right)}\left[p_{i}^{2}\left(t_{\mathrm{f}}\right)-p_{i}^{2}\left(t_{\mathrm{i}}\right)\right]}}{n_{0}^{\frac{1}{N}}\left(\mathcal{V}_{p}\left(t_{\mathrm{f}}\right)\mathcal{V}_{p}\left(t_{\mathrm{i}}\right)\right)^{\frac{1}{N}}}\left(\dfrac{\left\Vert \lambda\right\Vert }{\pi^{3}\epsilon^{3}}\right)^{\frac{N-1}{2N}},\label{eq:zio.006.Wp(k)}
\end{equation}
as suggested in § \ref{subsec:Keeping-N0-finite}, the factor $\tilde{N}_{0}$
remains finite. This amounts to defining the functional measure:
\begin{equation}
\int\left[d\vec{p}\right]...\overset{\mathrm{def}}{=}\lim_{\begin{array}{c}
\epsilon\rightarrow0\\
N\rightarrow+\infty
\end{array}}{\displaystyle \prod_{k=0}^{N}}\dfrac{e^{\frac{\zeta}{mN}\underset{i}{\sum}\lambda_{\left(i\right)}\left[p_{i}^{2}\left(t_{\mathrm{f}}\right)-p_{i}^{2}\left(t_{\mathrm{i}}\right)\right]}}{\left(\mathcal{V}_{p}\left(t_{\mathrm{f}}\right)\mathcal{V}_{p}\left(t_{\mathrm{i}}\right)\right)^{\frac{1}{N}}}\left(\dfrac{\left\Vert \lambda\right\Vert }{\pi^{3}\epsilon^{3}}\right)^{\frac{N-1}{2N}}\left({\displaystyle \prod_{i=1}^{3}}\int_{\mathbb{R}}dp_{i}\left(t_{k}\right)\right)...,\label{eq:zio.007.=00005Bdp=00005D}
\end{equation}
so that a concrete nature to the definition (\ref{eq:FH.integrali})
is given (in (\ref{eq:zio.007.=00005Bdp=00005D}) the finite constant
$n_{0}\equiv\tilde{N}_{0}$, used in § \ref{subsec:Keeping-N0-finite},
has been put equal to 1, that means: the necessity of a normalization
factor for $\mathcal{A}\left[\vec{x},\vec{p};t_{\mathrm{i}},t_{\mathrm{f}}\right)$
has been reabsorbed into the functional measure $\left[d\vec{p}\right]$).

\subsection{Leibniz systems\label{subsec:Leibniz-systems}}

Leibniz systems are dynamical systems whose ODEs may be expressed
in tensor terms as
\begin{equation}
\dfrac{d\psi^{I}}{dt}=T^{IJ}\left(\psi\right)\frac{\partial}{\partial\psi^{J}}F\left(\psi\right),\label{eq:J.0013.Leibniz.Systems}
\end{equation}
where the quantity $F$ is a function from $\mathbb{V}$ to $\mathbb{R}$
indicated as \emph{Leibniz-generator of the motion}, while $T\in\mathrm{T}\left(\mathbb{V}\right)\otimes\mathrm{T}\left(\mathbb{V}\right)$
is a tensor, referred to as \emph{Leibniz tensor} \cite{Leibniz.systems}.
Suppose to deal with a stochastic dynamics, the deterministic part
of which is (\ref{eq:J.0013.Leibniz.Systems}), i.e. some version
of (\ref{eq:Mate.001.Stochastic.Cauchy.Problem}) with
\begin{equation}
\Lambda^{I}\left(\psi\right)=T^{IJ}\left(\psi\right)\frac{\partial}{\partial\psi^{J}}F\left(\psi\right).\label{eq:J.0015.Lambda.Leibniz}
\end{equation}
Provided the SDE reads
\[
\dfrac{d\psi^{I}}{dt}=T^{IJ}\left(\psi\right)\frac{\partial F\left(\psi\right)}{\partial\psi^{J}}+g_{J}\left(t\right)\Gamma^{JI}\left(\psi\right)+f^{I}\left(t\right)
\]
and $f$ and $g$ are Gaussian time-$\delta$-correlated noises with
the statistics described in § \ref{subsec:delta.correlated}, one
must work with the functionals (\ref{eq:J.0009.A.psi.via.L}). The
calculations needed are shown in § \ref{sec:Calculations-for-Leibniz-Systems},
and one has:
\begin{equation}
\begin{cases}
 & \mathcal{A}\left[\psi;t_{\mathrm{i}},t_{\mathrm{f}}\right)=N_{0}\left(t_{\mathrm{i}},t_{\mathrm{f}}\right)e^{-i\int_{t_{\mathrm{i}}}^{t_{\mathrm{f}}}dt\mathcal{\mkern2mu\mathchar'40\mkern-7mu L}\left(\lambda,\mu,\dot{\psi},\psi\right)},\\
\\
 & \mathcal{\mkern2mu\mathchar'40\mkern-7mu L}\left(\lambda,\mu,\dot{\psi},\psi\right)=\frac{1}{4}i\beta^{\mathrm{T}}\left(\mu,\dot{\psi},\psi\right)\cdot\tau^{-1}\left(\lambda,\mu,\psi\right)\cdot\beta\left(\mu,\dot{\psi},\psi\right)+\\
\\
 & +{\displaystyle \sum_{I,J=1}^{n}}\left\{ g_{0}^{I}\frac{\partial\Gamma^{IJ}\left(\psi\right)}{\partial\psi^{J}}-\frac{i}{4\mu^{\left(I\right)}}\frac{\partial\Gamma^{IJ}\left(\psi\right)}{\partial\psi^{J}}\frac{\partial\Gamma^{IJ}\left(\psi\right)}{\partial\psi^{J}}-{\displaystyle \frac{i}{2}}\frac{\partial T^{IJ}\left(\psi\right)}{\partial\psi^{I}}\frac{\partial F\left(\psi\right)}{\partial\psi^{J}}-\frac{i}{2}T_{S}^{IJ}\left(\psi\right)\frac{\partial^{2}F\left(\psi\right)}{\partial\psi^{I}\partial\psi^{J}}\right\} ,\\
\\
 & \beta^{J}\left(\mu,\dot{\psi},\psi\right)\overset{\mathrm{def}}{=}\sum_{I=1}^{n}\left[\frac{\Gamma^{IJ}\left(\psi\right)}{2\mu^{\left(I\right)}}\frac{\partial}{\partial\psi^{H}}\Gamma^{IH}\left(\psi\right)+ig_{0}^{I}\Gamma^{IJ}\left(\psi\right)+i\delta^{IJ}\left(\dot{\psi}^{I}-T^{IK}\left(\psi\right)\frac{\partial F\left(\psi\right)}{\partial\psi^{K}}\right)\right],\\
\\
 & \tau^{JH}\left(\lambda,\mu,\psi\right)\overset{\mathrm{def}}{=}\frac{\delta^{JH}}{4\lambda^{\left(J\right)}}+\sum_{I=1}^{n}\frac{\Gamma^{IJ}\left(\psi\right)\Gamma^{IH}\left(\psi\right)}{4\mu^{\left(I\right)}}.
\end{cases}\label{eq:J.0017}
\end{equation}
The more tractable version of (\ref{eq:J.0017}) is the one with \emph{additive
noises only}: to obtain this case in the smoothest way possible one
may resort to (\ref{eq:J.0001.Gamma.eq.0}) and to the relationship
(\ref{eq:J.0007.mu.lambda.vs.sigma}). Then, one will write:
\begin{equation}
\begin{cases}
 & \mathcal{A}_{f}\left[\psi;t_{\mathrm{i}},t_{\mathrm{f}}\right)=N_{0}\left(t_{\mathrm{i}},t_{\mathrm{f}}\right)e^{-i\int_{t_{\mathrm{i}}}^{t_{\mathrm{f}}}dt\mathcal{\mkern2mu\mathchar'40\mkern-7mu L}\left(\lambda,\dot{\psi},\psi\right)},\\
\\
 & \mathcal{\mkern2mu\mathchar'40\mkern-7mu L}\left(\lambda,\dot{\psi},\psi\right)=-\sum_{I=1}^{n}\dfrac{i}{2\sigma^{2}\left(f^{I}\right)}\left(\dot{\psi}^{I}-T^{IJ}\left(\psi\right){\displaystyle \frac{\partial F\left(\psi\right)}{\partial\psi^{J}}}\right)^{2}+\\
\\
 & -{\displaystyle \frac{i}{2}}{\displaystyle \sum_{I,J=1}^{n}}\left[{\displaystyle \frac{\partial T^{IJ}\left(\psi\right)}{\partial\psi^{I}}}{\displaystyle \frac{\partial F\left(\psi\right)}{\partial\psi^{J}}}+T_{S}^{IJ}\left(\psi\right){\displaystyle \frac{\partial^{2}F\left(\psi\right)}{\partial\psi^{I}\partial\psi^{J}}}\right].
\end{cases}\label{eq:J.0018.Leibniz.additive}
\end{equation}

In the following two §§, the Hamiltonian and the metriplectic systems
are treated, as particularly relevant cases of Leibniz systems.

\subsubsection{Hamiltonian Systems\label{subsec:Hamiltonian-Systems}}

A Hamiltonian system is a DS as in (\ref{eq:J.0013.Leibniz.Systems}),
with $T^{IJ}$ being a \emph{Jocobi tensor} \cite{Morrison.Hamiltonian};
moreover, the dyamics generator $F$ is what they call \emph{Hamiltonian
of the system}, namely its \emph{energy}:
\[
F=H.
\]
In the relationships (\ref{eq:J.0017}) and (\ref{eq:J.0018.Leibniz.additive}),
this means in practical that $T^{IJ}=-T^{JI}$, and that $T_{S}^{IJ}=0$;
besides this, $T^{IJ}$ will be rather indicated as $J^{IJ}$, i.e.
the Jacobi tensor. All in all, one may write the functional formalism
for the statistical dynamics of a general Hamiltonian system, stirred
by both additive $f$ and multiplicative $g$ Gaussian time-$\delta$-correlated
noises, as follows:
\begin{equation}
\begin{cases}
 & \mathcal{A}\left[\psi;t_{\mathrm{i}},t_{\mathrm{f}}\right)=N_{0}\left(t_{\mathrm{i}},t_{\mathrm{f}}\right)e^{-i\int_{t_{\mathrm{i}}}^{t_{\mathrm{f}}}dt\mathcal{\mkern2mu\mathchar'40\mkern-7mu L}\left(\lambda,\mu,\dot{\psi},\psi\right)},\\
\\
 & \mathcal{\mkern2mu\mathchar'40\mkern-7mu L}\left(\lambda,\mu,\dot{\psi},\psi\right)=\frac{1}{4}i\beta\left(\mu,\dot{\psi},\psi\right)\cdot\tau^{-1}\left(\lambda,\mu,\psi\right)\cdot\beta^{\mathrm{T}}\left(\mu,\dot{\psi},\psi\right)+\\
\\
 & +{\displaystyle \sum_{I,J=1}^{n}}\left\{ g_{0}^{I}\frac{\partial\Gamma^{IJ}\left(\psi\right)}{\partial\psi^{J}}-\frac{i}{4\mu^{\left(I\right)}}\frac{\partial\Gamma^{IJ}\left(\psi\right)}{\partial\psi^{J}}\frac{\partial\Gamma^{IJ}\left(\psi\right)}{\partial\psi^{J}}-\frac{i}{2}\frac{\partial J^{IJ}\left(\psi\right)}{\partial\psi^{I}}\frac{\partial H\left(\psi\right)}{\partial\psi^{J}}\right\} ,\\
\\
 & \beta^{J}\left(\mu,\dot{\psi},\psi\right)\overset{\mathrm{def}}{=}\sum_{I=1}^{n}\left[\frac{\Gamma^{IJ}\left(\psi\right)}{2\mu^{\left(I\right)}}\frac{\partial}{\partial\psi^{H}}\Gamma^{IH}\left(\psi\right)+ig_{0}^{I}\Gamma^{IJ}\left(\psi\right)+i\delta^{IJ}\left(\dot{\psi}^{I}-J^{IK}\left(\psi\right)\frac{\partial H\left(\psi\right)}{\partial\psi^{K}}\right)\right],\\
\\
 & \tau^{JH}\left(\lambda,\mu,\psi\right)\overset{\mathrm{def}}{=}\frac{\delta^{JH}}{4\lambda^{\left(J\right)}}+\sum_{I=1}^{n}\frac{\Gamma^{IJ}\left(\psi\right)\Gamma^{IH}\left(\psi\right)}{4\mu^{\left(I\right)}}.
\end{cases}\label{eq:J.0019.Hamiltonian}
\end{equation}
If only additive noises appear, then one has to adapt directly (\ref{eq:J.0018.Leibniz.additive}),
i.e.:
\begin{equation}
\begin{cases}
 & \mathcal{A}_{f}\left[\psi;t_{\mathrm{i}},t_{\mathrm{f}}\right)=N_{0}\left(t_{\mathrm{i}},t_{\mathrm{f}}\right)e^{-i\int_{t_{\mathrm{i}}}^{t_{\mathrm{f}}}dt\mathcal{\mkern2mu\mathchar'40\mkern-7mu L}\left(\lambda,\dot{\psi},\psi\right)},\\
\\
 & \mathcal{\mkern2mu\mathchar'40\mkern-7mu L}\left(\lambda,\dot{\psi},\psi\right)=-\sum_{I=1}^{n}\dfrac{i}{2\sigma^{2}\left(f^{I}\right)}\left(\dot{\psi}^{I}-J^{IJ}\left(\psi\right)\frac{\partial H\left(\psi\right)}{\partial\psi^{J}}\right)^{2}+\\
\\
 & -\frac{i}{2}\sum_{I,J=1}^{n}\frac{\partial J^{IJ}\left(\psi\right)}{\partial\psi^{I}}\frac{\partial H\left(\psi\right)}{\partial\psi^{J}}.
\end{cases}\label{eq:J.0020.Hamiltonian.additive}
\end{equation}

\subsubsection{Metriplectic Systems\label{subsec:Metriplectic-Systems}}

Metriplectic systems are Leibniz systems in which the tensor $T$
is the sum of a Jacobi tensor $J$ and a semimetric tensor $G$. Moreover,
the generator $F$ takes the form of \emph{free energy} function,
namely the sum of a Hamiltonian $H$ and an entropy $S$, weighted
by some coefficient $\alpha$: $F=H+\alpha S$. The two tensors $J$
and $G$ must have the particular following relationships with the
gradients of $H$ and $S$:
\begin{equation}
J^{AB}\frac{\partial S}{\partial\psi^{B}}=0,\ G^{AB}\frac{\partial H}{\partial\psi^{B}}=0.\label{eq:J.0021.compatibility.conditions.metriplectic}
\end{equation}
As explained thoroughly in \cite{Mate.Metriplectic.Entropy} and references
therein, a metriplectic system is basically the algebrization of an
energetically closed, otherwise Hamiltonian system, to which some
dissipation is added, due to degrees of freedom whose entropy is represented
by $S$. All in all, one has energy conservation $\dot{H}=0$ and
entropy growth $\dot{S}\ge0$, thanks to the ODEs:
\begin{equation}
\frac{d\psi^{A}}{dt}=J^{AB}\left(\psi\right)\frac{\partial H\left(\psi\right)}{\partial\psi^{B}}+\alpha G^{AB}\left(\psi\right)\frac{\partial S\left(\psi\right)}{\partial\psi^{B}}.\label{eq:J.0022.classical.metriplectic}
\end{equation}
The stochastic version of the metriplectic system is simply written
as:
\begin{equation}
\frac{d\psi^{A}}{dt}=J^{AB}\left(\psi\right)\frac{\partial H\left(\psi\right)}{\partial\psi^{B}}+\alpha G^{AB}\left(\psi\right)\frac{\partial S\left(\psi\right)}{\partial\psi^{B}}+g_{B}\Gamma^{BA}\left(\psi\right)+f^{A}.\label{eq:J.023.stochastic.metriplectic}
\end{equation}

Of course, if the noises $f$ and $g$ have completely general statistics,
then all the formalism in § \ref{subsec:Stochastic-Lagrangian} is
applicable, but we are interested to the simpler particular case of
time-$\delta$-correlated Gaussian noises, so that what reported in
the general part of this § \ref{subsec:Leibniz-systems} is sufficient.
Particular reference is made to the relationship
\[
-\frac{i}{2}\frac{\partial}{\partial\psi^{I}}\left[T^{IJ}\left(\psi\right)\frac{\partial F\left(\psi\right)}{\partial\psi^{J}}\right]=-\frac{i}{2}\frac{\partial}{\partial\psi^{A}}\left[J^{AB}\left(\psi\right)\frac{\partial H\left(\psi\right)}{\partial\psi^{B}}+\alpha G^{AB}\left(\psi\right)\frac{\partial S\left(\psi\right)}{\partial\psi^{B}}\right]:
\]
as the compatibility conditions (\ref{eq:J.0021.compatibility.conditions.metriplectic})
are considered, the quantity $-\frac{i}{2}\frac{\partial}{\partial\psi^{I}}\left[T^{IJ}\left(\psi\right)\frac{\partial F\left(\psi\right)}{\partial\psi^{J}}\right]$
will be calculated as:
\begin{equation}
\begin{array}{c}
-\frac{i}{2}\frac{\partial}{\partial\psi^{I}}\left[T^{IJ}\left(\psi\right)\frac{\partial F\left(\psi\right)}{\partial\psi^{J}}\right]=\\
\\
=-\frac{i}{2}\frac{\partial J^{AB}\left(\psi\right)}{\partial\psi^{A}}\frac{\partial H\left(\psi\right)}{\partial\psi^{B}}-\frac{i\alpha}{2}\left[\frac{\partial G^{AB}\left(\psi\right)}{\partial\psi^{A}}\frac{\partial S\left(\psi\right)}{\partial\psi^{B}}+G^{AB}\left(\psi\right)\frac{\partial^{2}S\left(\psi\right)}{\partial\psi^{A}\partial\psi^{B}}\right].
\end{array}\label{eq:J.024}
\end{equation}

When the additive noise $f$ and the multiplicative one $g$ are both
present, the kernel of the \emph{metriplectic stochastic system} reads:
\begin{equation}
\begin{cases}
 & \mathcal{A}\left[\psi;t_{\mathrm{i}},t_{\mathrm{f}}\right)=N_{0}\left(t_{\mathrm{i}},t_{\mathrm{f}}\right)e^{-i\int_{t_{\mathrm{i}}}^{t_{\mathrm{f}}}dt\mathcal{\mkern2mu\mathchar'40\mkern-7mu L}\left(\lambda,\mu,\dot{\psi},\psi\right)},\\
\\
 & \mathcal{\mkern2mu\mathchar'40\mkern-7mu L}\left(\lambda,\mu,\dot{\psi},\psi\right)=\frac{1}{4}i\beta^{\mathrm{T}}\left(\mu,\dot{\psi},\psi\right)\cdot\tau^{-1}\left(\lambda,\mu,\psi\right)\cdot\beta\left(\mu,\dot{\psi},\psi\right)+\\
\\
 & +{\displaystyle \sum_{I,J=1}^{n}}\left\{ g_{0}^{I}\frac{\partial\Gamma^{IJ}\left(\psi\right)}{\partial\psi^{J}}-\frac{i}{4\mu^{\left(I\right)}}\frac{\partial\Gamma^{IJ}\left(\psi\right)}{\partial\psi^{J}}\frac{\partial\Gamma^{IJ}\left(\psi\right)}{\partial\psi^{J}}+\right.\\
\\
 & \left.-\frac{i}{2}\frac{\partial J^{AB}\left(\psi\right)}{\partial\psi^{A}}\frac{\partial H\left(\psi\right)}{\partial\psi^{B}}-\frac{i\alpha}{2}\left[\frac{\partial G^{AB}\left(\psi\right)}{\partial\psi^{A}}\frac{\partial S\left(\psi\right)}{\partial\psi^{B}}+G^{AB}\left(\psi\right)\frac{\partial^{2}S\left(\psi\right)}{\partial\psi^{A}\partial\psi^{B}}\right]\right\} ,\\
\\
 & \beta^{J}\left(\mu,\dot{\psi},\psi\right)\overset{\mathrm{def}}{=}\sum_{I=1}^{n}\left[\frac{\Gamma^{IJ}\left(\psi\right)}{2\mu^{\left(I\right)}}\frac{\partial}{\partial\psi^{H}}\Gamma^{IH}\left(\psi\right)+ig_{0}^{I}\Gamma^{IJ}\left(\psi\right)+\right.\\
\\
 & \left.+i\delta^{IJ}\left(\dot{\psi}^{I}-J^{AB}\left(\psi\right)\frac{\partial H\left(\psi\right)}{\partial\psi^{B}}-\alpha G^{AB}\left(\psi\right)\frac{\partial S\left(\psi\right)}{\partial\psi^{B}}\right)\right],\\
\\
 & \tau^{JH}\left(\lambda,\mu,\psi\right)\overset{\mathrm{def}}{=}\frac{\delta^{JH}}{4\lambda^{\left(J\right)}}+\sum_{I=1}^{n}\frac{\Gamma^{IJ}\left(\psi\right)\Gamma^{IH}\left(\psi\right)}{4\mu^{\left(I\right)}}.
\end{cases}\label{eq:J.025.metriplectic.fg}
\end{equation}
The kernel for the metriplectic system with \emph{purely additive,
Gaussian, time-$\delta$-correlated noise} $f$ will instead read:
\begin{equation}
\begin{cases}
 & \mathcal{A}_{f}\left[\psi;t_{\mathrm{i}},t_{\mathrm{f}}\right)=N_{0}\left(t_{\mathrm{i}},t_{\mathrm{f}}\right)e^{-i\int_{t_{\mathrm{i}}}^{t_{\mathrm{f}}}dt\mathcal{\mkern2mu\mathchar'40\mkern-7mu L}\left(\lambda,\dot{\psi},\psi\right)},\\
\\
 & \mathcal{\mkern2mu\mathchar'40\mkern-7mu L}\left(\lambda,\dot{\psi},\psi\right)=-\sum_{I,B=1}^{n}\dfrac{i}{2\sigma^{2}\left(f^{I}\right)}\left(\dot{\psi}^{I}-J^{IB}\left(\psi\right)\frac{\partial H\left(\psi\right)}{\partial\psi^{B}}-\alpha G^{IB}\left(\psi\right)\frac{\partial S\left(\psi\right)}{\partial\psi^{B}}\right)^{2}+\\
\\
 & -{\displaystyle \frac{i}{2}}{\displaystyle \sum_{A,B=1}^{n}}\left\{ \frac{\partial J^{AB}\left(\psi\right)}{\partial\psi^{A}}\frac{\partial H\left(\psi\right)}{\partial\psi^{B}}+\left[\frac{\partial G^{AB}\left(\psi\right)}{\partial\psi^{A}}\frac{\partial S\left(\psi\right)}{\partial\psi^{B}}+G^{AB}\left(\psi\right)\frac{\partial^{2}S\left(\psi\right)}{\partial\psi^{A}\partial\psi^{B}}\right]\right\} .
\end{cases}\label{eq:J.026.metriplectic.f}
\end{equation}

\section{Conclusions}

The whole science about ``stochastic processes'' appears to advance
a criticism to the DP described in § \ref{sec:Introduction}, which
is, instead, an upgrade of it. Indeed, \emph{non-deterministic finite
dimensional systems}, to which one resorts because of the only statistical
knowledge of some elements of the initial value problem, are promoted
to \emph{deterministic infinite-dimensional systems}, as the mathematical
quantities used to described them are not their state variables $\psi$
but, rather, statistical distributions of $\psi$.

In statistical dynamics the central tool for such a theory are \emph{master
equations} \cite{master.wiki}, i.e. the evolution equations for probability
distributions \cite{Haken.Synergetics.I}. An alternative way to describe
the same problem, appearing even more powerful to the Author of this
paper, is that of \emph{functional formalism} \cite{Phythian.77},
in which the goal is to calculate the probability that a certain particular
history $\hat{\psi}\left(t\right)$ takes place between $t_{\mathrm{i}}$
and $t_{\mathrm{f}}$, what we refer to as RPF in this work. The program
of functional formalism in Statistical Dynamics is, then, to start
from the knowledge of the noise history probability $\mathcal{P}\left[\gamma;t_{\mathrm{i}},t_{\mathrm{f}}\right)$
and of the equations of motion of $\psi$, and find the history probability
$\mathcal{A}\left[\psi;t_{\mathrm{i}},t_{\mathrm{f}}\right)$ of the
state variables $\psi$ of the system, namely studying the map (\ref{eq:P2A.thanks.Phi}).

Under the particular condition that the SDE of the system is in the
form of the ODE in (\ref{eq:Cauchy.problem.with.SDE}), i.e. the Langevin-Phythian
equation
\[
\dot{\psi}=\Lambda\left(\psi\right)+g^{\mathrm{T}}\cdot\Gamma\left(\psi\right)+f,
\]
($f$ and $g$ being noises of assigned RPF $\mathcal{P}\left[f,g;t_{\mathrm{i}},t_{\mathrm{f}}\right)$),
it is possible to obtain $\mathcal{A}\left[\psi;t_{\mathrm{i}},t_{\mathrm{f}}\right)$
in the form of imaginary exponential of a time-local function $\mathcal{\mkern2mu\mathchar'40\mkern-7mu L}\left(\dot{\psi}\left(t\right),\psi\left(t\right),t\right)$,
referred to as the \emph{stochastic Lagrangian}:
\begin{equation}
\mathcal{A}\left[\psi;t_{\mathrm{i}},t_{\mathrm{f}}\right)=N_{0}\left(t_{\mathrm{i}},t_{\mathrm{f}}\right)\exp\left[-i\int_{t_{\mathrm{i}}}^{t_{\mathrm{f}}}dt\mathcal{\mkern2mu\mathchar'40\mkern-7mu L}\left(\dot{\psi}\left(t\right),\psi\left(t\right),t\right)\right].\label{eq:conclusions.001}
\end{equation}
As the whole statistical dynamics of such a system is constructed
by using $\mathcal{A}\left[\psi;t_{\mathrm{i}},t_{\mathrm{f}}\right)$,
to some extent one could say it is encoded in $\mathcal{\mkern2mu\mathchar'40\mkern-7mu L}\left(\dot{\psi},\psi,t\right)$,
precisely as it happens for Quantum Mechanics when Feynman's path
integral approach is adopted \cite{Feynman.Hibbs.book}. Once all
the ingredients in (\ref{eq:conclusions.001}) are found, one will
use them to calculate ``anything'' of the ensemble statistics of
the system, e.g. averages, transition probabilities or correlations:
\begin{equation}
\begin{cases}
 & \left\langle F\right\rangle =\int\left[d\psi\right]F\left[\psi\right]\mathcal{A}\left[\psi;t_{\mathrm{i}},t_{\mathrm{f}}\right),\\
\\
 & P_{\psi_{\mathrm{i}}\rightarrow\psi_{\mathrm{f}}}\left(t_{\mathrm{i}},t_{\mathrm{f}}\right)=\int_{\begin{array}{c}
\psi\left(t_{\mathrm{i}}\right)=\psi_{\mathrm{i}}\\
\psi\left(t_{\mathrm{f}}\right)=\psi_{\mathrm{f}}
\end{array}}\left[d\psi\right]\mathcal{A}\left[\psi;t_{\mathrm{i}},t_{\mathrm{f}}\right),\\
\\
 & \mathcal{C}\left(t_{1},t_{2},...,t_{m}\right)=\int\left[d\psi\right]\prod_{j=1}^{m}\psi\left(t_{j}\right)\mathcal{A}\left[\psi;t_{\mathrm{i}},t_{\mathrm{f}}\right).
\end{cases}\label{eq:conclusions.002}
\end{equation}
This promises to be a very powerful tool.

Two big difficulties exist in obtaining things in (\ref{eq:conclusions.001}),
and using them as in (\ref{eq:conclusions.002}): first of all, as
underlined in § \ref{subsec:Stochastic-Lagrangian}, obtaining the
RPF of $\psi$ getting rid of the auxiliary variables $\chi$ is in
general a rather hard task, that appears to be ``simple'' only if
$f$ and $g$ are time-$\delta$-correlated Gaussian noises, see §
\ref{subsec:delta.correlated}; this impeds to obtain $\mathcal{A}\left[\psi;t_{\mathrm{i}},t_{\mathrm{f}}\right)$
from the more immediately defined $A\left[\chi,\psi;t_{\mathrm{i}},t_{\mathrm{f}}\right)$
(this difficulty is partly circumvented by the recipes in \cite{Jouvet.Phythian},
where the kernel $A\left[\chi,\psi;t_{\mathrm{i}},t_{\mathrm{f}}\right)$
is used to calculate suskeptibilities and other statistical observables).
Moreover, in all the statistical quantities as those in (\ref{eq:conclusions.002}),
there is the necessity of writing in a closed form \emph{the functional
measure} $\left[d\psi\right]$, that is \textbf{the celebrated big
difficulty} of any branch of Physics making use of path integrals!
Here, we have made the explicit calculation of the functional measure
$\int\left[d\psi\right]...\overset{\mathrm{def}}{=}...$ only in the
very simple case in which a point particle of Newton's mechanics undergoes
the action of a deterministic viscous force and a stochastic one $\vec{\xi}$
(of course, Gaussian and time-$\delta$-correlated!), so that its
SDE reads: $\frac{d\vec{p}}{dt}=-\frac{\zeta}{m}\vec{p}+\vec{\xi}$,
see § \ref{subsec:Point-Particle-with-noise} from page \pageref{paragraph.measure}
on, and all the big calculations in §§ \pageref{subsec:J0-and-Ik}
and \pageref{subsec:Keeping-N0-finite}.

As we have shown in §§ \pageref{subsec:Point-Particle-with-noise}
and \pageref{subsec:Leibniz-systems}, the functional formalism of
the stochastic Lagrangian appears to be applicable to a very wide
range of physical problems, from Newton's mechanics plus noise, to
Leibniz systems, included Hamiltonian or metriplectic systems, see
§§ \pageref{subsec:Hamiltonian-Systems} and \pageref{subsec:Metriplectic-Systems}.
Fluid dynamical and plasma physics examples are treated in the references
quoted here, and for sure much wider fields will be covered in the
future.\bigskip{}
\bigskip{}

\textbf{Acknowledgements}

At the end of this work, I want to thank Giuseppe Consolini (INAF,
Italy) and Tom Chang (MIT, Kavli Institute, USA), who always encourage
me not to stop fantasizing about mathematical objects to describe
the beauty of the world we're gifted to be living in.

Special thanks also go to Giorgio Longhi (University of Florence,
Italy) and Ruggero Vaia (CNR-ISC, Italy) for making me curious of
hacking functional integrals.

Last but not least, I feel indebted with Polish and Romanian people,
who are using the beautiful letters ``\L '' and ``\c{S}'' respectively,
that I could use for the stochastic Lagrangian $\mathcal{\mkern2mu\mathchar'40\mkern-7mu L}$
and stochastic action $\mathcal{\cedilla{S}}$: mathematics is also
a way to acknowledge the wonderful attitude of mankind to create symbols
mimicking our voices.

\appendix

\section{Calculation for § \ref{subsec:delta.correlated}}

\subsection{Calculation of $C\left[\chi,\Gamma;t_{\mathrm{i}},t_{\mathrm{f}}\right)$}

First of all, let us calculate the expression for $C\left[\chi,\Gamma;t_{\mathrm{i}},t_{\mathrm{f}}\right)$:

\medskip{}

$C\left[\chi,\Gamma;t_{\mathrm{i}},t_{\mathrm{f}}\right)=$\medskip{}

$=\left\langle e^{i\int_{t_{\mathrm{i}}}^{t_{\mathrm{f}}}dt\left[f^{I}\chi_{I}+g_{I}\Gamma^{IJ}\left(\psi\right)\chi_{J}+g_{I}\frac{\partial}{\partial\psi^{J}}\Gamma^{IJ}\left(\psi\right)\right]}\right\rangle _{f,g}=$\medskip{}

$=\left({\displaystyle \prod_{t\in\left[t_{\mathrm{i}},t_{\mathrm{f}}\right]}}\sqrt{\frac{2^{n-1}\left\Vert \lambda\left(t\right)\right\Vert dt^{n}}{\pi}}\int_{\mathbb{R}^{n}}df\left(t\right)\right)\left({\displaystyle \prod_{t\in\left[t_{\mathrm{i}},t_{\mathrm{f}}\right]}}\sqrt{\frac{2^{n-1}\left\Vert \mu\left(t\right)\right\Vert dt^{n}}{\pi}}\int_{\mathbb{R}^{n}}dg\left(t\right)\right)$\medskip{}

$e^{i\int_{t_{\mathrm{i}}}^{t_{\mathrm{f}}}dt\left[f^{I}\chi_{I}+g_{I}\Gamma^{IJ}\left(\psi\right)\chi_{J}+g_{I}\frac{\partial}{\partial\psi^{J}}\Gamma^{IJ}\left(\psi\right)\right]}*$\medskip{}

$*e^{-\int_{t_{\mathrm{i}}}^{t_{\mathrm{f}}}dt\lambda_{IJ}\left(t\right)\left(f^{I}\left(t\right)-f_{0}^{I}\left(t\right)\right)\left(f^{J}\left(t\right)-f_{0}^{J}\left(t\right)\right)}e^{-\int_{t_{\mathrm{i}}}^{t_{\mathrm{f}}}dt\mu_{IJ}\left(t\right)\left(g^{I}\left(t\right)-g_{0}^{I}\left(t\right)\right)\left(g^{J}\left(t\right)-g_{0}^{J}\left(t\right)\right)}=$\medskip{}

$=\left({\displaystyle \prod_{t\in\left[t_{\mathrm{i}},t_{\mathrm{f}}\right]}}\sqrt{\frac{2^{n-1}\left\Vert \lambda\left(t\right)\right\Vert dt^{n}}{\pi}}\int_{\mathbb{R}^{n}}df\left(t\right)\right)\left({\displaystyle \prod_{t\in\left[t_{\mathrm{i}},t_{\mathrm{f}}\right]}}\sqrt{\frac{2^{n-1}\left\Vert \mu\left(t\right)\right\Vert dt^{n}}{\pi}}\int_{\mathbb{R}^{n}}dg\left(t\right)\right)$\medskip{}

$e^{-\int_{t_{\mathrm{i}}}^{t_{\mathrm{f}}}dt\left[\lambda_{IJ}\left(t\right)\left(f^{I}\left(t\right)-f_{0}^{I}\left(t\right)\right)\left(f^{J}\left(t\right)-f_{0}^{J}\left(t\right)\right)-i\chi_{I}f^{I}\right]}*$\medskip{}

$*e^{-\int_{t_{\mathrm{i}}}^{t_{\mathrm{f}}}dt\left\{ \mu_{IJ}\left(t\right)\left(g^{I}\left(t\right)-g_{0}^{I}\left(t\right)\right)\left(g^{J}\left(t\right)-g_{0}^{J}\left(t\right)\right)-i\left[\Gamma^{IJ}\left(\psi\right)\chi_{J}+\frac{\partial}{\partial\psi^{J}}\Gamma^{IJ}\left(\psi\right)\right]g_{I}\right\} }.$\medskip{}

Without loss of generality one may put
\begin{equation}
\lambda_{IJ}=\lambda^{\left(I\right)}\delta_{IJ},\ \mu_{IJ}=\mu^{\left(I\right)}\delta_{IJ},\label{eq:diagonal.correlations}
\end{equation}
so that: and continue the calculation as:\medskip{}

$C\left[\chi,\Gamma;t_{\mathrm{i}},t_{\mathrm{f}}\right)=$\medskip{}

$=\left({\displaystyle \prod_{t\in\left[t_{\mathrm{i}},t_{\mathrm{f}}\right]}}\sqrt{\frac{2^{n-1}\left\Vert \lambda\left(t\right)\right\Vert dt^{n}}{\pi}}\int_{\mathbb{R}^{n}}df\left(t\right)\right)\left({\displaystyle \prod_{t\in\left[t_{\mathrm{i}},t_{\mathrm{f}}\right]}}\sqrt{\frac{2^{n-1}\left\Vert \mu\left(t\right)\right\Vert dt^{n}}{\pi}}\int_{\mathbb{R}^{n}}dg\left(t\right)\right)$\medskip{}

${\displaystyle \prod_{I=1}^{n}}e^{-\int_{t_{\mathrm{i}}}^{t_{\mathrm{f}}}dt\left[\lambda^{\left(I\right)}\left(t\right)\left(f^{I}\left(t\right)-f_{0}^{I}\left(t\right)\right)^{2}-i\chi_{I}f^{I}\right]}*$\medskip{}

$*{\displaystyle \prod_{I=1}^{n}}e^{-\int_{t_{\mathrm{i}}}^{t_{\mathrm{f}}}dt\left\{ \mu^{\left(I\right)}\left(g^{I}\left(t\right)-g_{0}^{I}\left(t\right)\right)^{2}-i\left[\Gamma^{IJ}\left(\psi\right)\chi_{J}+\frac{\partial}{\partial\psi^{J}}\Gamma^{IJ}\left(\psi\right)\right]g_{I}\right\} }=$\medskip{}

$=\left({\displaystyle \prod_{t\in\left[t_{\mathrm{i}},t_{\mathrm{f}}\right]}}\sqrt{\frac{2^{n-1}\left\Vert \lambda\left(t\right)\right\Vert dt^{n}}{\pi}}\int_{\mathbb{R}^{n}}df\left(t\right)\right)\left({\displaystyle \prod_{t\in\left[t_{\mathrm{i}},t_{\mathrm{f}}\right]}}\sqrt{\frac{2^{n-1}\left\Vert \mu\left(t\right)\right\Vert dt^{n}}{\pi}}\int_{\mathbb{R}^{n}}dg\left(t\right)\right)$\medskip{}

${\displaystyle \prod_{I=1}^{n}}e^{-\int_{t_{\mathrm{i}}}^{t_{\mathrm{f}}}dt\left[\lambda^{\left(I\right)}\left(t\right)\left(f^{I}\left(t\right)-f_{0}^{I}\left(t\right)\right)^{2}-i\chi_{I}f^{I}\right]}*$\medskip{}

$*{\displaystyle \prod_{I=1}^{n}}e^{-\int_{t_{\mathrm{i}}}^{t_{\mathrm{f}}}dt\left\{ \mu^{\left(I\right)}\left(g^{I}\left(t\right)-g_{0}^{I}\left(t\right)\right)^{2}-i\left[\Gamma^{IJ}\left(\psi\right)\chi_{J}+\frac{\partial}{\partial\psi^{J}}\Gamma^{IJ}\left(\psi\right)\right]g_{I}\right\} }.$\medskip{}

It is of use to give the definition:
\[
u^{I}\left(\chi,\psi\right)\overset{\mathrm{def}}{=}\Gamma^{IJ}\left(\psi\right)\chi_{J}+\frac{\partial}{\partial\psi^{J}}\Gamma^{IJ}\left(\psi\right)
\]
and to continue as:\medskip{}

$C\left[\chi,\Gamma;t_{\mathrm{i}},t_{\mathrm{f}}\right)=$\medskip{}

$=\left({\displaystyle \prod_{t\in\left[t_{\mathrm{i}},t_{\mathrm{f}}\right]}}\sqrt{\frac{2^{n-1}\left\Vert \lambda\left(t\right)\right\Vert dt^{n}}{\pi^{n}}}\int_{\mathbb{R}^{n}}df\left(t\right)\right)\left({\displaystyle \prod_{t\in\left[t_{\mathrm{i}},t_{\mathrm{f}}\right]}}\sqrt{\frac{2^{n-1}\left\Vert \mu\left(t\right)\right\Vert dt^{n}}{\pi^{n}}}\int_{\mathbb{R}^{n}}dg\left(t\right)\right)$\medskip{}

${\displaystyle \prod_{I=1}^{n}}e^{-\int_{t_{\mathrm{i}}}^{t_{\mathrm{f}}}dt\left[\lambda^{\left(I\right)}\left(t\right)\left(f^{I}\left(t\right)-f_{0}^{I}\left(t\right)\right)^{2}-i\chi_{I}f^{I}\right]}*$\medskip{}

$*{\displaystyle \prod_{I=1}^{n}}e^{-\int_{t_{\mathrm{i}}}^{t_{\mathrm{f}}}dt\left[\mu^{\left(I\right)}\left(g^{I}\left(t\right)-g_{0}^{I}\left(t\right)\right)^{2}-iu^{I}\left(\chi,\psi\right)g_{I}\right]}=$\medskip{}

$=\left({\displaystyle \prod_{t\in\left[t_{\mathrm{i}},t_{\mathrm{f}}\right]}}\sqrt{\frac{2^{n-1}\left\Vert \lambda\left(t\right)\right\Vert dt^{n}}{\pi^{n}}}\int_{\mathbb{R}^{n}}df\left(t\right)\right)\left({\displaystyle \prod_{t\in\left[t_{\mathrm{i}},t_{\mathrm{f}}\right]}}\sqrt{\frac{2^{n-1}\left\Vert \mu\left(t\right)\right\Vert dt^{n}}{\pi^{n}}}\int_{\mathbb{R}^{n}}dg\left(t\right)\right)$\medskip{}

$*{\displaystyle \prod_{I=1}^{n}}e^{-\int_{t_{\mathrm{i}}}^{t_{\mathrm{f}}}\lambda^{\left(I\right)}\left(t\right)\left(f_{0}^{I}\left(t\right)\right)^{2}dt}e^{-\int_{t_{\mathrm{i}}}^{t_{\mathrm{f}}}dt\left[\lambda^{\left(I\right)}\left(t\right)\left(f^{I}\left(t\right)\right)^{2}-\left(2\lambda^{\left(I\right)}\left(t\right)f_{0}^{I}\left(t\right)-i\chi_{I}\right)f^{I}\left(t\right)\right]}*$\medskip{}

$*{\displaystyle \prod_{I=1}^{n}}e^{-\int_{t_{\mathrm{i}}}^{t_{\mathrm{f}}}\mu^{\left(I\right)}\left(t\right)\left(g_{0}^{I}\left(t\right)\right)^{2}dt}e^{-\int_{t_{\mathrm{i}}}^{t_{\mathrm{f}}}dt\left[\mu^{\left(I\right)}\left(t\right)\left(g^{I}\left(t\right)\right)^{2}-\left(2\mu^{\left(I\right)}\left(t\right)g_{0}^{I}\left(t\right)-iu_{I}\right)g^{I}\left(t\right)\right]}=$\medskip{}

$={\displaystyle \prod_{I=1}^{n}}e^{-\int_{t_{\mathrm{i}}}^{t_{\mathrm{f}}}\left[\lambda^{\left(I\right)}\left(t\right)\left(f_{0}^{I}\left(t\right)\right)^{2}+\mu^{\left(I\right)}\left(t\right)\left(g_{0}^{I}\left(t\right)\right)^{2}\right]dt}*$\medskip{}

$*\left({\displaystyle \prod_{t\in\left[t_{\mathrm{i}},t_{\mathrm{f}}\right]}}\frac{2^{\left(n-1\right)}}{\pi^{n}}\sqrt{\left\Vert \lambda\left(t\right)\right\Vert \left\Vert \mu\left(t\right)\right\Vert dt^{2n}}\right)*$\medskip{}

$*{\displaystyle \prod_{I=1}^{n}}\int_{\mathbb{R}}df^{I}\left(t\right)e^{-dt\lambda^{\left(I\right)}\left(t\right)\left(f^{I}\left(t\right)\right)^{2}+dt\left(2\lambda^{\left(I\right)}\left(t\right)f_{0}^{I}\left(t\right)-i\chi_{I}\right)f^{I}\left(t\right)}*$\medskip{}

$*{\displaystyle \prod_{I=1}^{n}}\int_{\mathbb{R}}dg^{I}\left(t\right)e^{-dt\mu^{\left(I\right)}\left(t\right)\left(g^{I}\left(t\right)\right)^{2}+dt\left(2\mu^{\left(I\right)}\left(t\right)g_{0}^{I}\left(t\right)-iu_{I}\right)g^{I}\left(t\right)}.$\medskip{}

Pivoting on the relationship
\begin{equation}
\int e^{-Ax^{2}+Bx}dx=\dfrac{1}{2}\sqrt{\dfrac{\pi}{A}}e^{\frac{B^{2}}{4A}}\mathrm{erf}\left(\sqrt{A}x-\dfrac{B}{2\sqrt{A}}\right)+C,\label{eq:useful.001}
\end{equation}
i.e.
\begin{equation}
\int_{\mathbb{R}}e^{-Ax^{2}+Bx}dx=\sqrt{\dfrac{\pi}{A}}e^{\frac{B^{2}}{4A}},\label{eq:useful.002}
\end{equation}
one has:\medskip{}

$C\left[\chi,\Gamma;t_{\mathrm{i}},t_{\mathrm{f}}\right)=$\medskip{}

$={\displaystyle \prod_{I=1}^{n}}e^{-\int_{t_{\mathrm{i}}}^{t_{\mathrm{f}}}\left[\lambda^{\left(I\right)}\left(t\right)\left(f_{0}^{I}\left(t\right)\right)^{2}+\mu^{\left(I\right)}\left(t\right)\left(g_{0}^{I}\left(t\right)\right)^{2}\right]dt}*$\medskip{}

$*\left({\displaystyle \prod_{t\in\left[t_{\mathrm{i}},t_{\mathrm{f}}\right]}}\frac{2^{\left(n-1\right)}}{\pi^{n}}\sqrt{\left\Vert \lambda\left(t\right)\right\Vert \left\Vert \mu\left(t\right)\right\Vert dt^{2n}}\right)*$\medskip{}

$*{\displaystyle \prod_{I=1}^{n}}\sqrt{\frac{\pi}{\lambda^{\left(I\right)}\left(t\right)dt}}\exp\left(\frac{\left(2\lambda^{\left(I\right)}\left(t\right)f_{0}^{I}\left(t\right)-i\chi_{I}\right)^{2}}{4\lambda^{\left(I\right)}\left(t\right)}dt\right)*$\medskip{}

$*{\displaystyle \prod_{I=1}^{n}}\sqrt{\frac{\pi}{\mu^{\left(I\right)}\left(t\right)dt}}\exp\left(\frac{\left(2\mu^{\left(I\right)}\left(t\right)g_{0}^{I}\left(t\right)-iu_{I}\right)^{2}}{4\mu^{\left(I\right)}\left(t\right)}dt\right)=$\medskip{}

$={\displaystyle \prod_{I=1}^{n}}e^{-\int_{t_{\mathrm{i}}}^{t_{\mathrm{f}}}\left[\lambda^{\left(I\right)}\left(t\right)\left(f_{0}^{I}\left(t\right)\right)^{2}+\mu^{\left(I\right)}\left(t\right)\left(g_{0}^{I}\left(t\right)\right)^{2}\right]dt}*$\medskip{}

$*\left({\displaystyle \prod_{t\in\left[t_{\mathrm{i}},t_{\mathrm{f}}\right]}}\frac{2^{\left(n-1\right)}}{\pi^{n}}\sqrt{\left\Vert \lambda\left(t\right)\right\Vert \left\Vert \mu\left(t\right)\right\Vert dt^{2n}}\right)*$\medskip{}

$*\sqrt{\frac{\pi^{n}}{\left\Vert \lambda\left(t\right)\right\Vert dt^{n}}}{\displaystyle \prod_{I=1}^{n}}\exp\left(\frac{\left(2\lambda^{\left(I\right)}\left(t\right)f_{0}^{I}\left(t\right)-i\chi_{I}\right)^{2}}{4\lambda^{\left(I\right)}\left(t\right)}dt\right)*$\medskip{}

$*\sqrt{\frac{\pi^{n}}{\left\Vert \mu\left(t\right)\right\Vert dt^{n}}}{\displaystyle \prod_{I=1}^{n}}\exp\left(\frac{\left(2\mu^{\left(I\right)}\left(t\right)g_{0}^{I}\left(t\right)-iu_{I}\right)^{2}}{4\mu^{\left(I\right)}\left(t\right)}dt\right)=$\medskip{}

$={\displaystyle \prod_{I=1}^{n}}e^{-\int_{t_{\mathrm{i}}}^{t_{\mathrm{f}}}\left[\lambda^{\left(I\right)}\left(t\right)\left(f_{0}^{I}\left(t\right)\right)^{2}+\mu^{\left(I\right)}\left(t\right)\left(g_{0}^{I}\left(t\right)\right)^{2}\right]dt}*$\medskip{}

$*{\displaystyle \prod_{t\in\left[t_{\mathrm{i}},t_{\mathrm{f}}\right]}}2^{\left(n-1\right)}{\displaystyle \prod_{I=1}^{n}}\exp\left(\frac{\left(2\lambda^{\left(I\right)}\left(t\right)f_{0}^{I}\left(t\right)-i\chi_{I}\right)^{2}}{4\lambda^{\left(I\right)}\left(t\right)}dt\right)*$\medskip{}

$*{\displaystyle \prod_{I=1}^{n}}\exp\left(\frac{\left[2\mu^{\left(I\right)}\left(t\right)g_{0}^{I}\left(t\right)-i\left(\Gamma^{IJ}\left(\psi\right)\chi_{J}+\frac{\partial}{\partial\psi^{J}}\Gamma^{IJ}\left(\psi\right)\right)\right]^{2}}{4\mu^{\left(I\right)}\left(t\right)}dt\right)=$\medskip{}

$=2^{\left(n-1\right)}{\displaystyle \prod_{I=1}^{n}}e^{-\int_{t_{\mathrm{i}}}^{t_{\mathrm{f}}}\left[\lambda^{\left(I\right)}\left(t\right)\left(f_{0}^{I}\left(t\right)\right)^{2}+\mu^{\left(I\right)}\left(t\right)\left(g_{0}^{I}\left(t\right)\right)^{2}\right]dt}*$\medskip{}

$*{\displaystyle \prod_{I=1}^{n}}\exp\left(\int_{t_{\mathrm{i}}}^{t_{\mathrm{f}}}dt\frac{\left(2\lambda^{\left(I\right)}\left(t\right)f_{0}^{I}\left(t\right)-i\chi_{I}\right)^{2}}{4\lambda^{\left(I\right)}\left(t\right)}\right)*$\medskip{}

$*{\displaystyle \prod_{I=1}^{n}}\exp\left(\int_{t_{\mathrm{i}}}^{t_{\mathrm{f}}}dt\frac{\left[2\mu^{\left(I\right)}\left(t\right)g_{0}^{I}\left(t\right)-i\left(\Gamma^{IJ}\left(\psi\right)\chi_{J}+\frac{\partial}{\partial\psi^{J}}\Gamma^{IJ}\left(\psi\right)\right)\right]^{2}}{4\mu^{\left(I\right)}\left(t\right)}\right)=$\medskip{}

$=2^{\left(n-1\right)}{\displaystyle \prod_{I=1}^{n}}e^{-\int_{t_{\mathrm{i}}}^{t_{\mathrm{f}}}\left[\lambda^{\left(I\right)}\left(t\right)\left(f_{0}^{I}\left(t\right)\right)^{2}+\mu^{\left(I\right)}\left(t\right)\left(g_{0}^{I}\left(t\right)\right)^{2}\right]dt}*$\medskip{}

$*{\displaystyle \prod_{I=1}^{n}}e^{\int_{t_{\mathrm{i}}}^{t_{\mathrm{f}}}dt\left[-\frac{1}{4\lambda^{\left(I\right)}\left(t\right)}\chi_{I}^{2}\left(t\right)-if_{0}^{I}\left(t\right)\chi_{I}\left(t\right)+\lambda^{\left(I\right)}\left(t\right)\left(f_{0}^{I}\left(t\right)\right)^{2}\right]}*$\medskip{}

$*{\displaystyle \prod_{I=1}^{n}}\exp\left(\int_{t_{\mathrm{i}}}^{t_{\mathrm{f}}}dt\frac{\left[2\mu^{\left(I\right)}\left(t\right)g_{0}^{I}\left(t\right)-i\left(\Gamma^{IJ}\left(\psi\right)\chi_{J}+\frac{\partial}{\partial\psi^{J}}\Gamma^{IJ}\left(\psi\right)\right)\right]^{2}}{4\mu^{\left(I\right)}\left(t\right)}\right)=$\medskip{}

$=2^{\left(n-1\right)}{\displaystyle \prod_{I=1}^{n}}e^{\int_{t_{\mathrm{i}}}^{t_{\mathrm{f}}}dt\left[-\frac{1}{4\lambda^{\left(I\right)}\left(t\right)}\chi_{I}^{2}\left(t\right)-if_{0}^{I}\left(t\right)\chi_{I}\left(t\right)\right]}*$\medskip{}

$*{\displaystyle \prod_{I=1}^{n}}e^{\int_{t_{\mathrm{i}}}^{t_{\mathrm{f}}}dt\left[-\frac{1}{4\mu^{\left(I\right)}\left(t\right)}\left(\Gamma^{IJ}\left(\psi\right)\chi_{J}+\frac{\partial}{\partial\psi^{J}}\Gamma^{IJ}\left(\psi\right)\right)^{2}-ig_{0}^{I}\left(t\right)\left(\Gamma^{IJ}\left(\psi\right)\chi_{J}+\frac{\partial}{\partial\psi^{J}}\Gamma^{IJ}\left(\psi\right)\right)\right]}=$\medskip{}

$=2^{\left(n-1\right)}\exp\left\{ \int_{t_{\mathrm{i}}}^{t_{\mathrm{f}}}dt\sum_{I=1}^{n}\left[-\frac{1}{4\lambda^{\left(I\right)}\left(t\right)}\chi_{I}^{2}\left(t\right)-\frac{1}{4\mu^{\left(I\right)}\left(t\right)}\Gamma^{IJ}\left(\psi\right)\Gamma^{IH}\left(\psi\right)\chi_{J}\chi_{H}+\right.\right.$\medskip{}

$\left.\left.-\left(\frac{1}{2\mu^{\left(I\right)}\left(t\right)}\Gamma^{IJ}\left(\psi\right)\frac{\partial}{\partial\psi^{H}}\Gamma^{IH}\left(\psi\right)+ig_{0}^{I}\left(t\right)\Gamma^{IJ}\left(\psi\right)\right)\chi_{J}\right]\right\} *$\medskip{}

$*e^{-\int_{t_{\mathrm{i}}}^{t_{\mathrm{f}}}dt\sum_{I=1}^{n}\left[\frac{1}{4\mu^{\left(I\right)}\left(t\right)}\frac{\partial}{\partial\psi^{J}}\Gamma^{IJ}\left(\psi\right)\frac{\partial}{\partial\psi^{H}}\Gamma^{IH}\left(\psi\right)+ig_{0}^{I}\left(t\right)\frac{\partial}{\partial\psi^{J}}\Gamma^{IJ}\left(\psi\right)\right]}=$\medskip{}

$=e^{\int_{t_{\mathrm{i}}}^{t_{\mathrm{f}}}dt\sum_{I=1}^{n}\left[-\frac{1}{4\lambda^{\left(I\right)}}\chi_{I}^{2}-\frac{\Gamma^{IJ}\left(\psi\right)\Gamma^{IH}\left(\psi\right)}{4\mu^{\left(I\right)}}\chi_{J}\chi_{H}-\left(\frac{\Gamma^{IJ}\left(\psi\right)}{2\mu^{\left(I\right)}}\frac{\partial}{\partial\psi^{H}}\Gamma^{IH}\left(\psi\right)+ig_{0}^{I}\Gamma^{IJ}\left(\psi\right)\right)\chi_{J}\right]}*$\medskip{}

$*2^{\left(n-1\right)}e^{-\int_{t_{\mathrm{i}}}^{t_{\mathrm{f}}}dt\sum_{I=1}^{n}\left[\frac{1}{4\mu^{\left(I\right)}}\frac{\partial}{\partial\psi^{J}}\Gamma^{IJ}\left(\psi\right)\frac{\partial}{\partial\psi^{H}}\Gamma^{IH}\left(\psi\right)+ig_{0}^{I}\frac{\partial}{\partial\psi^{J}}\Gamma^{IJ}\left(\psi\right)\right]}.$\medskip{}

This leads to the expressions (\ref{eq:Mate.012}) and (\ref{eq:Mate013.C.deltacorr.J.times.M}).

\subsection{Calculation of \textmd{\normalsize{}$A\left[\psi,\chi;t_{\mathrm{i}},t_{\mathrm{f}}\right)$
and $\mathcal{A}\left[\psi;t_{\mathrm{i}},t_{\mathrm{f}}\right)$\label{subsec:A.2.A.chi.psi.A.psi}}}

Starting from (\ref{eq:Mate013.C.deltacorr.J.times.M}), let us first
of all place $C\left[\chi,\Gamma;t_{\mathrm{i}},t_{\mathrm{f}}\right)=J\left[\psi;t_{\mathrm{i}},t_{\mathrm{f}}\right)M\left[\psi,\chi;t_{\mathrm{i}},t_{\mathrm{f}}\right)$
in the definition of $A$, as follows:\medskip{}

$A\left[\psi,\chi;t_{\mathrm{i}},t_{\mathrm{f}}\right)=$\medskip{}

$=A_{0}\left(t_{\mathrm{i}},t_{\mathrm{f}}\right)C\left[\chi,\Gamma;t_{\mathrm{i}},t_{\mathrm{f}}\right)e^{-i\int_{t_{\mathrm{i}}}^{t_{\mathrm{f}}}dt\left[\dot{\psi}^{I}\chi_{I}-\Lambda^{I}\left(\psi\right)\chi_{I}-\frac{i}{2}\frac{\partial}{\partial\psi^{I}}\Lambda^{I}\left(\psi\right)\right]}=$\medskip{}

$=A_{0}\left(t_{\mathrm{i}},t_{\mathrm{f}}\right)J\left[\psi;t_{\mathrm{i}},t_{\mathrm{f}}\right)M\left[\psi,\chi;t_{\mathrm{i}},t_{\mathrm{f}}\right)e^{-i\int_{t_{\mathrm{i}}}^{t_{\mathrm{f}}}dt\left[\dot{\psi}^{I}\chi_{I}-\Lambda^{I}\left(\psi\right)\chi_{I}-\frac{i}{2}\frac{\partial}{\partial\psi^{I}}\Lambda^{I}\left(\psi\right)\right]}=$\medskip{}

$=A_{0}\left(t_{\mathrm{i}},t_{\mathrm{f}}\right)J\left[\psi;t_{\mathrm{i}},t_{\mathrm{f}}\right)M\left[\psi,\chi;t_{\mathrm{i}},t_{\mathrm{f}}\right)e^{-i\int_{t_{\mathrm{i}}}^{t_{\mathrm{f}}}dt\left[\dot{\psi}^{I}\chi_{I}-\Lambda^{I}\left(\psi\right)\chi_{I}\right]}e^{-\frac{1}{2}\int_{t_{\mathrm{i}}}^{t_{\mathrm{f}}}dt\frac{\partial}{\partial\psi^{I}}\Lambda^{I}\left(\psi\right)}=$\medskip{}

$=A_{0}\left(t_{\mathrm{i}},t_{\mathrm{f}}\right)J\left[\psi;t_{\mathrm{i}},t_{\mathrm{f}}\right)e^{-\frac{1}{2}\int_{t_{\mathrm{i}}}^{t_{\mathrm{f}}}dt\frac{\partial}{\partial\psi^{I}}\Lambda^{I}\left(\psi\right)}M\left[\psi,\chi;t_{\mathrm{i}},t_{\mathrm{f}}\right)e^{-i\int_{t_{\mathrm{i}}}^{t_{\mathrm{f}}}dt\left[\dot{\psi}^{I}\chi_{I}-\Lambda^{I}\left(\psi\right)\chi_{I}\right]}=$\medskip{}

$=A_{0}\left(t_{\mathrm{i}},t_{\mathrm{f}}\right)J\left[\psi;t_{\mathrm{i}},t_{\mathrm{f}}\right)e^{-\frac{1}{2}\int_{t_{\mathrm{i}}}^{t_{\mathrm{f}}}dt\frac{\partial}{\partial\psi^{I}}\Lambda^{I}\left(\psi\right)}*$\medskip{}

$*e^{\int_{t_{\mathrm{i}}}^{t_{\mathrm{f}}}dt\sum_{I=1}^{n}\left[-\frac{1}{4\lambda^{\left(I\right)}}\chi_{I}^{2}-\frac{\Gamma^{IJ}\left(\psi\right)\Gamma^{IH}\left(\psi\right)}{4\mu^{\left(I\right)}}\chi_{J}\chi_{H}-\left(\frac{\Gamma^{IJ}\left(\psi\right)}{2\mu^{\left(I\right)}}\frac{\partial}{\partial\psi^{H}}\Gamma^{IH}\left(\psi\right)+ig_{0}^{I}\Gamma^{IJ}\left(\psi\right)\right)\chi_{J}\right]}e^{-i\int_{t_{\mathrm{i}}}^{t_{\mathrm{f}}}dt\left[\dot{\psi}^{I}-\Lambda^{I}\left(\psi\right)\right]\chi_{I}}=$\medskip{}

$=A_{0}\left(t_{\mathrm{i}},t_{\mathrm{f}}\right)J\left[\psi;t_{\mathrm{i}},t_{\mathrm{f}}\right)e^{-\frac{1}{2}\int_{t_{\mathrm{i}}}^{t_{\mathrm{f}}}dt\frac{\partial}{\partial\psi^{I}}\Lambda^{I}\left(\psi\right)}*$\medskip{}

$*e^{\int_{t_{\mathrm{i}}}^{t_{\mathrm{f}}}dt\sum_{I=1}^{n}\left\{ -\frac{1}{4\lambda^{\left(I\right)}}\chi_{I}^{2}-\frac{\Gamma^{IJ}\left(\psi\right)\Gamma^{IH}\left(\psi\right)}{4\mu^{\left(I\right)}}\chi_{J}\chi_{H}-\left(\frac{\Gamma^{IJ}\left(\psi\right)}{2\mu^{\left(I\right)}}\frac{\partial}{\partial\psi^{H}}\Gamma^{IH}\left(\psi\right)+ig_{0}^{I}\Gamma^{IJ}\left(\psi\right)\right)\chi_{J}-i\left[\dot{\psi}^{I}-\Lambda^{I}\left(\psi\right)\right]\chi_{I}\right\} }=$\medskip{}

$=A_{0}\left(t_{\mathrm{i}},t_{\mathrm{f}}\right)J\left[\psi;t_{\mathrm{i}},t_{\mathrm{f}}\right)e^{-\frac{1}{2}\int_{t_{\mathrm{i}}}^{t_{\mathrm{f}}}dt\frac{\partial}{\partial\psi^{I}}\Lambda^{I}\left(\psi\right)}e^{-\frac{1}{4\lambda^{\left(I\right)}}\int_{t_{\mathrm{i}}}^{t_{\mathrm{f}}}dt\sum_{I=1}^{n}\chi_{I}^{2}}e^{-\int_{t_{\mathrm{i}}}^{t_{\mathrm{f}}}dt\sum_{I=1}^{n}\frac{\Gamma^{IJ}\left(\psi\right)\Gamma^{IH}\left(\psi\right)}{4\mu^{\left(I\right)}}\chi_{J}\chi_{H}}*$\medskip{}

$*e^{-\int_{t_{\mathrm{i}}}^{t_{\mathrm{f}}}dt\sum_{I=1}^{n}\left[\frac{\Gamma^{IJ}\left(\psi\right)}{2\mu^{\left(I\right)}}\frac{\partial}{\partial\psi^{H}}\Gamma^{IH}\left(\psi\right)+ig_{0}^{I}\Gamma^{IJ}\left(\psi\right)+i\delta^{IJ}\left(\dot{\psi}^{I}-\Lambda^{I}\left(\psi\right)\right)\right]\chi_{J}}=$\medskip{}

$=A_{0}\left(t_{\mathrm{i}},t_{\mathrm{f}}\right)J\left[\psi;t_{\mathrm{i}},t_{\mathrm{f}}\right)e^{-\frac{1}{2}\int_{t_{\mathrm{i}}}^{t_{\mathrm{f}}}dt\frac{\partial}{\partial\psi^{I}}\Lambda^{I}\left(\psi\right)}e^{-\int_{t_{\mathrm{i}}}^{t_{\mathrm{f}}}dt\left(\frac{\delta^{JH}}{4\lambda^{\left(J\right)}}+\sum_{I=1}^{n}\frac{\Gamma^{IJ}\left(\psi\right)\Gamma^{IH}\left(\psi\right)}{4\mu^{\left(I\right)}}\right)\chi_{J}\chi_{H}}*$\medskip{}

$*e^{-\int_{t_{\mathrm{i}}}^{t_{\mathrm{f}}}dt\sum_{I=1}^{n}\left[\frac{\Gamma^{IJ}\left(\psi\right)}{2\mu^{\left(I\right)}}\frac{\partial}{\partial\psi^{H}}\Gamma^{IH}\left(\psi\right)+ig_{0}^{I}\Gamma^{IJ}\left(\psi\right)+i\delta^{IJ}\left(\dot{\psi}^{I}-\Lambda^{I}\left(\psi\right)\right)\right]\chi_{J}}.$\medskip{}
As the tensor in (\ref{eq:Mate.014.def.tau}) is defined, with the
property (\ref{eq:Mate.022}), calculations may go ahead as follows:
\[
\begin{array}{c}
A\left[\psi,\chi;t_{\mathrm{i}},t_{\mathrm{f}}\right)\overset{(\ref{eq:Mate.014.def.tau})}{=}\\
\\
\overset{(\ref{eq:Mate.014.def.tau})}{=}A_{0}\left(t_{\mathrm{i}},t_{\mathrm{f}}\right)J\left[\psi;t_{\mathrm{i}},t_{\mathrm{f}}\right)e^{-\frac{1}{2}\int_{t_{\mathrm{i}}}^{t_{\mathrm{f}}}dt\frac{\partial}{\partial\psi^{I}}\Lambda^{I}\left(\psi\right)}e^{-\int_{t_{\mathrm{i}}}^{t_{\mathrm{f}}}dt\tau^{JH}\left(\lambda,\mu,\psi\right)\chi_{J}\chi_{H}}*\\
\\
*e^{-\int_{t_{\mathrm{i}}}^{t_{\mathrm{f}}}dt\sum_{I=1}^{n}\left[\frac{\Gamma^{IJ}\left(\psi\right)}{2\mu^{\left(I\right)}}\frac{\partial}{\partial\psi^{H}}\Gamma^{IH}\left(\psi\right)+ig_{0}^{I}\Gamma^{IJ}\left(\psi\right)+i\delta^{IJ}\left(\dot{\psi}^{I}-\Lambda^{I}\left(\psi\right)\right)\right]\chi_{J}},
\end{array}
\]

Due to the aforementioned nice property, the passage (\ref{eq:A.chi.psi.2.A.psi})
can be obtained rather easily, by making a functional integration
of the $A\left[\psi,\chi;t_{\mathrm{i}},t_{\mathrm{f}}\right)$ in
(\ref{eq:Mate.016}) in $\left[d\chi\right]$. This is done as follows:\medskip{}

$\mathcal{A}\left[\psi;t_{\mathrm{i}},t_{\mathrm{f}}\right)=\int\left[d\chi\right]A\left[\psi,\chi;t_{\mathrm{i}},t_{\mathrm{f}}\right)=$\medskip{}

$=\int\left[d\chi\right]A_{0}\left(t_{\mathrm{i}},t_{\mathrm{f}}\right)J\left[\psi;t_{\mathrm{i}},t_{\mathrm{f}}\right)e^{-\frac{1}{2}\int_{t_{\mathrm{i}}}^{t_{\mathrm{f}}}dt\frac{\partial}{\partial\psi^{I}}\Lambda^{I}\left(\psi\right)}e^{-\int_{t_{\mathrm{i}}}^{t_{\mathrm{f}}}dt\left[\tau^{JH}\left(\lambda,\mu,\psi\right)\chi_{J}\chi_{H}+\beta^{J}\left(\mu,\dot{\psi},\psi\right)\chi_{J}\right]}=$\medskip{}

$=A_{0}\left(t_{\mathrm{i}},t_{\mathrm{f}}\right)J\left[\psi;t_{\mathrm{i}},t_{\mathrm{f}}\right)e^{-\frac{1}{2}\int_{t_{\mathrm{i}}}^{t_{\mathrm{f}}}dt\frac{\partial}{\partial\psi^{I}}\Lambda^{I}\left(\psi\right)}\int\left[d\chi\right]e^{\int_{t_{\mathrm{i}}}^{t_{\mathrm{f}}}dt\left[-\chi^{\mathrm{T}}\cdot\tau\left(\lambda,\mu,\psi\right)\cdot\chi-\beta\left(\mu,\dot{\psi},\psi\right)\cdot\chi\right]}=$\medskip{}

$=A_{0}\left(t_{\mathrm{i}},t_{\mathrm{f}}\right)J\left[\psi;t_{\mathrm{i}},t_{\mathrm{f}}\right)e^{-\frac{1}{2}\int_{t_{\mathrm{i}}}^{t_{\mathrm{f}}}dt\frac{\partial}{\partial\psi^{I}}\Lambda^{I}\left(\psi\right)}*$\medskip{}

$*{\displaystyle \prod_{t\in\left[t_{\mathrm{i}},t_{\mathrm{f}}\right]}}\int_{\mathbb{R}^{n}}d^{n}\chi e^{\sum_{t\in\left[t_{\mathrm{i}},t_{\mathrm{f}}\right]}dt\left[-\chi^{\mathrm{T}}\left(t\right)\cdot\tau\left(\lambda\left(t\right),\mu\left(t\right),\psi\left(t\right)\right)\cdot\chi\left(t\right)-\beta\left(\mu\left(t\right),\dot{\psi}\left(t\right),\psi\left(t\right)\right)\cdot\chi\left(t\right)\right]}=$\medskip{}

$=A_{0}\left(t_{\mathrm{i}},t_{\mathrm{f}}\right)J\left[\psi;t_{\mathrm{i}},t_{\mathrm{f}}\right)e^{-\frac{1}{2}\int_{t_{\mathrm{i}}}^{t_{\mathrm{f}}}dt\frac{\partial}{\partial\psi^{I}}\Lambda^{I}\left(\psi\right)}\mathcal{N}\left[\psi,\mu,\lambda;t_{\mathrm{i}},t_{\mathrm{f}}\right)*$\medskip{}

$*{\displaystyle \prod_{t\in\left[t_{\mathrm{i}},t_{\mathrm{f}}\right]}}\int_{\mathbb{R}^{n}}d^{n}\chi\left(t\right)e^{\sum_{t\in\left[t_{\mathrm{i}},t_{\mathrm{f}}\right]}dt\left[-\chi^{\mathrm{T}}\left(t\right)\cdot\tau\left(\lambda\left(t\right),\mu\left(t\right),\psi\left(t\right)\right)\cdot\chi\left(t\right)-\beta\left(\mu\left(t\right),\dot{\psi}\left(t\right),\psi\left(t\right)\right)\cdot\chi\left(t\right)\right]}=...$\medskip{}

The factor $\mathcal{N}\left[\psi,\mu,\lambda;t_{\mathrm{i}},t_{\mathrm{f}}\right)$
just introduced in the foregoing passages is a correction that should
avoid mathematical non-senses when the calculations
\[
{\displaystyle \prod_{t\in\left[t_{\mathrm{i}},t_{\mathrm{f}}\right]}}\int_{\mathbb{R}^{n}}d^{n}\chi\left(t\right)e^{\sum_{t\in\left[t_{\mathrm{i}},t_{\mathrm{f}}\right]}dt\left[-\chi^{\mathrm{T}}\left(t\right)\cdot\tau\left(\lambda\left(t\right),\mu\left(t\right),\psi\left(t\right)\right)\cdot\chi\left(t\right)-\beta\left(\mu\left(t\right),\dot{\psi}\left(t\right),\psi\left(t\right)\right)\cdot\chi\left(t\right)\right]}
\]
are performed, i.e. part of the functional measure $\left[d\chi\right]$.
One may continue as:
\begin{equation}
\begin{array}{c}
\mathcal{A}\left[\psi;t_{\mathrm{i}},t_{\mathrm{f}}\right)=A_{0}\left(t_{\mathrm{i}},t_{\mathrm{f}}\right)J\left[\psi;t_{\mathrm{i}},t_{\mathrm{f}}\right)e^{-\frac{1}{2}\int_{t_{\mathrm{i}}}^{t_{\mathrm{f}}}dt\frac{\partial}{\partial\psi^{I}}\Lambda^{I}\left(\psi\right)}\mathcal{N}\left[\psi,\mu,\lambda;t_{\mathrm{i}},t_{\mathrm{f}}\right)*\\
\\
*{\displaystyle \prod_{t\in\left[t_{\mathrm{i}},t_{\mathrm{f}}\right]}}\int_{\mathbb{R}^{n}}d^{n}\chi\left(t\right)e^{-\chi^{\mathrm{T}}\left(t\right)\cdot\tau\left(\lambda\left(t\right),\mu\left(t\right),\psi\left(t\right)\right)dt\cdot\chi\left(t\right)-\beta\left(\mu\left(t\right),\dot{\psi}\left(t\right),\psi\left(t\right)\right)dt\cdot\chi\left(t\right)}
\end{array}\label{eq:Mate.018}
\end{equation}
The relationship on which one relies is obtained from the theory of
$n$-dimensional multi-variate Gaussian processes, with non-diagonal
covariance matrix
\begin{equation}
\int_{\mathbb{R}^{n}}d^{n}xe^{-x^{\mathrm{T}}\cdot A\cdot x+B\cdot x}=\sqrt{\dfrac{\pi^{n}}{\left\Vert A\right\Vert }}e^{\frac{1}{4}B\cdot A^{-1}\cdot B^{\mathrm{T}}}:\label{eq:Mate.017.int.Gaussian.n.dim}
\end{equation}
in order to use (\ref{eq:Mate.017.int.Gaussian.n.dim}) in (\ref{eq:Mate.018})
one has simply to make the following identifications:
\[
\tau\left(\lambda\left(t\right),\mu\left(t\right),\psi\left(t\right)\right)dt=A,\ -\beta\left(\mu\left(t\right),\dot{\psi}\left(t\right),\psi\left(t\right)\right)dt=B,
\]
so that one may write:
\[
\int_{\mathbb{R}^{n}}d^{n}\chi e^{-\chi^{\mathrm{T}}\cdot\tau\left(\lambda,\mu,\psi\right)dt\cdot\chi-\beta\left(\mu,\dot{\psi},\psi\right)dt\cdot\chi}=\sqrt{\dfrac{\pi^{n}}{\left\Vert \tau\left(\lambda,\mu,\psi\right)\right\Vert dt^{n}}}e^{\frac{1}{4}\beta\left(\mu,\dot{\psi},\psi\right)\cdot\tau^{-1}\left(\lambda,\mu,\psi\right)\cdot\beta^{\mathrm{T}}\left(\mu,\dot{\psi},\psi\right)dt}.
\]
This is placed into (\ref{eq:Mate.018}), and the calculation goes
ahead:\medskip{}

$\mathcal{A}\left[\psi;t_{\mathrm{i}},t_{\mathrm{f}}\right)=A_{0}\left(t_{\mathrm{i}},t_{\mathrm{f}}\right)J\left[\psi;t_{\mathrm{i}},t_{\mathrm{f}}\right)e^{-\frac{1}{2}\int_{t_{\mathrm{i}}}^{t_{\mathrm{f}}}dt\frac{\partial}{\partial\psi^{I}}\Lambda^{I}\left(\psi\right)}\mathcal{N}\left[\psi,\mu,\lambda;t_{\mathrm{i}},t_{\mathrm{f}}\right)*$\medskip{}

$*{\displaystyle \prod_{t\in\left[t_{\mathrm{i}},t_{\mathrm{f}}\right]}}\int_{\mathbb{R}^{n}}d^{n}\chi\left(t\right)e^{-\chi^{\mathrm{T}}\left(t\right)\cdot\tau\left(\lambda\left(t\right),\mu\left(t\right),\psi\left(t\right)\right)dt\cdot\chi\left(t\right)-\beta\left(\mu\left(t\right),\dot{\psi}\left(t\right),\psi\left(t\right)\right)dt\cdot\chi\left(t\right)}=$\medskip{}

$=A_{0}\left(t_{\mathrm{i}},t_{\mathrm{f}}\right)J\left[\psi;t_{\mathrm{i}},t_{\mathrm{f}}\right)e^{-\frac{1}{2}\int_{t_{\mathrm{i}}}^{t_{\mathrm{f}}}dt\frac{\partial}{\partial\psi^{I}}\Lambda^{I}\left(\psi\right)}\mathcal{N}\left[\psi,\mu,\lambda;t_{\mathrm{i}},t_{\mathrm{f}}\right)*$\medskip{}

$*{\displaystyle \prod_{t\in\left[t_{\mathrm{i}},t_{\mathrm{f}}\right]}}\sqrt{\dfrac{\pi^{n}}{\left\Vert \tau\left(\lambda\left(t\right),\mu\left(t\right),\psi\left(t\right)\right)\right\Vert dt^{n}}}e^{\frac{1}{4}\beta\left(\mu\left(t\right),\dot{\psi}\left(t\right),\psi\left(t\right)\right)\cdot\tau^{-1}\left(\lambda\left(t\right),\mu\left(t\right),\psi\left(t\right)\right)\cdot\beta^{\mathrm{T}}\left(\mu\left(t\right),\dot{\psi}\left(t\right),\psi\left(t\right)\right)dt}=$\medskip{}

$=A_{0}\left(t_{\mathrm{i}},t_{\mathrm{f}}\right)J\left[\psi;t_{\mathrm{i}},t_{\mathrm{f}}\right)e^{-\frac{1}{2}\int_{t_{\mathrm{i}}}^{t_{\mathrm{f}}}dt\frac{\partial}{\partial\psi^{I}}\Lambda^{I}\left(\psi\right)}\mathcal{N}\left[\psi,\mu,\lambda;t_{\mathrm{i}},t_{\mathrm{f}}\right){\displaystyle \prod_{t\in\left[t_{\mathrm{i}},t_{\mathrm{f}}\right]}}\sqrt{\dfrac{\pi^{n}}{\left\Vert \tau\left(\lambda\left(t\right),\mu\left(t\right),\psi\left(t\right)\right)\right\Vert dt^{n}}}*$\medskip{}

$*{\displaystyle \prod_{t\in\left[t_{\mathrm{i}},t_{\mathrm{f}}\right]}}e^{\frac{1}{4}\beta\left(\mu\left(t\right),\dot{\psi}\left(t\right),\psi\left(t\right)\right)\cdot\tau^{-1}\left(\lambda\left(t\right),\mu\left(t\right),\psi\left(t\right)\right)\cdot\beta^{\mathrm{T}}\left(\mu\left(t\right),\dot{\psi}\left(t\right),\psi\left(t\right)\right)dt}=$\medskip{}

$=A_{0}\left(t_{\mathrm{i}},t_{\mathrm{f}}\right)J\left[\psi;t_{\mathrm{i}},t_{\mathrm{f}}\right)e^{-\frac{1}{2}\int_{t_{\mathrm{i}}}^{t_{\mathrm{f}}}dt\frac{\partial}{\partial\psi^{I}}\Lambda^{I}\left(\psi\right)}\mathcal{N}\left[\psi,\mu,\lambda;t_{\mathrm{i}},t_{\mathrm{f}}\right){\displaystyle \prod_{t\in\left[t_{\mathrm{i}},t_{\mathrm{f}}\right]}}\sqrt{\dfrac{\pi^{n}}{\left\Vert \tau\left(\lambda\left(t\right),\mu\left(t\right),\psi\left(t\right)\right)\right\Vert dt^{n}}}*$\medskip{}

$*e^{\frac{1}{4}\sum_{t\in\left[t_{\mathrm{i}},t_{\mathrm{f}}\right]}\beta\left(\mu\left(t\right),\dot{\psi}\left(t\right),\psi\left(t\right)\right)\cdot\tau^{-1}\left(\lambda\left(t\right),\mu\left(t\right),\psi\left(t\right)\right)\cdot\beta^{\mathrm{T}}\left(\mu\left(t\right),\dot{\psi}\left(t\right),\psi\left(t\right)\right)dt}=$\medskip{}

$=A_{0}\left(t_{\mathrm{i}},t_{\mathrm{f}}\right)J\left[\psi;t_{\mathrm{i}},t_{\mathrm{f}}\right)e^{-\frac{1}{2}\int_{t_{\mathrm{i}}}^{t_{\mathrm{f}}}dt\frac{\partial}{\partial\psi^{I}}\Lambda^{I}\left(\psi\right)}\mathcal{N}\left[\psi,\mu,\lambda;t_{\mathrm{i}},t_{\mathrm{f}}\right){\displaystyle \prod_{t\in\left[t_{\mathrm{i}},t_{\mathrm{f}}\right]}}\sqrt{\dfrac{\pi^{n}}{\left\Vert \tau\left(\lambda\left(t\right),\mu\left(t\right),\psi\left(t\right)\right)\right\Vert dt^{n}}}*$\medskip{}

$*e^{\frac{1}{4}\int_{t_{\mathrm{i}}}^{t_{\mathrm{f}}}\beta\left(\mu\left(t\right),\dot{\psi}\left(t\right),\psi\left(t\right)\right)\cdot\tau^{-1}\left(\lambda\left(t\right),\mu\left(t\right),\psi\left(t\right)\right)\cdot\beta^{\mathrm{T}}\left(\mu\left(t\right),\dot{\psi}\left(t\right),\psi\left(t\right)\right)dt}.$\medskip{}

The factor $\mathcal{N}\left[\psi,\mu,\lambda;t_{\mathrm{i}},t_{\mathrm{f}}\right)$
can be defined as
\[
\mathcal{N}^{-1}\left[\psi,\mu,\lambda;t_{\mathrm{i}},t_{\mathrm{f}}\right)\overset{\mathrm{def}}{=}{\displaystyle \prod_{t\in\left[t_{\mathrm{i}},t_{\mathrm{f}}\right]}}\sqrt{\dfrac{\pi^{n}}{\left\Vert \tau\left(\lambda\left(t\right),\mu\left(t\right),\psi\left(t\right)\right)\right\Vert dt^{n}}},
\]
that turns our result into:
\begin{equation}
\begin{array}{c}
\mathcal{A}\left[\psi;t_{\mathrm{i}},t_{\mathrm{f}}\right)=A_{0}\left(t_{\mathrm{i}},t_{\mathrm{f}}\right)J\left[\psi;t_{\mathrm{i}},t_{\mathrm{f}}\right)e^{-\frac{1}{2}\int_{t_{\mathrm{i}}}^{t_{\mathrm{f}}}dt\frac{\partial}{\partial\psi^{I}}\Lambda^{I}\left(\psi\right)}*\\
\\
*e^{\frac{1}{4}\int_{t_{\mathrm{i}}}^{t_{\mathrm{f}}}\beta\left(\mu\left(t\right),\dot{\psi}\left(t\right),\psi\left(t\right)\right)\cdot\tau^{-1}\left(\lambda\left(t\right),\mu\left(t\right),\psi\left(t\right)\right)\cdot\beta^{\mathrm{T}}\left(\mu\left(t\right),\dot{\psi}\left(t\right),\psi\left(t\right)\right)dt}.
\end{array}\label{eq:Mate.019}
\end{equation}

It is time to conclude our calculation rendering the expression $\beta\cdot\tau^{-1}\cdot\beta^{\mathrm{T}}$
more explicit. Considering (\ref{eq:Mate.014.def.tau}) and (\ref{eq:Mate.015.def.beta}),
once one has defined
\begin{equation}
\alpha^{IJ}\left(\mu,\psi\right)\overset{\mathrm{def}}{=}\frac{\Gamma^{IJ}\left(\psi\right)}{\mu^{\left(I\right)}}\label{eq:Mate.021.def.alpha}
\end{equation}
one has:\medskip{}

$\frac{1}{4}\beta\cdot\tau^{-1}\cdot\beta^{\mathrm{T}}=\frac{1}{4}\beta_{J}\beta_{H}\left(\tau^{-1}\right)^{JH}\overset{(\ref{eq:Mate.021.def.alpha})}{=}$\medskip{}

$\overset{(\ref{eq:Mate.021.def.alpha})}{=}\frac{1}{4}\left[\frac{\alpha_{IJ}\left(\psi\right)}{2}\frac{\partial\Gamma^{IK}\left(\psi\right)}{\partial\psi^{K}}+ig_{0}^{I}\Gamma_{IJ}\left(\psi\right)+i\left(\dot{\psi}_{J}-\Lambda_{J}\left(\psi\right)\right)\right]*$\medskip{}

$*\left[\frac{\alpha_{PH}\left(\psi\right)}{2}\frac{\partial\Gamma^{PN}\left(\psi\right)}{\partial\psi^{N}}+ig_{0}^{M}\Gamma_{MH}\left(\psi\right)+i\left(\dot{\psi}_{H}-\Lambda_{H}\left(\psi\right)\right)\right]\left(\tau^{-1}\right)^{JH}=$\medskip{}

$=\frac{1}{4}\left[i\left(\dot{\psi}_{J}-\Lambda_{J}\left(\psi\right)\right)+ig_{0}^{I}\Gamma_{IJ}\left(\psi\right)+\frac{\alpha_{IJ}\left(\psi\right)}{2}\frac{\partial\Gamma^{IK}\left(\psi\right)}{\partial\psi^{K}}\right]*$\medskip{}

$*\left[i\left(\dot{\psi}_{H}-\Lambda_{H}\left(\psi\right)\right)+ig_{0}^{M}\Gamma_{MH}\left(\psi\right)+\frac{\alpha_{PH}\left(\psi\right)}{2}\frac{\partial\Gamma^{PN}\left(\psi\right)}{\partial\psi^{N}}\right]\left(\tau^{-1}\right)^{JH}=$\medskip{}

$=\frac{1}{4}i\left[\left(\dot{\psi}_{J}-\Lambda_{J}\left(\psi\right)\right)+g_{0}^{I}\Gamma_{IJ}\left(\psi\right)-i\frac{\alpha_{IJ}\left(\psi\right)}{2}\frac{\partial\Gamma^{IK}\left(\psi\right)}{\partial\psi^{K}}\right]*$\medskip{}

$*\left[i\left(\dot{\psi}_{H}-\Lambda_{H}\left(\psi\right)\right)+ig_{0}^{M}\Gamma_{MH}\left(\psi\right)+\frac{\alpha_{PH}\left(\psi\right)}{2}\frac{\partial\Gamma^{PN}\left(\psi\right)}{\partial\psi^{N}}\right]\left(\tau^{-1}\right)^{JH}=$\medskip{}

$=\frac{1}{4}i^{2}\left[\dot{\psi}_{J}-\Lambda_{J}\left(\psi\right)+g_{0}^{I}\Gamma_{IJ}\left(\psi\right)-i\frac{\alpha_{IJ}\left(\psi\right)}{2}\frac{\partial\Gamma^{IK}\left(\psi\right)}{\partial\psi^{K}}\right]*$\medskip{}

$*\left[\dot{\psi}_{H}-\Lambda_{H}\left(\psi\right)+g_{0}^{M}\Gamma_{MH}\left(\psi\right)-i\frac{\alpha_{PH}\left(\psi\right)}{2}\frac{\partial\Gamma^{PN}\left(\psi\right)}{\partial\psi^{N}}\right]\left(\tau^{-1}\right)^{JH}=$\medskip{}

$=-\frac{1}{4}\left[\dot{\psi}_{J}-\Lambda_{J}\left(\psi\right)+g_{0}^{I}\Gamma_{IJ}\left(\psi\right)-i\frac{\alpha_{IJ}\left(\psi\right)}{2}\frac{\partial\Gamma^{IK}\left(\psi\right)}{\partial\psi^{K}}\right]*$\medskip{}

$*\left[\dot{\psi}_{H}-\Lambda_{H}\left(\psi\right)+g_{0}^{M}\Gamma_{MH}\left(\psi\right)-i\frac{\alpha_{PH}\left(\psi\right)}{2}\frac{\partial\Gamma^{PN}\left(\psi\right)}{\partial\psi^{N}}\right]\left(\tau^{-1}\right)^{JH}=$\medskip{}

$=-\frac{1}{4}\left(\tau^{-1}\right)^{JH}\left(\dot{\psi}_{J}+E_{J}\left(\psi\right)\right)\left(\dot{\psi}_{H}+E_{H}\left(\psi\right)\right),$\medskip{}

\noindent where one has defined:
\[
E_{J}\left(\psi\right)\overset{\mathrm{def}}{=}g_{0}^{I}\Gamma_{IJ}\left(\psi\right)-\Lambda_{J}\left(\psi\right)-i\frac{\alpha_{IJ}\left(\psi\right)}{2}\frac{\partial\Gamma^{IK}\left(\psi\right)}{\partial\psi^{K}}.
\]
In this way, the expression (\ref{eq:Mate.023}) ends up reading:
\[
\begin{array}{c}
\mathcal{A}\left[\psi;t_{\mathrm{i}},t_{\mathrm{f}}\right)=N_{0}\left(t_{\mathrm{i}},t_{\mathrm{f}}\right)e^{-\frac{1}{4}\int_{t_{\mathrm{i}}}^{t_{\mathrm{f}}}dt\left(\tau^{-1}\right)^{JH}\left(\dot{\psi}_{J}+E_{J}\left(\psi\right)\right)\left(\dot{\psi}_{H}+E_{H}\left(\psi\right)\right)}*\\
\\
*e^{-\int_{t_{\mathrm{i}}}^{t_{\mathrm{f}}}dt\sum_{I,J=1}^{n}\left[\frac{1}{4\mu^{\left(I\right)}}\frac{\partial\Gamma^{IJ}\left(\psi\right)}{\partial\psi^{J}}\frac{\partial\Gamma^{IJ}\left(\psi\right)}{\partial\psi^{J}}+\frac{1}{2}\frac{\partial\Lambda^{I}\left(\psi\right)}{\partial\psi^{I}}+ig_{0}^{I}\frac{\partial\Gamma^{IJ}\left(\psi\right)}{\partial\psi^{J}}\right]};
\end{array}
\]
about this, as the matrix $\tau$ in (\ref{eq:Mate.014.def.tau})
appears to be positive definite, see also (\ref{eq:Mate.022}), what
we note in $\mathcal{A}\left[\psi;t_{\mathrm{i}},t_{\mathrm{f}}\right)$
is, first of all, that it contains a negative square term in the velocities
$-\frac{1}{4}\left(\tau^{-1}\right)^{JH}\dot{\psi}_{J}\dot{\psi}_{H}$,
that will help in the convergence of $\mathcal{A}\left[\psi;t_{\mathrm{i}},t_{\mathrm{f}}\right)$
and the determination of the functional measure $\left[d\psi\right]$.

\section{Calculations for § \ref{subsec:Point-Particle-with-noise}}

\subsection{Calculation of the Stochastic Lagrangian\label{subsec:Point.Particle.SL}}

What we need to go ahead with the calculations in (\ref{eq:J.0011.SL.Brown.002})
is to determine $\frac{1}{2}\sum_{h=1}^{3}\frac{\partial\Phi^{h}}{\partial p^{h}}$,
that is done as follows:\medskip{}

${\displaystyle \frac{1}{2}}{\displaystyle \sum_{h=1}^{3}}{\displaystyle \frac{\partial\Phi^{h}}{\partial p^{h}}}={\displaystyle \frac{1}{2}}{\displaystyle \sum_{h=1}^{3}}{\displaystyle \frac{\partial}{\partial p^{h}}}\left[-\dfrac{\partial}{\partial x_{h}}V\left(\vec{x}\right)+{\displaystyle \frac{q}{m}}\epsilon^{hij}p_{i}B_{j}\left(\vec{x}\right)-{\displaystyle \frac{\zeta}{m}}p^{h}\right]=$\medskip{}

$={\displaystyle \frac{1}{2}}{\displaystyle \sum_{h=1}^{3}}\left[{\displaystyle \frac{q}{m}}\epsilon^{hij}{\displaystyle \frac{\partial p_{i}}{\partial p^{h}}}B_{j}\left(\vec{x}\right)-{\displaystyle \frac{\zeta}{m}}{\displaystyle \frac{\partial p^{h}}{\partial p^{h}}}\right]={\displaystyle \frac{1}{2}}{\displaystyle \frac{q}{m}}{\displaystyle \sum_{h=1}^{3}}\epsilon^{hij}\delta_{hi}B_{j}\left(\vec{x}\right)-{\displaystyle \frac{1}{2}}{\displaystyle \sum_{h=1}^{3}}{\displaystyle \frac{\zeta}{m}}=$\medskip{}

$\overset{\mathrm{symm}}{=}{\displaystyle \frac{1}{2}}{\displaystyle \frac{q}{m}}\cdot0-{\displaystyle \frac{3}{2}}{\displaystyle \frac{\zeta}{m}}=-{\displaystyle \frac{3}{2}}{\displaystyle \frac{\zeta}{m}}.$\medskip{}

Calculations for discretization

With (\ref{eq:FH.p.dot}) and (\ref{eq:FH.p.mid}) one performs the
following calculations:

\medskip{}

$\dot{p}_{i}^{2}\left(k\right)+{\displaystyle \frac{\zeta^{2}}{m^{2}}}p_{i}^{2}\left(k\right)\cong$

\medskip{}

$\cong{\displaystyle \frac{1}{\epsilon^{2}}}\left[p_{i}\left(k+1\right)-p_{i}\left(k\right)\right]^{2}+{\displaystyle \frac{\zeta^{2}}{4m^{2}}}\left[p_{i}\left(k+1\right)+p_{i}\left(k\right)\right]^{2}=$

\medskip{}

$\cong{\displaystyle \frac{1}{\epsilon^{2}}}\left[p_{i}^{2}\left(k+1\right)+p_{i}^{2}\left(k\right)-2p_{i}\left(k+1\right)p_{i}\left(k\right)\right]+$

\medskip{}

$+{\displaystyle \frac{\zeta^{2}}{4m^{2}}}\left[p_{i}^{2}\left(k+1\right)+p_{i}^{2}\left(k\right)+2p_{i}\left(k+1\right)p_{i}\left(k\right)\right]=$

\medskip{}

$={\displaystyle \frac{1}{\epsilon^{2}}}p_{i}^{2}\left(k+1\right)+{\displaystyle \frac{1}{\epsilon^{2}}}p_{i}^{2}\left(k\right)-{\displaystyle \frac{2}{\epsilon^{2}}}p_{i}\left(k+1\right)p_{i}\left(k\right)+$

\medskip{}

$+{\displaystyle \frac{\mu^{2}}{4m^{2}}}p_{i}^{2}\left(k+1\right)+{\displaystyle \frac{\mu^{2}}{4m^{2}}}p_{i}^{2}\left(k\right)+{\displaystyle \frac{\mu^{2}}{2m^{2}}}p_{i}\left(k+1\right)p_{i}\left(k\right)=$

\medskip{}

$={\displaystyle \frac{1}{\epsilon^{2}}}p_{i}^{2}\left(k+1\right)+{\displaystyle \frac{\zeta^{2}}{4m^{2}}}p_{i}^{2}\left(k+1\right)+{\displaystyle \frac{1}{\epsilon^{2}}}p_{i}^{2}\left(k\right)+{\displaystyle \frac{\zeta^{2}}{4m^{2}}}p_{i}^{2}\left(k\right)+$

\medskip{}

$+{\displaystyle \frac{\zeta^{2}}{2m^{2}}}p_{i}\left(k+1\right)p_{i}\left(k\right)-{\displaystyle \frac{2}{\epsilon^{2}}}p_{i}\left(k+1\right)p_{i}\left(k\right)=$

\medskip{}

$=\left({\displaystyle \frac{1}{\epsilon^{2}}}+{\displaystyle \frac{\zeta^{2}}{4m^{2}}}\right)p_{i}^{2}\left(k+1\right)+\left({\displaystyle \frac{1}{\epsilon^{2}}}+{\displaystyle \frac{\zeta^{2}}{4m^{2}}}\right)p_{i}^{2}\left(k\right)+$

\medskip{}

$+\left({\displaystyle \frac{\zeta^{2}}{2m^{2}}}-{\displaystyle \frac{2}{\epsilon^{2}}}\right)p_{i}\left(k+1\right)p_{i}\left(k\right)=$

\medskip{}

$=\left({\displaystyle \frac{4m^{2}+\epsilon^{2}\zeta^{2}}{4m^{2}}}\right)p_{i}^{2}\left(k+1\right)+\left({\displaystyle \frac{4m^{2}+\epsilon^{2}\zeta^{2}}{4m^{2}}}\right)p_{i}^{2}\left(k\right)+$

\medskip{}

$+\left({\displaystyle \frac{\zeta^{2}\epsilon^{2}-4m^{2}}{2m^{2}\epsilon^{2}}}\right)p_{i}\left(k+1\right)p_{i}\left(k\right).$

\medskip{}

\noindent After all, one has:
\begin{equation}
\mathcal{\cedilla{S}}\cong-i\sum_{k=0}^{N-1}\epsilon\mathcal{\mkern2mu\mathchar'40\mkern-7mu L}\left(k\right).\label{eq:mea1.001}
\end{equation}

\subsection{Calculating $\mathcal{J}_{0}$ and $\mathcal{I}_{k}$\label{subsec:J0-and-Ik}}

In order to make the calculations needed in (\ref{eq:Sstoca.004}),
we start by computing $\mathcal{J}_{0}$:

\medskip{}

$\mathcal{J}_{0}=\mathcal{W}_{p}\left(0\right){\displaystyle \prod_{i=1}^{3}}\int_{\mathbb{R}}dp_{i}\left(t_{\mathrm{i}}\right)e^{\frac{\zeta}{m}\lambda_{\left(i\right)}p_{i}^{2}\left(t_{\mathrm{i}}\right)}e^{-\frac{\lambda_{\left(i\right)}}{\epsilon}\left[p_{i}^{2}\left(1\right)+p_{i}^{2}\left(t_{\mathrm{i}}\right)-2p_{i}\left(1\right)p_{i}\left(t_{\mathrm{i}}\right)\right]}=$

\medskip{}

$=\mathcal{W}_{p}\left(0\right){\displaystyle \prod_{i=1}^{3}}\int_{\mathbb{R}}dp_{i}\left(t_{\mathrm{i}}\right)e^{\frac{\zeta}{m}\lambda_{\left(i\right)}p_{i}^{2}\left(t_{\mathrm{i}}\right)-\frac{\lambda_{\left(i\right)}}{\epsilon}p_{i}^{2}\left(1\right)-\frac{\lambda_{\left(i\right)}}{\epsilon}p_{i}^{2}\left(t_{\mathrm{i}}\right)+\frac{\lambda_{\left(i\right)}}{\epsilon}2p_{i}\left(1\right)p_{i}\left(t_{\mathrm{i}}\right)}=$

\medskip{}

$=\mathcal{W}_{p}\left(0\right){\displaystyle \prod_{i=1}^{3}}\int_{\mathbb{R}}dp_{i}\left(t_{\mathrm{i}}\right)e^{-\frac{\lambda_{\left(i\right)}}{\epsilon}p_{i}^{2}\left(1\right)}e^{\frac{\zeta}{m}\lambda_{\left(i\right)}p_{i}^{2}\left(t_{\mathrm{i}}\right)-\frac{\lambda_{\left(i\right)}}{\epsilon}p_{i}^{2}\left(t_{\mathrm{i}}\right)+\frac{\lambda_{\left(i\right)}}{\epsilon}2p_{i}\left(1\right)p_{i}\left(t_{\mathrm{i}}\right)}=$

\medskip{}

$=\mathcal{W}_{p}\left(0\right){\displaystyle \prod_{i=1}^{3}}e^{-\frac{\lambda_{\left(i\right)}}{\epsilon}p_{i}^{2}\left(1\right)}\int_{\mathbb{R}}dp_{i}\left(t_{\mathrm{i}}\right)e^{\frac{\mu}{m}\lambda_{\left(i\right)}p_{i}^{2}\left(t_{\mathrm{i}}\right)-\frac{\lambda_{\left(i\right)}}{\epsilon}p_{i}^{2}\left(t_{\mathrm{i}}\right)+\frac{\lambda_{\left(i\right)}}{\epsilon}2p_{i}\left(1\right)p_{i}\left(t_{\mathrm{i}}\right)}=$

\medskip{}

$=\mathcal{W}_{p}\left(0\right){\displaystyle \prod_{i=1}^{3}}e^{-\frac{\lambda_{\left(i\right)}}{\epsilon}p_{i}^{2}\left(1\right)}\int_{\mathbb{R}}dp_{i}\left(t_{\mathrm{i}}\right)e^{-\lambda_{\left(i\right)}\left(\frac{m-\epsilon\zeta}{\epsilon m}\right)p_{i}^{2}\left(t_{\mathrm{i}}\right)+\frac{2\lambda_{\left(i\right)}p_{i}\left(1\right)}{\epsilon}p_{i}\left(t_{\mathrm{i}}\right)}=$

\medskip{}

$=\mathcal{W}_{p}\left(0\right){\displaystyle \prod_{i=1}^{3}}e^{-\frac{\lambda_{\left(i\right)}}{\epsilon}p_{i}^{2}\left(1\right)}\int_{\mathbb{R}}d\rho e^{-\lambda_{\left(i\right)}\left(\frac{m-\epsilon\mu}{\epsilon m}\right)\rho^{2}+\frac{2\lambda_{\left(i\right)}p_{i}\left(1\right)}{\epsilon}\rho}\overset{(\ref{eq:useful.002})}{=}$

\medskip{}

$\overset{(\ref{eq:useful.002})}{=}\mathcal{W}_{p}\left(0\right){\displaystyle \prod_{i=1}^{3}}e^{-\frac{\lambda_{\left(i\right)}}{\epsilon}p_{i}^{2}\left(1\right)}\sqrt{\dfrac{\pi m\epsilon}{\lambda_{\left(i\right)}\left(m-\epsilon\zeta\right)}}e^{\frac{\lambda_{\left(i\right)}mp_{i}^{2}\left(1\right)}{\epsilon\left(m-\epsilon\zeta\right)}}=$

\medskip{}

$=\mathcal{W}_{p}\left(0\right){\displaystyle \prod_{i=1}^{3}}e^{-\frac{\lambda_{\left(i\right)}}{\epsilon}p_{i}^{2}\left(1\right)}e^{\frac{\lambda_{\left(i\right)}}{\epsilon}\left(\frac{m}{m-\epsilon\zeta}\right)p_{i}^{2}\left(1\right)}\cdot{\displaystyle \prod_{i=1}^{3}}\sqrt{\dfrac{\pi m\epsilon}{\lambda_{\left(i\right)}\left(m-\epsilon\zeta\right)}}\overset{o\left(\epsilon\right)}{=}$

\medskip{}

$\overset{o\left(\epsilon\right)}{=}\mathcal{W}_{p}\left(0\right)\sqrt{\left\Vert \lambda\right\Vert ^{-1}}\left(\pi\epsilon\right)^{\frac{3}{2}}{\displaystyle \prod_{i=1}^{3}}e^{-\frac{\lambda_{\left(i\right)}}{\epsilon}p_{i}^{2}\left(1\right)}e^{\frac{\lambda_{\left(i\right)}}{\epsilon}p_{i}^{2}\left(1\right)}=$

\medskip{}

$=\mathcal{W}_{p}\left(0\right)\sqrt{\left\Vert \lambda\right\Vert ^{-1}}=\mathcal{W}_{p}\left(0\right)\sqrt{\dfrac{\pi^{3}\epsilon^{3}}{\left\Vert \lambda\right\Vert }}.$

\medskip{}

Let us, then, calculate the expression of $\mathcal{I}_{k}$ as defined
in (\ref{eq:I.k.def}) and then compute the product $\prod_{k=1}^{N-1}\mathcal{I}_{k}$:

\medskip{}

$\mathcal{I}_{k}=\mathcal{W}_{p}\left(k\right){\displaystyle \prod_{i=1}^{3}}\int_{\mathbb{R}}dp_{i}\left(t_{k}\right)e^{-\frac{\lambda_{\left(i\right)}}{\epsilon}\left[p_{i}^{2}\left(k+1\right)+p_{i}^{2}\left(k\right)-2p_{i}\left(k+1\right)p_{i}\left(k\right)\right]}=$

\medskip{}

$=\mathcal{W}_{p}\left(k\right){\displaystyle \prod_{i=1}^{3}}\int_{\mathbb{R}}dp_{i}\left(t_{k}\right)e^{-\frac{\lambda_{\left(i\right)}}{\epsilon}p_{i}^{2}\left(k+1\right)-\frac{\lambda_{\left(i\right)}}{\epsilon}p_{i}^{2}\left(k\right)+\frac{2\lambda_{\left(i\right)}}{\epsilon}p_{i}\left(k+1\right)p_{i}\left(k\right)}=$

\medskip{}

$=\mathcal{W}_{p}\left(k\right){\displaystyle \prod_{i=1}^{3}}\int_{\mathbb{R}}dp_{i}\left(t_{k}\right)e^{-\frac{\lambda_{\left(i\right)}}{\epsilon}p_{i}^{2}\left(k+1\right)}e^{-\frac{\lambda_{\left(i\right)}}{\epsilon}p_{i}^{2}\left(k\right)+\frac{2\lambda_{\left(i\right)}}{\epsilon}p_{i}\left(k+1\right)p_{i}\left(k\right)}=$

\medskip{}

$=e^{-\frac{\lambda_{\left(i\right)}}{\epsilon}p_{i}^{2}\left(k+1\right)}\mathcal{W}_{p}\left(k\right){\displaystyle \prod_{i=1}^{3}}\int_{\mathbb{R}}dp_{i}\left(t_{k}\right)e^{-\frac{\lambda_{\left(i\right)}}{\epsilon}p_{i}^{2}\left(k\right)+\frac{2\lambda_{\left(i\right)}}{\epsilon}p_{i}\left(k+1\right)p_{i}\left(k\right)}=$

\medskip{}

$=e^{-\frac{\lambda_{\left(i\right)}}{\epsilon}p_{i}^{2}\left(k+1\right)}\mathcal{W}_{p}\left(k\right){\displaystyle \prod_{i=1}^{3}}\int_{\mathbb{R}}d\rho e^{-\frac{\lambda_{\left(i\right)}}{\epsilon}\rho^{2}+\frac{2\lambda_{\left(i\right)}p_{i}\left(k+1\right)}{\epsilon}\rho}\overset{(\ref{eq:useful.002})}{=}$

\medskip{}

$\overset{(\ref{eq:useful.002})}{=}e^{-\frac{\lambda_{\left(i\right)}}{\epsilon}p_{i}^{2}\left(k+1\right)}\mathcal{W}_{p}\left(k\right){\displaystyle \prod_{i=1}^{3}}\sqrt{\dfrac{\pi\epsilon}{\lambda_{\left(i\right)}}}e^{\frac{\lambda_{\left(i\right)}p_{i}^{2}\left(k+1\right)}{\epsilon}}=$

\medskip{}

$=e^{-\frac{\lambda_{\left(i\right)}}{\epsilon}p_{i}^{2}\left(k+1\right)}e^{\frac{\lambda_{\left(i\right)}p_{i}^{2}\left(k+1\right)}{\epsilon}}\mathcal{W}_{p}\left(k\right){\displaystyle \prod_{i=1}^{3}}\sqrt{\dfrac{\pi\epsilon}{\lambda_{\left(i\right)}}}=\mathcal{W}_{p}\left(k\right)\sqrt{\dfrac{\pi^{3}\epsilon^{3}}{\left\Vert \lambda\right\Vert }}.$

\subsection{Keeping $\tilde{N}_{0}\left(\zeta;t_{\mathrm{i}},t_{\mathrm{f}}\right)$
finite\label{subsec:Keeping-N0-finite}}

In order to keep $\tilde{N}_{0}\left(\zeta;t_{\mathrm{i}},t_{\mathrm{f}}\right)$
finite in (\ref{eq:zio.004}), the reasoning is the following one:

\medskip{}

$\tilde{N}_{0}\left(\zeta;t_{\mathrm{i}},t_{\mathrm{f}}\right)={\displaystyle \lim_{\begin{array}{c}
\epsilon\rightarrow0\\
N\rightarrow+\infty
\end{array}}}\dfrac{e^{\frac{\zeta}{m}\underset{i}{\sum}\lambda_{\left(i\right)}\left[p_{i}^{2}\left(t_{\mathrm{f}}\right)-p_{i}^{2}\left(t_{\mathrm{i}}\right)\right]}}{{\displaystyle \prod_{k=0}^{N}}\mathcal{W}_{p}\left(k\right)\mathcal{V}_{p}\left(t_{\mathrm{f}}\right)\mathcal{V}_{p}\left(t_{\mathrm{i}}\right)}\left(\dfrac{\left\Vert \lambda\right\Vert }{\pi^{3}\epsilon^{3}}\right)^{\frac{N-1}{2}}\equiv n_{0},$

\medskip{}

${\displaystyle \lim_{\begin{array}{c}
\epsilon\rightarrow0\\
N\rightarrow+\infty
\end{array}}}\dfrac{e^{\frac{\zeta}{m}\underset{i}{\sum}\lambda_{\left(i\right)}\left[p_{i}^{2}\left(t_{\mathrm{f}}\right)-p_{i}^{2}\left(t_{\mathrm{i}}\right)\right]}}{{\displaystyle \prod_{k=0}^{N}}\mathcal{W}_{p}\left(k\right)\mathcal{V}_{p}\left(t_{\mathrm{f}}\right)\mathcal{V}_{p}\left(t_{\mathrm{i}}\right)}\left(\dfrac{\left\Vert \lambda\right\Vert }{\pi^{3}\epsilon^{3}}\right)^{\frac{N-1}{2}}=n_{0},$

\medskip{}

${\displaystyle \lim_{\begin{array}{c}
\epsilon\rightarrow0\\
N\rightarrow+\infty
\end{array}}}\left({\displaystyle \prod_{k=0}^{N}}\mathcal{W}_{p}\left(k\right)\right)^{-1}{\displaystyle \lim_{\begin{array}{c}
\epsilon\rightarrow0\\
N\rightarrow+\infty
\end{array}}}\dfrac{e^{\frac{\zeta}{m}\underset{i}{\sum}\lambda_{\left(i\right)}\left[p_{i}^{2}\left(t_{\mathrm{f}}\right)-p_{i}^{2}\left(t_{\mathrm{i}}\right)\right]}}{\mathcal{V}_{p}\left(t_{\mathrm{f}}\right)\mathcal{V}_{p}\left(t_{\mathrm{i}}\right)}\left(\dfrac{\left\Vert \lambda\right\Vert }{\pi^{3}\epsilon^{3}}\right)^{\frac{N-1}{2}}=n_{0},$

\medskip{}

${\displaystyle \lim_{\begin{array}{c}
\epsilon\rightarrow0\\
N\rightarrow+\infty
\end{array}}}\dfrac{e^{\frac{\zeta}{m}\underset{i}{\sum}\lambda_{\left(i\right)}\left[p_{i}^{2}\left(t_{\mathrm{f}}\right)-p_{i}^{2}\left(t_{\mathrm{i}}\right)\right]}}{\mathcal{V}_{p}\left(t_{\mathrm{f}}\right)\mathcal{V}_{p}\left(t_{\mathrm{i}}\right)}\left(\dfrac{\left\Vert \lambda\right\Vert }{\pi^{3}\epsilon^{3}}\right)^{\frac{N-1}{2}}=n_{0}{\displaystyle \lim_{\begin{array}{c}
\epsilon\rightarrow0\\
N\rightarrow+\infty
\end{array}}}{\displaystyle \prod_{k=0}^{N}}\mathcal{W}_{p}\left(k\right),$

\medskip{}

${\displaystyle \lim_{\begin{array}{c}
\epsilon\rightarrow0\\
N\rightarrow+\infty
\end{array}}}\dfrac{e^{\frac{\zeta}{m}\underset{i}{\sum}\lambda_{\left(i\right)}\left[p_{i}^{2}\left(t_{\mathrm{f}}\right)-p_{i}^{2}\left(t_{\mathrm{i}}\right)\right]}}{n_{0}\mathcal{V}_{p}\left(t_{\mathrm{f}}\right)\mathcal{V}_{p}\left(t_{\mathrm{i}}\right)}\left(\dfrac{\left\Vert \lambda\right\Vert }{\pi^{3}\epsilon^{3}}\right)^{\frac{N-1}{2}}={\displaystyle \lim_{\begin{array}{c}
\epsilon\rightarrow0\\
N\rightarrow+\infty
\end{array}}}{\displaystyle \prod_{k=0}^{N}}\mathcal{W}_{p}\left(k\right).$

\section{Calculations for § \ref{subsec:Leibniz-systems}\label{sec:Calculations-for-Leibniz-Systems}}

In order to obtain the relationships (\ref{eq:J.0017}) one starts
from:
\begin{equation}
\begin{cases}
 & \mathcal{A}\left[\psi;t_{\mathrm{i}},t_{\mathrm{f}}\right)=N_{0}\left(t_{\mathrm{i}},t_{\mathrm{f}}\right)e^{-i\int_{t_{\mathrm{i}}}^{t_{\mathrm{f}}}dt\mathcal{\mkern2mu\mathchar'40\mkern-7mu L}\left(\lambda,\mu,\dot{\psi},\psi\right)},\\
\\
 & \mathcal{\mkern2mu\mathchar'40\mkern-7mu L}\left(\lambda,\mu,\dot{\psi},\psi\right)=\frac{1}{4}i\beta\left(\mu,\dot{\psi},\psi\right)\cdot\tau^{-1}\left(\lambda,\mu,\psi\right)\cdot\beta^{\mathrm{T}}\left(\mu,\dot{\psi},\psi\right)+\\
\\
 & +{\displaystyle \sum_{I,J=1}^{n}}\left\{ g_{0}^{I}\frac{\partial\Gamma^{IJ}\left(\psi\right)}{\partial\psi^{J}}-\frac{i}{4\mu^{\left(I\right)}}\frac{\partial\Gamma^{IJ}\left(\psi\right)}{\partial\psi^{J}}\frac{\partial\Gamma^{IJ}\left(\psi\right)}{\partial\psi^{J}}-\frac{i}{2}\frac{\partial}{\partial\psi^{I}}\left[T^{IJ}\left(\psi\right)\frac{\partial F\left(\psi\right)}{\partial\psi^{J}}\right]\right\} ,\\
\\
 & \beta^{J}\left(\mu,\dot{\psi},\psi\right)\overset{\mathrm{def}}{=}\sum_{I=1}^{n}\left[\frac{\Gamma^{IJ}\left(\psi\right)}{2\mu^{\left(I\right)}}\frac{\partial}{\partial\psi^{H}}\Gamma^{IH}\left(\psi\right)+ig_{0}^{I}\Gamma^{IJ}\left(\psi\right)+i\delta^{IJ}\left(\dot{\psi}^{I}-T^{IJ}\left(\psi\right)\frac{\partial F\left(\psi\right)}{\partial\psi^{J}}\right)\right],\\
\\
 & \tau^{JH}\left(\lambda,\mu,\psi\right)\overset{\mathrm{def}}{=}\frac{\delta^{JH}}{4\lambda^{\left(J\right)}}+\sum_{I=1}^{n}\frac{\Gamma^{IJ}\left(\psi\right)\Gamma^{IH}\left(\psi\right)}{4\mu^{\left(I\right)}}.
\end{cases}\label{eq:J.0016.kernel.Leibniz}
\end{equation}

The only qualitative implication of having a Leibniz system in (\ref{eq:J.0016.kernel.Leibniz})
is the calculation of the term $-\frac{i}{2}\frac{\partial}{\partial\psi^{I}}\left[T^{IJ}\left(\psi\right)\frac{\partial F\left(\psi\right)}{\partial\psi^{J}}\right]$:
this is performed as follows:\medskip{}

$-{\displaystyle \frac{i}{2}}{\displaystyle \frac{\partial}{\partial\psi^{I}}}\left[T^{IK}\left(\psi\right){\displaystyle \frac{\partial F\left(\psi\right)}{\partial\psi^{K}}}\right]=$\medskip{}

$=-{\displaystyle \frac{i}{2}}{\displaystyle \frac{\partial T^{IK}\left(\psi\right)}{\partial\psi^{I}}}{\displaystyle \frac{\partial F\left(\psi\right)}{\partial\psi^{K}}}-{\displaystyle \frac{i}{2}}T^{IK}\left(\psi\right){\displaystyle \frac{\partial^{2}F\left(\psi\right)}{\partial\psi^{I}\partial\psi^{K}}}=$\medskip{}

$=-{\displaystyle \frac{i}{2}}{\displaystyle \frac{\partial T^{IK}\left(\psi\right)}{\partial\psi^{I}}}{\displaystyle \frac{\partial F\left(\psi\right)}{\partial\psi^{K}}}-{\displaystyle \frac{i}{2}}T_{S}^{IK}\left(\psi\right){\displaystyle \frac{\partial^{2}F\left(\psi\right)}{\partial\psi^{I}\partial\psi^{K}}},$\medskip{}

\noindent where what really matters in the second addendum is the
symmetric part of the Leibniz tensor, due to the symmetric nature
of the Hessian $\frac{\partial^{2}F\left(\psi\right)}{\partial\psi^{I}\partial\psi^{K}}$.

\end{document}